# Testing SETI Message Designs

Michael W. Busch
*Division of Geological and Planetary Sciences, California Institute of Technology*

Rachel M. Reddick
*Department of Physics, Stanford University*

*Abstract*

*Much work in SETI has focused on detecting radio broadcasts due to extraterrestrial intelligence, but there have been limited efforts to transmit messages over interstellar distances. As a check if such messages can be interpreted once received, we conducted a blind test. One of us coded a 75-kilobit message, which the other then attempted to decipher. The decryption was accurate, supporting the message design as a general structure for communicating with aliens capable of detecting narrow-band radio transmissions.*

## 1. Introduction

The search for extraterrestrial intelligence consists of two parts: conventional receive-only 'passive SETI' and 'active SETI', where signals are transmitted with the goal of an extraterrestrial observing eventually detecting them. Passive SETI is now highly developed, and our ability to detect and identify artificial radio beacons continues to improve (e.g., Siemion et al. 2008). Active SETI has been comparatively neglected, in both theory and implementation.

While the mechanics of transmitting radio signals across interstellar distances are well understood (e.g., NAIC 1975, Zaitsev 2006), there has been little effort spent on ensuring that a transmitted message will be understandable to an alien listener. Possible counter-examples include the Arecibo Message (NAIC 1975) and the Cosmic Call Message transmitted from Evpatoria (Zaitsev 2006). To the best of our knowledge, neither was tested for decipherability.

Regardless of if active SETI is desirable, designing a message that is deliberately easy to interpret with a minimum of additional information is an interesting theoretical exercise. To this end, one of us (Busch) developed a coding scheme and a possible message and provided the other (Reddick) with the encoded data, in a blind test of the effort required for decryption.

## 2. The Test Message

The coder based the encryption scheme on the general-purpose binary languages proposed by several authors (e.g. Freudenthal 1960, McConnell 2001), to avoid the potential for bias inherent in pictorial representations (Vakoch 2000). Blocks of code (in this case, 8-quad words) represent numbers, mathematical operators/verbs, variables/nouns, or delimiters, and are assembled into a series of statements. The message totaled 113960 quads (provided in Supplementary Material), but the content of the message was on average repeated three times, so that the true length is ~75 kilobits.

We assumed that for an interstellar beacon, a watcher would be able to identify the signal as artificial due to its low bandwidth, frequency modulation, and periodic Doppler shifts due to the Earth's rotation and motion around the Sun. Such a signal would be detected by almost all SETI data-analysis programs (e.g., SETI@home, Anderson et al. 2008), provided it was strong enough to be detected and at an appropriate carrier frequency. In addition, the coder assumed the watchers would be able to locate the Sun as the likely source of the message and derive its approximate mass, age, and photospheric composition from spectroscopy, and had at least an equivalent knowledge of mathematics, astronomy, and physics.

The watcher was provided with a version of the message which was missing a randomly selected amount of material from the beginning (10-20% of the total) and ~2% of later quads, to represent the initial detection of the beacon and intermittent instrument downtime. We did compromise the blind nature of this test in one way. The watcher was forewarned that the coding scheme was not one of those used previous active SETI messages, which have relied on images with a prime number of pixels in each row or column (NAIC 1975, Zaitsev 2006).

### 3. Initial Decryption: Pattern Recognition and Mathematics

To decrypt the message, the watcher used mostly pencil-and-paper for analysis and search-and-replace to replace deciphered blocks of the message. The lack of a need for high-power computing reflects the relatively small amount of data in the message. The watcher first noticed that the strings '00000000' and '00010000' occur in the message far more frequently than other strings, that they occur in almost exactly equal numbers, and that the combination 'XXXXXXXX 00010000 00000000 XXXXXXXX' occurs often while the reverse never does. From this, she correctly deduced that '00000000' = '(' and '00010000' = ')' – that is, these strings are delimiters separating individual pieces of code from each other.

After decoding the delimiters, the watcher observed statements such as:

```
( 10000000  01000001  10000000 )
( 10000003  01000001  10000003 )
( 10022133  01000001  10022133 )
( 10000000  01000001  20000000 )
( 10000001  01000100  20000001 )
( 10000001  01000100  10000000 )
( 10031242  01000100  10031243 )
```

and recognized '01000001' = '=' and '01000100' = '≠'. She then determined the notation for integers ('1XXXXXXX' for positive, '2XXXXXXX' for negative). Additional statements illustrated variables ('30XXXXXX'), and assigning values to them ('01111110').

The next set of code the watcher decrypted consisted of fundamental arithmetic: functions for addition, multiplication, subtraction, division, and exponentiation. With division is the symbol for '.', the notation for floating point numbers, and with these a representation of the first 'nouns': '31000001' = e and '31000002' = π. The set of statements defining exponentiation provided additional floating-point numbers, and also the definition of 'i'. As a final test of arithmetic, the coder also included a block coding for the quadratic formula, which also introduced the 'or' operator ('01222210').

To provide additional perspectives on the decryption, we provided the message independently to five undergraduates, who each spent no more than an hour attempting to decode it, again working without any pattern-recognition software. Three of the students correctly identified the delimiters, with two also identifying '=', '≠', and the notation for integers. It would have been preferable for the students to spend additional time on the exercise.

## 4. Defining Physical Constants and 'Nouns'

After defining arithmetic, the messages recommended but not transmitted by McConnell (2001) propose defining a new class of data: images. Those transmitted by Zaitsev (2006) are entirely graphical. With even primitive graphics, it is easy to convey some types of information: geometry, physics, chemical structures, human biology. However, given the brevity of the test message, there was insufficient space for images. Therefore, the next relevant block of the message attempts to define physically relevant quantities by relatively simple formulae.

The nuclear fine structure constant, the gravitational coupling constant, the proton-to-electron mass ratio, and neutron-to-electron mass ratio are dimensionless numbers that should be universally recognized, given that the watcher is proficient in physics. To define a system of units, the message contains a series of formulae relating the various Planck units to each other and to these dimensionless numbers. The watcher found this transition from mathematics to physics to be the most difficult portion of decryption.

After recognizing the units, the watcher found the remaining portions of the message readily understandable. The proton, neutron, and electron are defined as nouns equal to statements providing their masses, charges, and, for the proton, charge radius. A partial chart of the nuclides defines the stable isotopes of hydrogen, helium, carbon, nitrogen, oxygen, and neon and the most common isotopes of sodium, magnesium, silicon, sulfur, argon, and iron, providing their masses and number of protons, neutrons, and electrons. For hydrogen, it also includes the radius, nuclear and electric binding energies, and a selection of spectral lines. Given the nouns for the atomic species, chemical definitions follow naturally. In a slight subversion of the intended decryption, the watcher recognized the chemical formulae for molecular hydrogen, water, and oxygen, and used them to more rapidly decrypt this section of the message.

At this point, almost all of the symbols and over 75% of the content of the message had been decrypted. While we have described the decryption in a linear fashion, the message is not structured this way. Individual blocks of code (series of statements) are separated from each other and repeated a varying number of times. For example, the blocks defining =, ≠, and + are given five times, while the chart of the nuclides and the chemical formulae are given only twice. Overall, the message is three times longer than the individual blocks of code. This redundancy serves the obvious purpose of allowing the message to be decrypted regardless of when the watcher started observing.

The last two blocks of code contain the truly interesting information. One block defines a new noun '31130000', which describes something with a mass of $1.989e30$ kg, a temperature of 5778 K, a radius of $6.955e8$ m, and a power of $3.846e26$ W. The block also describes the mass in terms of fractions of hydrogen; deuterium; the isotopes of helium, carbon, nitrogen, oxygen, and neon; and amounts of iron, sulfur, silicon, and magnesium. Together, this composition information defines the Sun, which would presumably be the subject of scrutiny when it is in the same direction as the source of the message.  The watcher did know the composition of the Sun beforehand, which slightly compromises the results.

The Sun block also refers to eight subsections of the last block, which provide masses, radii, temperatures, and a set of distances and times that follow Kepler's third law.  These blocks define the planets.  One block contains two sets of composition information: one typical of rocky terrestrial planets and one a list of gases dominated by oxygen.  In addition to defining the noun for 'atmosphere', this indicates certain chemical disequilibria on the object.  Its orbit also matches the Doppler behavior of the beacon.  So the message defines the Earth and provides a minimalist description of terrestrial life.  The watcher recognized all of this and accurately decoded it after a total of approximately twelve hours of work.

## 5. Comparison to Other Message Designs

As a comparison, the watcher was also provided with one of the Cosmic Call messages (Zaitsev 2006).  Deciphering this message was trivial, since it is a pictorial message based on a grid and the watcher merely had to adjust the width of the display window of her text editor.  As mentioned above, the coder used a general-purpose language rather than a pictorial representation to avoid such biases.  There is a more relevant comparison between the two message designs, however.

The information contained in our test message is roughly equivalent to that contained in the first twelve images of the 2002 Cosmic Call message.  These images total 193 kilobits, or ~2.5x the length of our message.  In addition to avoiding a possible human bias towards interpreting images, constructed languages convey information more densely than purely pictorial messages.

## 6. Conclusions

In a sharply limited number of bits and assuming only a common knowledge of radio and stellar astronomy and the physics and mathematics required to build a radio telescope, we can establish a common vocabulary and describe the solar system in considerable detail. Tentatively, those who receive it can reliably decrypt such a message. Finally, a general-purpose constructed language is significantly more information-dense than a series of prime-number-by-prime-number images, an advantage for beacons to be detected over large volumes.

We ask two additional questions. How much data is required for a blind decryption of more complicated ideas? And, given that there are no technical or apparent theoretical limitations on communicating intelligibly with unknown extraterrestrial watchers, should active SETI be developed on a large scale?

**Acknowledgements**

We thank C.J. White and B.R. Lawrence of Caltech and three anonymous contributors for their efforts in decrypting the test message.

**References:**

Anderson, D.P., Cobb, J., Korpela, E., Lebofsky, M., Werthimer, D., 2002, *SETI@home: an experiment in public-resource computing*, ACM Communications, **45**, 56-61.

Freudental, H., 1960, *Lincos: design of a language for cosmic intercourse*, North Holland Publishing, Amsterdam.

McConnell, B.S., 2001, *Beyond contact: a guide to SETI and communicating with alien civilizations,* O'Reilly Media, Sebastopol, CA.

NAIC – the staff of the National Astronomy and Ionosphere Center, 1975, *The Arecibo message of November 1974,* Icarus, **26**, 462-466.

Siemion, A., and 12 colleagues, 2009, *New SETI sky surveys for radio pulses,*

http://arxiv.org/abs/0811.3046, **to appear in Acta Astronautica.**

Vakoch, D.A., 2000, *The conventionality of pictorial representation in interstellar messages*, Acta Astrononautica, **46**, 733-736.

Zaitsev, A., 2006, *Messaging to extra-terrestrial intelligence*, arXiv:physics/0610031v1

# Supplementary Material 1: Test Message, Decoded

```
( ___0  =  ___0 )
( ___1  =  ___1 )
( ___2  =  ___2 )
( ___3  =  ___3 )
( _671  =  _671 )
( ___0  =  -__0 )
( ___1  ≠  -__1 )
( ___1  ≠  ___0 )
( _870  =  _870 )
( _870  ≠  _871 )

( ___2  >=  ___1 )
( ___1  >=  ___1 )
( __17  >=  ___1 )
( _341  >=  _341 )
( 1064  >=  _341 )
( ___1  <=  ___1 )
( ___0  <=  ___1 )
( -_17  <=  ___1 )
( -_55  <=  ___0 )
( __65  <=  __65 )
( -135  <=  ___0 )

( + ( ___0 , ___0 ) = ___0 )
( + ( ___0 , ___1 ) = ___1 )
( + ( ___0 , _987 ) = _987 )
( + ( ___1 , ___1 ) = ___2 )
( + ( __54 , ___1 ) = __55 )
( + ( _159 , _651 ) = _810 )
( + ( ___2 , ___2 ) = ___4 )
( + ( ___2 , -__2 ) = ___0 )
( + ( + ( ___5 , __41 ) , ___6 ) = __52 )
( + ( + ( __41 , ___6 ) , ___5 ) = __52 )
( + ( _563 , _741 ) = + ( _741 , _563 ) )

( ( x ⇐ ___1 )
  ( x = ___1 )
  ( + ( x , ___1 ) = ___2 )
  ( x ≠ _105 )
)
( ( x ⇐ _105 )
  ( x ≠ ___1 )
  ( + ( x , ___1 ) = _106 )
  ( x = _105 )
)
( ( x ⇐ _561 )
  ( y ⇐ x )
  ( y = _561 )
  ( + ( y , -_11 ) = _550 )
  ( y ⇐ + ( x , 50 ) )
  ( y = _611 )
)

( * ( ___0 , ___0 ) = ___0 )
( * ( ___1 , ___0 ) = ___0 )
```

```
( * ( ___1 , ___2 ) = ___2 )
( * ( ___2 , ___2 ) = ___4 )
( * ( ___3 , ___4 ) = __12 )
( * ( x , ___0 ) = ___0 )
( * ( x , ___2 ) = + ( x , x ) )
( * ( x , ___2 ) = * ( ___2 , x ) )
( * ( x , y ) = * ( y , x ) )
( * ( + ( x , y ) , z ) = + ( * ( x , z ) , * ( y , z ) ) )
( * ( * ( x , y ) , z ) = * ( * ( x , z) , y ) )

( ___2 >= ___1 )
( ___1 >= ___1 )
( __17 >= ___1 )
( _341 >= _341 )
( 1064 >= _341 )
( ___1 <= ___1 )
( ___0 <= ___1 )
( -_17 <= ___1 )
( -_55 <= ___0 )
( __65 <= __65 )
( -135 <= ___0 )

( - ( x , y ) = + ( x , * ( -__1 , y ) ) )
( - ( ___4 , -__5 ) = 9 )
( - ( __14 , ___9 ) = 5 )
( - ( - ( x , y ) , z ) = - ( x , + ( y , z ) ) )
( - ( -__67 , __71 ) = -_138 )

( / ( __4 , __4 ) = __1 )
( / ( __4 , __2 ) = __2 )
( / ( * ( x , y ) , y ) = x )
( / ( * ( x , y ) , x ) = y )
( / ( / ( x , y ) , z ) = / ( x , * ( y , z ) ) )
( / ( / ( x , y ) , / ( z , w ) ) = / ( * ( x , w ) , * ( y , z ) ) )
( / ( __63 , __9 ) = __7 )
( / ( _125 , _25 ) = __5 )
( / ( ___3 , __2 ) = __1 . __5 )
( / ( ___1 , __3 ) = __0 . _333 )
( / ( __15 , __4 ) = __3 . _75 )
( / ( ___8 , __7 ) = __1 . 1/7 )

( * ( ___1 . _3333 , _33 ) = _44 )
( * ( 36 , ___6 . _1/6 ) = 222 )
( + ( ___2 . _3333 , __5 . _6667 ) = __8 )
( + ( ___7 . _3/14 , _11 . _4/14 ) = _18 . 5 )

( ^ ( _1 , x ) = 1 )
( ^ ( _4 , 0 . 5 ) = 2 )
( ^ ( _9 , 0 . 5 ) = 3 )
( ^ ( * ( x , x ) , 0. 5 ) = x )
( ^ ( x , 2 ) = * ( x , x ) )
( ^ ( x , 3 ) = * ( * ( x , x ) , x ) )
( ^ ( ^ ( x , y ) , z ) = ^ ( x , * ( y , z ) ) )
( ^ ( _3 , _4 ) = _81 )
( ^ ( _4 , 1 . 5 ) = 8 )
( ^ ( _2 , 0 . 5 ) = 1 . 414 )
( ^ ( _5 , 0 . 5 ) = 2 . 2 )
```

```
( ^ ( -_1 , 0 . 5 ) = i ( 1 ) )
( ^ ( -_4 , 0 . 5 ) = i ( 2 ) )
( * ( i ( 1 ) , i ( 1 ) ) = -__1 )
( * ( + ( x , i ( y ) ) , + ( z , i ( w ) ) ) =
    + ( * ( x , z ) - * ( y , w ) , i ( + ( * ( y , z ) ,
      * ( x , w ) ) ) ) )

( ( ^ ( x , 2 ) ⇐ 1 )
  ( x = 1 ) || (x = -__1)
)

( ( * ( + ( x , 1 ) , + ( x , -__1 ) ) ⇐ 0 )
  ( x = 1 ) || (x = -__1)
)

( ( y ⇐ 0 )
  ( x = / ( + ( * ( -___1 , b ) , ^ ( - ( * ( b , b ) , * ( 4 ,
          * ( a , c )) , 0 . 5 ) ) , * ( 2 , a ) ) )
  ||
  ( x = / ( - ( * ( -___1 , b ) , ^ ( - ( * ( b , b ) , * ( 4 ,
          * ( a , c )) , 0 . 5 ) ) , * ( 2 , a ) ) )
  ( y = + ( * ( a , * ( x , x ) ) , + ( * ( b , x ) , c ) ) )
)

( ^ ( x , * ( -__1 , y ) ) = / ( 1 , ^ ( x , y ) ) )
( ^ ( i ( 1 ) , 0 . 5 ) = + ( ^ ( 2 , -0 . 5 ) , i ( ^ ( 2 ,
    -0 . 5 ) ) ) )
( ^ ( e , i ( π ) ) = -___1 )
( e = 2 . 78183 )
( π = 3 . 14159 )
( ^ ( e , i ( / ( π , 2 ) ) ) = i ( -___1 ) )
( ^ ( e , 2 ) = 7 . 38906 )
( ^ ( π , e ) = 22 . 45916 )

( ___0 = ___0 )
( ___1 = ___1 )
( ___2 = ___2 )
( ___3 = ___3 )
( _671 = _671 )
( ___0 = -__0 )
( ___1 ≠ -__1 )
( ___1 ≠ ___0 )
( _870 = _870 )
( _870 ≠ _871 )

( * ( ___0 , ___0 ) = ___0 )
( * ( ___1 , ___0 ) = ___0 )
( * ( ___1 , ___2 ) = ___2 )
( * ( ___2 , ___2 ) = ___4 )
( * ( ___3 , ___4 ) = __12 )
( * ( x , ___0 ) = ___0)
( * ( x , ___2 ) = + ( x , x ) )
( * ( x , ___2 ) = * ( ___2 , x ) )
( * ( x , y ) = * ( y , x ) )
( * ( + ( x , y ) , z ) = + ( * ( x , z ) , * ( y , z ) ) )
( * ( * ( x , y ) , z ) = * ( * ( x , z) , y ) )
```

```
( ___0 = ___0 )
( ___1 = ___1 )
( ___2 = ___2 )
( ___3 = ___3 )
( _671 = _671 )
( ___0 = -__0 )
( ___1 ≠ -__1 )
( ___1 ≠ ___0 )
( _870 = _870 )
( _870 ≠ _871 )

( + ( ___0 , ___0 ) = ___0 )
( + ( ___0 , ___1 ) = ___1 )
( + ( ___0 , _987 ) = _987 )
( + ( ___1 , ___1 ) = ___2 )
( + ( __54 , ___1 ) = __55 )
( + ( _159 , _651 ) = _810 )
( + ( ___2 , ___2 ) = ___4 )
( + ( ___2 , -__2 ) = ___0 )
( + ( + ( ___5 , __41 ) , ___6 ) = __52 )
( + ( + ( __41 , ___6 ) , ___5 ) = __52 )
( + ( _563 , _741 ) = + ( _741 , _563 ) )

( / ( __4 , __4 ) = __1 )
( / ( __4 , __2 ) = __2 )
( / ( * ( x , y ) , y ) = x )
( / ( * ( x , y ) , x ) = y )
( / ( / ( x , y ) , z ) = / ( x , * ( y , z ) ) )
( / ( / ( x , y ) , / ( z , w ) ) = / ( * ( x , w ) , * ( y
    , z ) ) )
( / ( __63 , __9 ) = __7 )
( / ( _125 , _25 ) = __5 )
( / ( ___3 , __2 ) = __1 . __5 )
( / ( ___1 , __3 ) = __0 . _333 )
( / ( __15 , __4 ) = __3 . _75 )
( / ( ___8 , __7 ) = __1 . 1/7 )

( * ( ___1 . _3333 , _33 ) = _44 )
( * ( 36 , ___6 . _1/6 ) = 222 )
( + ( ___2 . _3333 , __5 . _6667 ) = __8 )
( + ( ___7 . _3/14 , _11 . _4/14 ) = _18 . 5 )

( ( x ⇐ ___1 )
  ( x = ___1 )
  ( + ( x , ___1 ) = ___2 )
  ( x ≠ _105 )
)
( ( x ⇐ _105 )
  ( x ≠ ___1 )
  ( + ( x , ___1 ) = _106 )
  ( x = _105 )
)
( ( x ⇐ _561 )
  ( y ⇐ x )
  ( y = _561 )
  ( + ( y , -_11 ) = _550 )
  ( y ⇐ + ( x , 50 ) )
```

```
    ( y = _611 )
)

( * ( ___1 . _3333 , _33 ) = _44 )
( * ( 36 , ___6 . _1/6 ) = 222 )
( + ( ___2 . _3333 , __5 . _6667 ) = __8 )
( + ( ___7 . _3/14 , _11 . _4/14 ) = _18 . 5 )

( ^ ( _1 , x ) = 1 )
( ^ ( _4 , 0 . 5 ) = 2 )
( ^ ( _9 , 0 . 5 ) = 3 )
( ^ ( * ( x , x ) , 0 . 5 ) = x )
( ^ ( x , 2 ) = * ( x , x ) )
( ^ ( x , 3 ) = * ( * ( x , x ) , x ) )
( ^ ( ^ ( x , y ) , z ) = ^ ( x , * ( y , z ) ) )
( ^ ( _3 , _4 ) = _81 )
( ^ ( _4 , 1 . 5 ) = 8 )
( ^ ( _2 , 0 . 5 ) = 1 . 414 )
( ^ ( _5 , 0 . 5 ) = 2 . 2 )

( ^ ( -_1 , 0 . 5 ) = i ( 1 ) )
( ^ ( -_4 , 0 . 5 ) = i ( 2 ) )
( * ( i ( 1 ) , i ( 1 ) ) = -__1 )
( * ( + ( x , i ( y ) ) , + ( z , i ( w ) ) ) =
     + ( * ( x , z ) - * ( y , w ) , i ( + ( * ( y , z ) ,
        * ( x , w ) ) ) ) )

( ( ^ ( x , 2 ) ⇐ 1 )
  ( x = 1 ) || (x = -__1)
)

( ( * ( + ( x , 1 ) , + ( x , -__1 ) ) ⇐ 0 )
  ( x = 1 ) || (x = -__1)
)

( ( y ⇐ 0 )
  ( x = / ( + ( * ( -__1 , b ) , ^ ( - ( * ( b , b ) , * ( 4 ,
          * ( a , c )) , 0 . 5 ) ) , * ( 2 , a ) ) )
  ||
  ( x = / ( - ( * ( -__1 , b ) , ^ ( - ( * ( b , b ) , * ( 4 ,
          * ( a , c )) , 0 . 5 ) ) , * ( 2 , a ) ) )
  ( y = + ( * ( a , * ( x , x ) ) , + ( * ( b , x ) , c ) ) )
)

( ___0 = ___0 )
( ___1 = ___1 )
( ___2 = ___2 )
( ___3 = ___3 )
( _671 = _671 )
( ___0 = -__0 )
( ___1 ≠ -__1 )
( ___1 ≠ ___0 )
( _870 = _870 )
( _870 ≠ _871 )

( - ( x , y ) = + ( x , * ( -__1 , y ) ) )
( - ( ___4 , -__5 ) = 9 )
( - ( __14 , ___9 ) = 5 )
```

```
( - ( - ( x , y ) , z ) = - ( x , + ( y , z ) ) )
( - ( -__67 , __71 ) = -_138 )

( ^ ( x , * ( -__1 , y ) ) = / ( 1 , ^ ( x , y ) ) )
( ^ ( i ( 1 ) , 0 . 5 ) = + ( ^ ( 2 , -0 . 5 ) , i ( ^ ( 2 ,
    -0 . 5 ) ) ) )
( ^ ( e , i ( π ) ) = -__1 )
( e = 2 . 78183 )
( π = 3 . 14159 )
( ^ ( e , i ( / ( π , 2 ) ) ) = i ( -__1 ) )
( ^ ( e , 2 ) = 7 . 38906 )
( ^ ( π , e ) = 22 . 45916 )

( α = * ( 7 . 29735 , 0 . 001 ) )
( α = / ( 1 , 137 . 035999 ) )
( μ = 1836 . 15267 )
( αg = * ( 1 . 7518 , ^ ( 10 , -_45 ) ) )
( m_e = * ( ^ ( αg , 0 . 5 ) , m_pl ) )
( m_p = * ( μ , m_e ) )
( m_p = * ( 1836 . 15267 , m_e ) )
( m_n = * ( 1838 . 68366 , m_e ) )
( m_n = * ( 1 . 00138 , m_p ) )
( m_p = * ( * ( 7 . 68513 , ^ ( 10 , -_20 ) ) , m_pl ) )
( m_pl = ^ ( * ( h , / ( c , G ) ) , 0 . 5 ) )
( α = / ( * ( q , q ) , * ( h_b , * ( c , * ( 4 , * ( π , e_0 ) )
    ) ) ) )
( l_pl = ^ ( / ( * ( h_b , G ) , ^ ( c , 3 ) ) , 0 . 5 ) )
( l_pl = / ( * ( 2 , * ( G , m_pl ) ) , ^ ( c , 2 ) ) )
( q_pl = * ( m_pl , * ( 2 , * ( π , ^ ( * ( G , e_0 ) , 0 . 5 ) )
    ) ) )
( q_pl = * ( 11 . 70624 , q ) )
( e_pl = ^ ( * ( h , / ( ^ ( c , 5 ) , G ) ) , 0 . 5 ) )
( e_pl = * ( m_pl , * ( c , c ) ) )
( T_pl = / ( e_pl , k ) )
( P_pl = / ( ^ ( c , 5 ) , G ) )

( Sun =
  (
  ( m = * ( * ( 9 . 1384 , ^ ( 10 , 37 ) ) , m_pl ) )
  ( T = * ( * ( 3 . 9371 , ^ ( 10 , -29 ) ) , T_pl ) )
  ( P = * ( * ( 1 . 0600 , ^ ( 10 , -26 ) ) , P_pl ) )
  ( r = * ( * ( 4 . 3032 , ^ ( 10 , 43 ) ) , l_pl ) )
  ( age = * ( * ( 2 . 675 , ^ ( 10 , 60 ) ) , t_pl ) )
  ( Orbited_by ( Mercury , Venus , Earth , Mars , Jupiter ,
                Saturn , Uranus , Neptune ) )
  ( m = * ( m , + ( * ( 0 . 7345 , 1_H ) , * ( 0 . 2485 , 4_He ) ,
                    * ( 0 . 0077 , 16_O ) , * ( 0 . 0029 , 12_C ) ,
                    * ( 0 . 0015 , 56_Fe ) , * ( 0 . 0011 , 32_S ) ,
                    * ( 0 . 0009 , 14_N ) , * ( 0 . 0007 , 28_Si ) ,
                    * ( 0 . 00011 , 2_H ) , * ( 0 . 0012 , 20_Ne ) ,
                    * ( * ( 3 . 26 , ^ ( 10 , -5 ) ) , 13_C ) ,
                    * ( * ( 1 . 57 , ^ ( 10 , -5 ) ) , 18_O ) ,
                    * ( * ( 2 . 86 , ^ ( 10 , -6 ) ) , 17_O ) ,
                    * ( * ( 3 . 3 , ^ ( 10 , -6 ) ) , 15_N ) ,
                    * ( * ( 1 . 2 , ^ ( 10 , -4 ) ) , 21_Ne ) ,
                    * ( * ( 3 . 5 , ^ ( 10 , -6 ) ) , 22_Ne ) ,
                    0.0008 ) )
  )
```

)

( + ( x , y , z ) = + ( + ( x , y ) , z ) )
( + ( x , y , z , w ) = + ( + ( x , y ) , + ( z , w ) ) )

( proton =
  (
  ( m = m_p )
  ( charge = * ( -1 , q ) )
  ( r = * ( * ( 5 . 4138 , ^ ( 10 , 19 ) ) , l_pl ) )
  )
)

( electron =
  (
  ( m = m_e )
  ( charge = q )
  ( r = 0 )
  )
)

( neutron =
  (
  ( m = m_n )
  ( charge = 0 )
  )
)

( 1_H =
  (
  ( + ( proton , electron ) )
  ( m = * ( * ( 7 . 6815 ) , ^ ( 10 , -20 ) ) , m_pl )
  ( r_ground = * ( * ( 3 . 2744 ) , ^ ( 10 , 24 ) ) , l_pl )
  ( e_binding = * ( * ( 1 . 1139 ) , ^ ( 10 , -27 ) ) , e_pl )
  ( n_binding = 0 )
  ( Lines = ( * ( * ( 1 . 30587 , ^ ( 10 , 34 ) ) , l_pl ) ,
              ( * ( * ( 4 . 06063 , ^ ( 10 , 28 ) ) , l_pl ) ,
              ( * ( * ( 3 . 00758 , ^ ( 10 , 28 ) ) , l_pl ) ,
              ( * ( * ( 2 . 68584 , ^ ( 10 , 28 ) ) , l_pl ) ,
              ( * ( * ( 2 . 53797 , ^ ( 10 , 28 ) ) , l_pl )
              ) )
  )
)

( 2_H =
  (
  ( + ( + ( proton , neutron ) , electron ) )
  ( m = * ( * ( 1 . 5347 ) , ^ ( 10 , -19 ) ) , m_pl )
  ( n_binding = * ( * ( 1 . 82034 ) , ^ ( 10 , -22 ) ) , e_pl )
  )
)

( 3_He =
  (
  ( + ( + ( * ( 2 , proton ) , neutron ) , * ( 2 , electron ) ) )
  ( m = * ( * ( 2 . 3011 ) , ^ ( 10 , -19 ) ) , m_pl )
  )
)

```
( 4_He =
  (
  ( + ( + ( * ( 2 , proton ) , * ( 2 , neutron ) , * ( 2 ,
    electron ) ) )
  ( m = * ( * ( 3 . 0538 ) , ^ ( 10 , -19 ) ) , m_pl )
  )
)

( 12_C =
  (
  ( + ( + ( * ( 6 , proton ) , * ( 6 , neutron ) , * ( 6 ,
    electron ) ) )
  ( m = * ( * ( 9 . 1555 , ^ ( 10 , -19 ) ) , m_pl ) )
  ( n_binding = * ( * ( 7 . 5487 , ^ ( 10 , -21 ) ) , e_pl ) )
  )
)

( 13_C =
  (
  ( + ( + ( * ( 6 , proton ) , * ( 7 , neutron ) , * ( 6 ,
    electron ) ) )
  ( m = * ( * ( 9 . 9211 , ^ ( 10 , -19 ) ) , m_pl ) )
  )
)

( 14_N =
  (
  ( + ( + ( * ( 7 , proton ) , * (7, neutron ) , * ( 7 ,
    electron ) ) )
  ( m = * ( * ( 1 . 06838 , ^ ( 10 , -18 ) ) , m_pl ) )
  )
)

( 15_N =
  (
  ( + ( + ( * ( 7 , proton ) , * (8, neutron ) , * ( 7 ,
    electron ) ) )
  ( m = * ( * ( 1 . 14445 , ^ ( 10 , -18 ) ) , m_pl ) )
  )
)

( 16_O =
  (
  ( + ( + ( * ( 8 , proton ) , * ( 8 , neutron ) , * ( 8 ,
    electron ) ) )
  ( m = * ( * ( 1 . 22035 , ^ ( 10 , -18 ) ) , m_pl ) )
  )
)

( 17_O =
  (
  ( + ( + ( * ( 8 , proton ) , * ( 9 , neutron ) , * ( 8 ,
    electron ) ) )
  ( m = * ( * ( 1 . 29703 , ^ ( 10 , -18 ) ) , m_pl ) )
  )
)
```

```
( 18_O =
  (
  ( + ( + ( * ( 8 , proton ) , * ( 10 , neutron ) , * ( 8 ,
    electron ) ) )
  ( m = * ( * ( 1 . 37327 , ^ ( 10 , -18 ) ) , m_pl ) )
  )
)

( 20_Ne =
  (
  ( + ( + ( * ( 10 , proton ) , * ( 10 , neutron ) , * ( 10 ,
    electron ) ) )
  ( m = * ( * ( 1 . 52535 , ^ ( 10 , -18 ) ) , m_pl ) )
  )
)

( 21_Ne =
  (
  ( + ( + ( * ( 10 , proton ) , * ( 11 , neutron ) , * ( 10 ,
    electron ) ) )
  ( m = * ( * ( 1 . 60175 , ^ ( 10 , -18 ) ) , m_pl ) )
  )
)

( 22_Ne =
  (
  ( + ( + ( * ( 10 , proton ) , * ( 12 , neutron ) , * ( 10 ,
    electron ) ) )
  ( m = * ( * ( 1 . 67786 , ^ ( 10 , -18 ) ) , m_pl ) )
  )
)

( 23_Na =
  (
  ( + ( + ( * ( 11 , proton ) , * ( 12 , neutron ) , * ( 11 ,
    electron ) ) )
  ( m = * ( * ( 1 . 75403 , ^ ( 10 , -18 ) ) , m_pl ) )
  )
)

( 24_Mg =
  (
  ( + ( + ( * ( 12 , proton ) , * ( 12 , neutron ) , * ( 12 ,
    electron ) ) )
  ( m = * ( * ( 1 . 82997 , ^ ( 10 , -18 ) ) , m_pl ) )
  )
)

( 28_Si =
  (
  ( + ( + ( * ( 14 , proton ) , * ( 14 , neutron ) , * ( 14 ,
    electron ) ) )
  ( m = * ( * ( 2 . 13453 , ^ ( 10 , -18 ) ) , m_pl ) )
  )
)

( 32_S =
  (
```

```
      ( + ( + ( * ( 16 , proton ) , * ( 16 , neutron ) , * ( 16 ,
        electron ) ) )
      ( m = * ( * ( 2 . 43935 , ^ ( 10 , -18 ) ) , m_pl ) )
      )
  )

  ( 40_Ar =
    (
    ( + ( + ( * ( 18 , proton ) , * ( 22 , neutron ) , * ( 18 ,
      electron ) ) )
    ( m = * ( * ( 3 . 04898 , ^ ( 10 , -18 ) ) , m_pl ) )
    )
  )

  ( 56_Fe =
    (
    ( + ( + ( * ( 26 , proton ) , * ( 30 , neutron ) , * ( 26 ,
      electron ) ) )
    ( m = * ( * ( 4 . 26762 , ^ ( 10 , -18 ) ) , m_pl ) )
    )
  )

  ( H2 =
    (
      ( * ( 2 , ( 1_H || 2_H ) ) )
      )
  )

  ( H2O =
    (
      ( + ( * ( 2 , ( 1_H || 2_H ) ) , ( 16_O || 17_O || 18_O ) ) )
      )
  )

  ( O2 =
    (
      ( * ( 2 , ( 16_O || 17_O || 18_O ) ) ) )
      )
  )

  ( O3 =
    (
      ( * ( 3 , ( 16_O || 17_O || 18_O ) ) ) )
      )
  )

  ( N2 =
    (
      ( * ( 2 , ( 14_N || 15_N ) ) ) )
      )
  )

  ( NO =
    (
      ( + ( ( 14_N || 15_N ) , ( 16_O || 17_O || 18_O ) ) ) )
      )
  )
```

```
( CO =
  (
    ( + ( ( 12_C || 13_C ) , ( 16_O || 17_O || 18_O ) ) )
  )
)

( CO2 =
  (
    ( + ( ( 12_C || 13_C ) , * ( 2, ( 16_O || 17_O || 18_O ) ) ) )
  )
)

( SO2 =
  (
    ( + ( 32_S , * ( 2, ( 16_O || 17_O || 18_O ) ) ) )
  )
)

( NH3 =
  (
    ( + ( ( 14_N || 15_N ) , * ( 3 , ( 1_H || 2_H ) ) ) )
  )
)

( CH4 =
  (
    ( + ( ( 12_C || 13_C ) , * ( 4 , ( 1_H || 2_H ) ) ) )
  )
)

( C2H6 =
  (
    ( + ( * ( 2 , ( 12_C || 13_C ) ) , * ( 6 , ( 1_H || 2_H ) ) ) )
  )
)

( NSH5 =
  (
    ( + ( ( 14_N || 15_N ) , 32_S , * ( 5 , ( 1_H || 2_H ) ) ) )
  )
)

( ___0  =  ___0 )
( ___1  =  ___1 )
( ___2  =  ___2 )
( ___3  =  ___3 )
( _671  =  _671 )
( ___0  = -__0 )
( ___1  ≠ -__1 )
( ___1  ≠  ___0 )
( _870  =  _870 )
( _870  ≠  _871 )

( ___2 >= ___1 )
( ___1 >= ___1 )
( __17 >= ___1 )
( _341 >= _341 )
( 1064 >= _341 )
```

```
(    ___1 <=  ___1 )
(    ___0 <=  ___1 )
(   -_17 <=  ___1 )
(   -_55 <=  ___0 )
(    __65 <=  __65 )
(   -135 <=  ___0 )

( + (  ___0 ,  ___0 ) =  ___0 )
( + (  ___0 ,  ___1 ) =  ___1 )
( + (  ___0 ,  _987 ) =  _987 )
( + (  ___1 ,  ___1 ) =  ___2 )
( + (  __54 ,  ___1 ) =  __55 )
( + ( _159 ,  _651 ) =  _810 )
( + (  ___2 ,  ___2 ) =  ___4 )
( + (  ___2 , -__2 ) =  ___0 )
( + ( + (  ___5 ,  __41 ) ,  ___6 ) =  __52 )
( + ( + (  __41 ,  ___6 ) ,  ___5 ) =  __52 )
( + ( _563 ,  _741 ) = + (  _741 ,  _563 ) )

( ( x ⇐  ___1 )
  ( x =  ___1 )
  ( + ( x ,  ___1 ) =  ___2 )
  ( x ≠  _105 )
)
( ( x ⇐  _105 )
  ( x ≠  ___1 )
  ( + ( x ,  ___1 ) =  _106 )
  ( x =  _105 )
)
( ( x ⇐  _561 )
  ( y ⇐ x )
  ( y =  _561 )
  ( + ( y , -_11 ) =  _550 )
  ( y ⇐ + ( x , 50 ) )
  ( y =  _611 )
)

( * (  ___0 ,  ___0 ) =  ___0 )
( * (  ___1 ,  ___0 ) =  ___0 )
( * (  ___1 ,  ___2 ) =  ___2 )
( * (  ___2 ,  ___2 ) =  ___4 )
( * (  ___3 ,  ___4 ) =  __12 )
( * ( x ,  ___0 ) =  ___0 )
( * ( x ,  ___2 ) = + ( x , x ) )
( * ( x ,  ___2 ) = * (  ___2 , x ) )
( * ( x , y ) = * ( y , x ) )
( * ( + ( x , y ) , z ) = + ( * ( x , z ) , * ( y , z ) ) )
( * ( * ( x , y ) , z ) = * ( * ( x , z) , y ) )

( α = * ( 7 . 29735 , 0 . 001 ) )
( α = / ( 1 , 137 . 035999 ) )
( μ = 1836 . 15267 )
( αg = * ( 1 . 7518 , ^ ( 10 , -_45 ) ) )
( m_e = * ( ^ ( αg , 0 . 5 ) , m_pl ) )
( m_p = * ( μ , m_e ) )
( m_p = * ( 1836 . 15267 , m_e ) )
( m_n = * ( 1838 . 68366 , m_e ) )
( m_n = * ( 1 . 00138 , m_p ) )
```

```
( m_p = * ( * ( 7 . 68513 , ^ ( 10 , -_20 ) ) , m_pl ) )
( m_pl = ^ ( * ( h , / ( c , G ) ) , 0 . 5 ) )
( α = / ( * ( q , q ) , * ( h_b , * ( c , * ( 4 , * ( π , e_0 ) )
    ) ) ) ) )
( l_pl = ^ ( / ( * ( h_b , G ) , ^ ( c , 3 ) ) , 0 . 5 ) )
( l_pl = / ( * ( 2 , * ( G , m_pl ) ) , ^ ( c , 2 ) ) )
( q_pl = * ( m_pl , * ( 2 , * ( π , ^ ( * ( G , e_0 ) , 0 . 5 ) )
    ) ) )
( q_pl = * ( 11 . 70624 , q ) )
( e_pl = ^ ( * ( h , / ( ^ ( c , 5 ) , G ) ) , 0 . 5 ) )
( e_pl = * ( m_pl , * ( c , c ) ) )
( t_pl = / ( l_pl , c ) )
( t_pl = ^ ( / ( * ( h_b , G ) , ^ ( c , 5 ) ) , 0 . 5 ) )

( - ( x , y ) = + ( x , * ( -__1 , y ) ) )
( - ( ___4 , -__5 ) = 9 )
( - ( __14 , ___9 ) = 5 )
( - ( - ( x , y ) , z ) = - ( x , + ( y , z ) ) )
( - ( -__67 , __71 ) = -_138 )

( / ( __4 , __4 ) = __1 )
( / ( __4 , __2 ) = __2 )
( / ( * ( x , y ) , y ) = x )
( / ( * ( x , y ) , x ) = y )
( / ( / ( x , y ) , z ) = / ( x , * ( y , z ) ) )
( / ( / ( x , y ) , / ( z , w ) ) = / ( * ( x , w ) , * ( y
    , z ) ) )
( / ( __63 , __9 ) = __7 )
( / ( _125 , _25 ) = __5 )
( / ( ___3 , __2 ) = __1 . __5 )
( / ( ___1 , __3 ) = __0 . _333 )
( / ( __15 , __4 ) = __3 . _75 )
( / ( ___8 , __7 ) = __1 . 1/7 )

( + ( x , y , z ) = + ( + ( x , y ) , z ) )
( + ( x , y , z , w ) = + ( + ( x , y ) , + ( z , w ) ) )

( * ( ___1 . _3333 , _33 ) = _44 )
( * ( 36 , ___6 . _1/6 ) = 222 )
( + ( ___2 . _3333 , __5 . _6667 ) = __8 )
( + ( ___7 . _3/14 , _11 . _4/14 ) = _18 . 5 )

( * ( ___1 . _3333 , _33 ) = _44 )
( * ( 36 , ___6 . _1/6 ) = 222 )
( + ( ___2 . _3333 , __5 . _6667 ) = __8 )
( + ( ___7 . _3/14 , _11 . _4/14 ) = _18 . 5 )

( ^ ( _1 , x ) = 1 )
( ^ ( _4 , 0 . 5 ) = 2 )
( ^ ( _9 , 0 . 5 ) = 3 )
( ^ ( * ( x , x ) , 0 . 5 ) = x )
( ^ ( x , 2 ) = * ( x , x ) )
( ^ ( x , 3 ) = * ( * ( x , x ) , x ) )
( ^ ( ^ ( x , y ) , z ) = ^ ( x , * ( y , z ) ) )
( ^ ( _3 , _4 ) = _81 )
( ^ ( _4 , 1 . 5 ) = 8 )
( ^ ( _2 , 0 . 5 ) = 1 . 414 )
( ^ ( _5 , 0 . 5 ) = 2 . 2 )
```

```
( ^ ( -_1 , 0 . 5 ) = i ( 1 ) )
( ^ ( -_4 , 0 . 5 ) = i ( 2 ) )
( * ( i ( 1 ) , i ( 1 ) ) = -__1 )
( * ( + ( x , i ( y ) ) , + ( z , i ( w ) ) ) =
    + ( * ( x , z ) - * ( y , w ) , i ( + ( * ( y , z ) ,
       * ( x , w ) ) ) ) )

( Sun =
 (
  ( m = * ( * ( 9 . 1384 , ^ ( 10 , 37 ) ) , m_pl ) )
  ( T = * ( * ( 3 . 9371 , ^ ( 10 , -29 ) ) , T_pl ) )
  ( P = * ( * ( 1 . 0600 , ^ ( 10 , -26 ) ) , P_pl ) )
  ( r = * ( * ( 4 . 3032 , ^ ( 10 , 43 ) ) , l_pl ) )
  ( age = * ( * ( 2 . 675 , ^ ( 10 , 60 ) ) , t_pl ) )
  ( Orbited_by ( Mercury , Venus , Earth , Mars , Jupiter ,
               Saturn , Uranus , Neptune ) )
 ( m = * ( m , + ( * ( 0 . 7345 , 1_H ) , * ( 0 . 2485 , 4_He ) ,
                 * ( 0 . 0077 , 16_O ) , * ( 0 . 0029 , 12_C ) ,
                 * ( 0 . 0015 , 56_Fe ) , * ( 0 . 0011 , 32_S ) ,
                 * ( 0 . 0009 , 14_N ) , * ( 0 . 0007 , 28_Si ) ,
                 * ( 0 . 00011 , 2_H ) , * ( 0 . 0012 , 20_Ne ) ,
                 * ( * ( 3 . 26 , ^ ( 10 , -5 ) ) , 13_C ) ,
                 * ( * ( 1 . 57 , ^ ( 10 , -5 ) ) , 18_O ) ,
                 * ( * ( 2 . 86 , ^ ( 10 , -6 ) ) , 17_O ) ,
                 * ( * ( 3 . 3 , ^ ( 10 , -6 ) ) , 15_N ) ,
                 * ( * ( 1 . 2 , ^ ( 10 , -4 ) ) , 21_Ne ) ,
                 * ( * ( 3 . 5 , ^ ( 10 , -6 ) ) , 22_Ne ) ,
                 0.0008 ) )
 )
)

( Mercury =
  (
    ( m = * ( * ( 1 . 51725 , ^ ( 10 , 31 ) ) , m_pl ) )
    ( T <= * ( * ( 4 . 94076 , ^ ( 10 , -30 ) ) , T_pl ) )
    ( T >= * ( * ( 5 . 64659 , ^ ( 10 , -31 ) ) , T_pl ) )
    ( r = * ( * ( 1 . 50948 , ^ ( 10 , 41 ) ) , l_pl ) )
    ( semimaj_a = * ( * ( 3 . 58293 , ^ ( 10 , -35 ) ) , l_pl ) )
    ( ecc = 0 . 20563 )
    ( year = * ( * ( 1 . 40979 , ^ ( 10 , 50 ) ) , t_pl ) )
    ( day = * ( * ( 9 . 39861 , ^ ( 10 , 49 ) ) , t_pl ) )
    ( Orbits ( Sun ) )
    ( atmosphere =
      (
        ( m = * ( * ( 9 . 21 , ^ ( 10 , 10 ) ) , m_pl ) )
        ( m = * ( m , + ( * ( 0 . 42 , O2 ) , * ( 0 . 29 , 23_Na ) ,
                        * ( 0 . 22 , H2 ) , * ( 0 . 06 , 4_He ) ,
                        0 . 01 )
             ) )
      )
    )
  )
)

( Venus =
  (
    ( m = * ( * ( 2 . 23714 , ^ ( 10 , 32 ) ) , m_pl ) )
```

```
        ( T = * ( * ( 5 . 1878 , ^ ( 10 , -30 ) ) , T_pl ) )
        ( r = * ( * ( 3 . 7443 , ^ ( 10 , 41 ) ) , l_pl ) )
        ( semimaj_a = * ( * ( 6 . 69505 , ^ ( 10 , 45 ) ) , l_pl ) )
        ( ecc = 0 . 00677 )
        ( year = * ( * ( 3 . 6011 , ^ ( 10 , 50 ) ) , t_pl ) )
        ( day = * ( * ( 3 . 8946 , ^ ( 10 , 50 ) ) , t_pl ) )
        ( Orbits ( Sun ) )
        ( atmosphere =
          (
            ( m = * ( * ( 2 . 2054 , ^ ( 10 , 28 ) ) , m_pl ) )
            ( m = * ( m , + ( * ( 0 . 965 , CO2 ) , * ( 0 . 035 , N2 ) ,
                             * ( * ( 1 . 5 , ^ ( 10 , -4 ) ) , SO2 ) ,
                             * ( * ( 7 , ^ ( 10 , -5 ) ) , 40_Ar ) ,
                             * ( * ( 1 . 7 , ^ ( 10 , -5 ) ) , CO ) ,
                             * ( * ( 1 . 2 , ^ ( 10 , -5 ) ) , 4_He ) ,
                             * ( * ( 7 , ^ ( 10 , -6 ) ) , 20_Ne )
                  ) ) )
          )
        )
      )
    )

    ( Earth =
      (
        ( m = * ( * ( 2 . 74494 , ^ ( 10 , 32 ) ) , m_pl ) )
        ( T <= * ( * ( 2 . 33628 , ^ ( 10 , -30 ) ) , T_pl ) )
        ( T >= * ( * ( 1 . 29872 , ^ ( 10 , -30 ) ) , T_pl ) )
        ( r <= * ( * ( 3 . 94623 , ^ ( 10 , 44 ) ) , l_pl ) )
        ( r >= * ( * ( 3 . 93305, ^ ( 10 , 44 ) ) , l_pl ) )
        ( semimaj_a = * ( * ( 9 . 25585 , ^ ( 10 , 45 ) ) , l_pl ) )
        ( ecc = 0 . 01671 )
        ( year = * ( * ( 5 . 85360 , ^ ( 10 , 50 ) ) , t_pl ) )
        ( day = * ( * ( 1 . 59822 , ^ ( 10 , 48 ) ) , t_pl ) )
        ( Orbits ( Sun ) )
        ( Orbited_by ( Moon ) )
        ( m = * ( m , + ( * ( 0 . 29453 , 56_Fe ) ,
                         * ( 0 . 300 , 16_O ) ,
                         * ( 0 . 139 , 28_Si ) , * ( 0 . 110 , 24_Mg ) ,
                         * ( 0 . 028 , 32_S ) , 0 . 147 )
                ) )
        ( atmosphere =
          (
            ( m <= * ( * ( 5 . 14725 , ^ ( 10 , 18 ) ) , m_pl ) )
            ( m >= * ( * ( 5 . 14875 , ^ ( 10 , 18 ) ) , m_pl ) )
          ( m = * ( m , + ( * ( 0 . 77771 , N2 ) , * ( 0 . 20862 , O2 ) ,
                           * ( 0 . 00930 , 40_Ar ) , * ( 0 . 004 , H20 ) ,
                           * ( * ( >= 3 . 83 , ^ ( 10 , -4 ) ) , CO2 ) ,
                           * ( * ( 1 . 811 , ^ ( 10 , -5 ) ) , 20_Ne ) ,
                           * ( * ( 5 . 22 , ^ ( 10 , -6 ) ) , 4_He ) ,
                           * ( * ( >= 1 . 745 , ^ ( 10 , -6 ) ) , CH4 ) ,
                           * ( * ( 3 . 5 , ^ ( 10 , -8 ) ) , O3 ) )
                  ) )
          )
        )
        ( Life )
        ( Humanity )
      )
    )
```

```
( Moon =
  (
    ( m = * ( * ( 3 . 37602 , ^ ( 10 , 30 ) ) , m_pl ) )
    ( T <= * ( * ( 2 . 75271 , ^ ( 10 , -30 ) ) , T_pl ) )
    ( T >= * ( * ( 4 . 94076 , ^ ( 10 , -31 ) ) , T_pl ) )
    ( r >= * ( * ( 1 . 07407 , ^ ( 10 , 41 ) ) , l_pl ) )
    ( r <= * ( * ( 1 . 07541 , ^ ( 10 , 41 ) ) , l_pl ) )
    ( semimaj_a = * ( * ( 2 . 3783 , ^ ( 10 , 43 ) ) , l_pl ) )
    ( ecc = 0 . 0549 )
    ( year = * ( * ( 4 . 37856 , ^ ( 10 , 49 ) ) , t_pl ) )
    ( day = * ( * ( 4 . 37856 , ^ ( 10 , 49 ) ) , t_pl ) )
    ( Orbits ( Earth ) )
    ( atmosphere =
      (
        ( m = * ( * ( 3 . 26601 , ^ ( 10 , 11 ) ) , m_pl ) )
      )
    )
  )
)

( Mars =
  (
    ( m = * ( * ( 2 . 94936 , ^ ( 10 , 31 ) ) , m_pl ) )
    ( T <= * ( * ( 1 . 89161 , ^ ( 10 , -30 ) ) , T_pl ) )
    ( T >= * ( * ( 1 . 31283 , ^ ( 10 , -30 ) ) , T_pl ) )
    ( r <= * ( * ( 2 . 10128 , ^ ( 10 , 41 ) ) , l_pl ) )
    ( r >= * ( * ( 2 . 08891 , ^ ( 10 , 41 ) ) , l_pl ) )
    ( semimaj_a = * ( * ( 1 . 41029 , ^ ( 10 , 46 ) ) , l_pl ) )
    ( ecc = 0 . 09341 )
    ( year = * ( * ( 1 . 10094 , ^ ( 10 , 51 ) ) , t_pl ) )
    ( day = * ( * ( 1 . 6442 , ^ ( 10 , 48 ) ) , t_pl ) )
    ( Orbits ( Sun ) )
    ( atmosphere =
      (
        ( m = * ( * ( 1 . 14866 , ^ ( 10 , 24 ) ) , m_pl ) )
        ( m = * ( m , + ( * ( 0 . 9532 , CO2 ) , * ( 0 . 017 , N2 ) ,
                    * ( 0 . 016 , 40_Ar ) , * ( 0 . 0013 , O2 ) ,
                    * ( 0 . 0007 , CO ) , * ( 0 . 0003 , H2O ) ,
                    * ( * ( 1 . 3 , ^ ( 10 , -4 ) ) , NO ) ,
                    * ( * ( 2 . 5 , ^ ( 10 , -6 ) ) , 20_Ne )
             ) ) )
      )
    )
  )
)

( Jupiter =
  (
    ( m = * ( * ( 8 . 72388 , ^ ( 10 , 34 ) ) , m_pl ) )
    ( T = * ( * ( 1 . 16461 , ^ ( 10 , -30 ) ) , T_pl ) )
    ( r <= * ( * ( 4 . 42332 , ^ ( 10 , 42 ) ) , l_pl ) )
    ( r >= * ( * ( 4 . 13636 , ^ ( 10 , 42 ) ) , l_pl ) )
    ( semimaj_a = * ( * ( 4 . 81699, ^ ( 10 , 46 ) ) , l_pl ) )
    ( ecc = 0 . 04839 )
    ( year = * ( * ( 6 . 94178 , ^ ( 10 , 51 ) ) , t_pl ) )
    ( day = * ( * ( 6 . 62742 , ^ ( 10 , 47 ) ) , t_pl ) )
    ( Orbits ( Sun ) )
```

```
      ( m = * ( m , + ( * ( 0 . 71 , H2 ) , * ( 0 . 24 , 4_He ) ,
                      0.05 ) ) )
  )
)

( Saturn =
  (
    ( m = * ( * ( 2 . 61211 , ^ ( 10 , 34 ) ) , m_pl ) )
    ( T = * ( * ( 9 . 45803 , ^ ( 10 , -31 ) ) , T_pl ) )
    ( r <= * ( * ( 3 . 72887 , ^ ( 10 , 42 ) ) , l_pl ) )
    ( r >= * ( * ( 3 . 36358 , ^ ( 10 , 42 ) ) , l_pl ) )
    ( semimaj_a = * ( * ( 8 . 86897, ^ ( 10 , 46 ) ) , l_pl ) )
    ( ecc = 0 . 05415 )
    ( year = * ( * ( 1 . 73599 , ^ ( 10 , 52 ) ) , t_pl ) )
    ( day = * ( * ( 7 . 11554 , ^ ( 10 , 47 ) ) , t_pl ) )
    ( Orbits ( Sun ) )
  )
)

( Uranus =
  (
    ( m = * ( * ( 3 . 99042 , ^ ( 10 , 33 ) ) , m_pl ) )
    ( T = * ( * ( 5 . 36426 , ^ ( 10 , -31 ) ) , T_pl ) )
    ( r <= * ( * ( 1 . 58137 , ^ ( 10 , 42 ) ) , l_pl ) )
    ( r >= * ( * ( 1 . 54512 , ^ ( 10 , 42 ) ) , l_pl ) )
    ( semimaj_a = * ( * ( 1 . 77985 , ^ ( 10 , 47 ) ) , l_pl ) )
    ( ecc = 0 . 04717 )
    ( year = * ( * ( 4 . 93586 , ^ ( 10 , 52 ) ) , t_pl ) )
    ( day = * ( * ( 1 . 1512 , ^ ( 10 , 48 ) ) , t_pl ) )
    ( Orbits ( Sun ) )
  )
)

( Neptune =
  (
    ( m = * ( * ( 4 . 70677 , ^ ( 10 , 33 ) ) , m_pl ) )
    ( T = * ( * ( 5 . 08193 , ^ ( 10 , -31 ) ) , T_pl ) )
    ( r <= * ( * ( 1 . 53219 , ^ ( 10 , 42 ) ) , l_pl ) )
    ( r >= * ( * ( 1 . 50602 , ^ ( 10 , 42 ) ) , l_pl ) )
    ( semimaj_ax = * ( * ( 2 . 78635 , ^ ( 10 , 47 ) ) , l_pl ) )
    ( ecc = 0 . 00859 )
    ( year = * ( * ( 9 . 64605 , ^ ( 10 , 52 ) ) , t_pl ) )
    ( day = * ( * ( 1 . 07583 , ^ ( 10 , 48 ) ) , t_pl ) )
    ( Orbits ( Sun ) )
    ( m = * ( m , 1 ) )
  )
)

( ( ^ ( x , 2 ) <= 1 )
  ( x = 1 ) || (x = -__1)
)

( ( * ( + ( x , 1 ) , + ( x , -__1 ) ) <= 0 )
  ( x = 1 ) || (x = -__1)
)

( ( y <= 0 )
  ( x = / ( + ( * ( -__1 , b ) , ^ ( - ( * ( b , b ) , * ( 4 ,
```

```
                   * ( a , c ) ) , 0 . 5 ) ) , * ( 2 , a ) ) )
      ||
      ( x = / ( - ( * ( -___1 , b ) , ^ ( - ( * ( b , b ) , * ( 4 ,
                   * ( a , c ) ) , 0 . 5 ) ) , * ( 2 , a ) ) )
      ( y = + ( * ( a , * ( x , x ) ) , + ( * ( b , x ) , c ) ) )
  )

  ( ^ ( x , * ( -___1 , y ) ) = / ( 1 , ^ ( x , y ) ) )
  ( ^ ( i ( 1 ) , 0 . 5 ) = + ( ^ ( 2 , -0 . 5 ) , i ( ^ ( 2 ,
         -0 . 5 ) ) ) )
  ( ^ ( e , i ( π ) ) = -___1 )
  ( e = 2 . 78183 )
  ( π = 3 . 14159 )
  ( ^ ( e , i ( / ( π , 2 ) ) ) = i ( -___1 ) )
  ( ^ ( e , 2 ) = 7 . 38906 )
  ( ^ ( π , e ) = 22 . 45916 )

  ( α = * ( 7 . 29735 , 0 . 001 ) )
  ( α = / ( 1 , 137 . 035999 ) )
  ( μ = 1836 . 15267 )
  ( αg = * ( 1 . 7518 , ^ ( 10 , -___45 ) ) )
  ( m_e = * ( ^ ( αg , 0 . 5 ) , m_pl ) )
  ( m_p = * ( μ , m_e ) )
  ( m_p = * ( 1836 . 15267 , m_e ) )
  ( m_n = * ( 1838 . 68366 , m_e ) )
  ( m_n = * ( 1 . 00138 , m_p ) )
  ( m_p = * ( * ( 7 . 68513 , ^ ( 10 , -___20 ) ) , m_pl ) )
  ( m_pl = ^ ( * ( h , / ( c , G ) ) , 0 . 5 ) )
  ( α = / ( * ( q , q ) , * ( h_b , * ( c , * ( 4 , * ( π , e_0 ) )
         ) ) ) ) )
  ( l_pl = ^ ( / ( * ( h_b , G ) , ^ ( c , 3 ) ) , 0 . 5 )
  ( l_pl = / ( * ( 2 , * ( G , m_pl ) ) , ^ ( c , 2 ) )
  ( q_pl = * ( m_pl , * ( 2 , * ( π , ^ ( * ( G , e_0 ) , 0 . 5 ) )
         ) ) )
  ( q_pl = * ( 11 . 70624 , q ) )
  ( e_pl = ^ ( * ( h , / ( ^ ( c , 5 ) , G ) ) , 0 . 5 ) )
  ( e_pl = * ( m_pl , * ( c , c ) ) )
  ( t_pl = / ( l_pl , c ) )
  ( t_pl = ^ ( / ( * ( h_b , G ) , ^ ( c , 5 ) ) , 0 . 5 )

   ( proton =
     (
     ( m = m_p )
     ( charge = * ( -1 , q ) )
     ( r = * ( * ( 5 . 4138 , ^ ( 10 , 19 ) ) , l_pl ) )
     )
  )

  ( electron =
     (
     ( m = m_e )
     ( charge = q )
     ( r = 0 )
     )
  )

  ( neutron =
     (
```

```
    ( m = m_n )
    ( charge = 0 )
    )
)

( + ( x , y , z ) = + ( + ( x , y ) , z ) )
( + ( x , y , z , w ) = + ( + ( x , y ) , + ( z , w ) ) )

( 1_H =
   (
   ( + ( proton , electron ) )
   ( m = * ( * ( 7 . 6815 ) , ^ ( 10 , -20 ) ) , m_pl )
   ( r_ground = * ( * ( 3 . 2744 ) , ^ ( 10 , 24 ) ) , l_pl )
   ( e_binding = * ( * ( 1 . 1139 ) , ^ ( 10 , -27 ) ) , e_pl )
   ( n_binding = 0 )
   ( Lines = ( * ( * ( 1 . 30587 , ^ ( 10 , 34 ) ) , l_pl ) ,
               ( * ( * ( 4 . 06063 , ^ ( 10 , 28 ) ) , l_pl ) ,
               ( * ( * ( 3 . 00758 , ^ ( 10 , 28 ) ) , l_pl ) ,
               ( * ( * ( 2 . 68584 , ^ ( 10 , 28 ) ) , l_pl ) ,
               ( * ( * ( 2 . 53797 , ^ ( 10 , 28 ) ) , l_pl )
               ) )
   )
)

( 2_H =
   (
   ( + ( + ( proton , neutron ) , electron ) )
   ( m = * ( * ( 1 . 5347 ) , ^ ( 10 , -19 ) ) , m_pl )
   ( n_binding = * ( * ( 1 . 82034 ) , ^ ( 10 , -22 ) ) , e_pl )
   )
)

( 3_He =
   (
   ( + ( + ( * ( 2 , proton ) , neutron ) , * ( 2 , electron ) ) )
   ( m = * ( * ( 2 . 3011 ) , ^ ( 10 , -19 ) ) , m_pl )
   )
)

( 4_He =
   (
   ( + ( + ( * ( 2 , proton ) , * ( 2 , neutron ) , * ( 2 ,
     electron ) ) )
   ( m = * ( * ( 3 . 0538 ) , ^ ( 10 , -19 ) ) , m_pl )
   )
)

( 12_C =
   (
   ( + ( + ( * ( 6 , proton ) , * ( 6 , neutron ) , * ( 6 ,
     electron ) ) )
   ( m = * ( * ( 9 . 1555 , ^ ( 10 , -19 ) ) , m_pl ) )
   ( n_binding = * ( * ( 7 . 5487 , ^ ( 10 , -21 ) ) , e_pl ) )
   )
)

( 13_C =
```

```
(
  ( + ( + ( * ( 6 , proton ) , * ( 7 , neutron ) , * ( 6 ,
    electron ) ) )
  ( m = * ( * ( 9 . 9211 , ^ ( 10 , -19 ) ) , m_pl ) )
  )
)

( 14_N =
  (
  ( + ( + ( * ( 7 , proton ) , * (7, neutron ) , * ( 7 ,
    electron ) ) )
  ( m = * ( * ( 1 . 06838 , ^ ( 10 , -18 ) ) , m_pl ) )
  )
)

( 15_N =
  (
  ( + ( + ( * ( 7 , proton ) , * (8, neutron ) , * ( 7 ,
    electron ) ) )
  ( m = * ( * ( 1 . 14445 , ^ ( 10 , -18 ) ) , m_pl ) )
  )
)

( 16_O =
  (
  ( + ( + ( * ( 8 , proton ) , * ( 8 , neutron ) , * ( 8 ,
    electron ) ) )
  ( m = * ( * ( 1 . 22035 , ^ ( 10 , -18 ) ) , m_pl ) )
  )
)

( 17_O =
  (
  ( + ( + ( * ( 8 , proton ) , * ( 9 , neutron ) , * ( 8 ,
    electron ) ) )
  ( m = * ( * ( 1 . 29703 , ^ ( 10 , -18 ) ) , m_pl ) )
  )
)

( 18_O =
  (
  ( + ( + ( * ( 8 , proton ) , * ( 10 , neutron ) , * ( 8 ,
    electron ) ) )
  ( m = * ( * ( 1 . 37327 , ^ ( 10 , -18 ) ) , m_pl ) )
  )
)

( 20_Ne =
  (
  ( + ( + ( * ( 10 , proton ) , * ( 10 , neutron ) , * ( 10 ,
    electron ) ) )
  ( m = * ( * ( 1 . 52535 , ^ ( 10 , -18 ) ) , m_pl ) )
  )
)

( 21_Ne =
  (
  ( + ( + ( * ( 10 , proton ) , * ( 11 , neutron ) , * ( 10 ,
```

```
    electron ) ) )
  ( m = * ( * ( 1 . 60175 , ^ ( 10 , -18 ) ) , m_pl ) )
  )
)

( 22_Ne =
  (
  ( + ( + ( * ( 10 , proton ) , * ( 12 , neutron ) , * ( 10 ,
    electron ) ) )
  ( m = * ( * ( 1 . 67786 , ^ ( 10 , -18 ) ) , m_pl ) )
  )
)

( 23_Na =
  (
  ( + ( + ( * ( 11 , proton ) , * ( 12 , neutron ) , * ( 11 ,
    electron ) ) )
  ( m = * ( * ( 1 . 75403 , ^ ( 10 , -18 ) ) , m_pl ) )
  )
)

( 24_Mg =
  (
  ( + ( + ( * ( 12 , proton ) , * ( 12 , neutron ) , * ( 12 ,
    electron ) ) )
  ( m = * ( * ( 1 . 82997 , ^ ( 10 , -18 ) ) , m_pl ) )
  )
)

( 28_Si =
  (
  ( + ( + ( * ( 14 , proton ) , * ( 14 , neutron ) , * ( 14 ,
    electron ) ) )
  ( m = * ( * ( 2 . 13453 , ^ ( 10 , -18 ) ) , m_pl ) )
  )
)

( 32_S =
  (
  ( + ( + ( * ( 16 , proton ) , * ( 16 , neutron ) , * ( 16 ,
    electron ) ) )
  ( m = * ( * ( 2 . 43935 , ^ ( 10 , -18 ) ) , m_pl ) )
  )
)

( 40_Ar =
  (
  ( + ( + ( * ( 18 , proton ) , * ( 22 , neutron ) , * ( 18 ,
    electron ) ) )
  ( m = * ( * ( 3 . 04898 , ^ ( 10 , -18 ) ) , m_pl ) )
  )
)

( 56_Fe =
  (
  ( + ( + ( * ( 26 , proton ) , * ( 30 , neutron ) , * ( 26 ,
    electron ) ) )
```

```
          ( m = * ( * ( 4 . 26762 , ^ ( 10 , -18 ) ) , m_pl ) )
        )
    )

    ( ___2 >= ___1 )
    ( ___1 >= ___1 )
    ( __17 >= ___1 )
    ( _341 >= _341 )
    ( 1064 >= _341 )
    ( ___1 <= ___1 )
    ( ___0 <= ___1 )
    ( -_17 <= ___1 )
    ( -_55 <= ___0 )
    ( __65 <= __65 )
    ( -135 <= ___0 )

    ( Mercury =
        (
            ( m = * ( * ( 1 . 51725 , ^ ( 10 , 31 ) ) , m_pl ) )
            ( T <= * ( * ( 4 . 94076 , ^ ( 10 , -30 ) ) , T_pl ) )
            ( T >= * ( * ( 5 . 64659 , ^ ( 10 , -31 ) ) , T_pl ) )
            ( r = * ( * ( 1 . 50948 , ^ ( 10 , 41 ) ) , l_pl ) )
            ( semimaj_a = * ( * ( 3 . 58293 , ^ ( 10 , -35 ) ) , l_pl ) )
            ( ecc = 0 . 20563 )
            ( year = * ( * ( 1 . 40979 , ^ ( 10 , 50 ) ) , t_pl ) )
            ( day = * ( * ( 9 . 39861 , ^ ( 10 , 49 ) ) , t_pl ) )
            ( Orbits ( Sun ) )
            ( atmosphere =
                (
                    ( m = * ( * ( 9 . 21 , ^ ( 10 , 10 ) ) , m_pl ) )
                    ( m = * ( m , + ( * ( 0 . 42 , O2 ) , * ( 0 . 29 , 23_Na ) ,
                                      * ( 0 . 22 , H2 ) , * ( 0 . 06 , 4_He ) ,
                                      0 . 01 )
                            ) )
                )
            )
        )
    )

    ( Venus =
        (
            ( m = * ( * ( 2 . 23714 , ^ ( 10 , 32 ) ) , m_pl ) )
            ( T = * ( * ( 5 . 1878 , ^ ( 10 , -30 ) ) , T_pl ) )
            ( r = * ( * ( 3 . 7443 , ^ ( 10 , 41 ) ) , l_pl ) )
            ( semimaj_a = * ( * ( 6 . 69505 , ^ ( 10 , 45 ) ) , l_pl ) )
            ( ecc = 0 . 00677 )
            ( year = * ( * ( 3 . 6011 , ^ ( 10 , 50 ) ) , t_pl ) )
            ( day = * ( * ( 3 . 8946 , ^ ( 10 , 50 ) ) , t_pl ) )
            ( Orbits ( Sun ) )
            ( atmosphere =
                (
                    ( m = * ( * ( 2 . 2054 , ^ ( 10 , 28 ) ) , m_pl ) )
                    ( m = * ( m , + ( * ( 0 . 965 , CO2 ) , * ( 0 . 035 , N2 ) ,
                                      * ( * ( 1 . 5 , ^ ( 10 , -4 ) ) , SO2 ) ,
                                      * ( * ( 7 , ^ ( 10 , -5 ) ) , 40_Ar ) ,
                                      * ( * ( 1 . 7 , ^ ( 10 , -5 ) ) , CO ) ,
                                      * ( * ( 1 . 2 , ^ ( 10 , -5 ) ) , 4_He ) ,
                                      * ( * ( 7 , ^ ( 10 , -6 ) ) , 20_Ne )
```

```
                ) ) )
             )
          )
       )
    )

    ( Earth =
       (
          ( m = * ( * ( 2 . 74494 , ^ ( 10 , 32 ) ) , m_pl ) )
          ( T <= * ( * ( 2 . 33628 , ^ ( 10 , -30 ) ) , T_pl ) )
          ( T >= * ( * ( 1 . 29872 , ^ ( 10 , -30 ) ) , T_pl ) )
          ( r <= * ( * ( 3 . 94623 , ^ ( 10 , 44 ) ) , l_pl ) )
          ( r >= * ( * ( 3 . 93305, ^ ( 10 , 44 ) ) , l_pl ) )
          ( semimaj_a = * ( * ( 9 . 25585 , ^ ( 10 , 45 ) ) , l_pl ) )
          ( ecc = 0 . 01671 )
          ( year = * ( * ( 5 . 85360 , ^ ( 10 , 50 ) ) , t_pl ) )
          ( day = * ( * ( 1 . 59822 , ^ ( 10 , 48 ) ) , t_pl ) )
          ( Orbits ( Sun ) )
          ( Orbited_by ( Moon ) )
          ( m = * ( m , + ( * ( 0 . 29453 , 56_Fe ) ,
                            * ( 0 . 300 , 16_O ) ,
                            * ( 0 . 139 , 28_Si ) , * ( 0 . 110 , 24_Mg ) ,
                            * ( 0 . 028 , 32_S  ) , 0 . 147 )
                      ) )
          ( atmosphere =
             (
                ( m <= * ( * ( 5 . 14725 , ^ ( 10 , 18 ) ) , m_pl ) )
                ( m >= * ( * ( 5 . 14875 , ^ ( 10 , 18 ) ) , m_pl ) )
          ( m = * ( m , + ( * ( 0 . 77771 , N2 ) , * ( 0 . 20862 , O2 ) ,
                            * ( 0 . 00930 , 40_Ar ) , * ( 0 . 004 , H20 ) ,
                            * ( * ( >= 3 . 83 , ^ ( 10 , -4 ) ) , CO2 ) ,
                            * ( * ( 1 . 811 , ^ ( 10 , -5 ) ) , 20_Ne ) ,
                            * ( * ( 5 . 22 , ^ ( 10 , -6 ) ) , 4_He ) ,
                            * ( * ( >= 1 . 745 , ^ ( 10 , -6 ) ) , CH4 ) ,
                            * ( * ( 3 . 5 , ^ ( 10 , -8 ) ) , O3 ) )
                       ) )
             )
          )
          ( Life )
          ( Humanity )
       )
    )

    ( Moon =
       (
          ( m = * ( * ( 3 . 37602 , ^ ( 10 , 30 ) ) , m_pl ) )
          ( T <= * ( * ( 2 . 75271 , ^ ( 10 , -30 ) ) , T_pl ) )
          ( T >= * ( * ( 4 . 94076 , ^ ( 10 , -31 ) ) , T_pl ) )
          ( r >= * ( * ( 1 . 07407 , ^ ( 10 , 41 ) ) , l_pl ) )
          ( r <= * ( * ( 1 . 07541 , ^ ( 10 , 41 ) ) , l_pl ) )
          ( semimaj_a = * ( * ( 2 . 3783 , ^ ( 10 , 43 ) ) , l_pl ) )
          ( ecc = 0 . 0549 )
          ( year = * ( * ( 4 . 37856 , ^ ( 10 , 49 ) ) , t_pl ) )
          ( day = * ( * ( 4 . 37856 , ^ ( 10 , 49 ) ) , t_pl ) )
          ( Orbits ( Earth ) )
          ( atmosphere =
             (
                ( m = * ( * ( 3 . 26601 , ^ ( 10 , 11 ) ) , m_pl ) )
```

```
          )
        )
      )
    )

    ( Mars =
      (
        ( m = * ( * ( 2 . 94936 , ^ ( 10 , 31 ) ) , m_pl ) )
        ( T <= * ( * ( 1 . 89161 , ^ ( 10 , -30 ) ) , T_pl ) )
        ( T >= * ( * ( 1 . 31283 , ^ ( 10 , -30 ) ) , T_pl ) )
        ( r <= * ( * ( 2 . 10128 , ^ ( 10 , 41 ) ) , l_pl ) )
        ( r >= * ( * ( 2 . 08891 , ^ ( 10 , 41 ) ) , l_pl ) )
        ( semimaj_a = * ( * ( 1 . 41029 , ^ ( 10 , 46 ) ) , l_pl ) )
        ( ecc = 0 . 09341 )
        ( year = * ( * ( 1 . 10094 , ^ ( 10 , 51 ) ) , t_pl ) )
        ( day = * ( * ( 1 . 6442 , ^ ( 10 , 48 ) ) , t_pl ) )
        ( Orbits ( Sun ) )
        ( atmosphere =
          (
            ( m = * ( * ( 1 . 14866 , ^ ( 10 , 24 ) ) , m_pl ) )
          ( m = * ( m , + ( * ( 0 . 9532 , CO2 ) , * ( 0 . 017 , N2 ) ,
                            * ( 0 . 016 , 40_Ar ) , * ( 0 . 0013 , O2 ) ,
                            * ( 0 . 0007 , CO ) , * ( 0 . 0003 , H2O ) ,
                            * ( * ( 1 . 3 , ^ ( 10 , -4 ) ) , NO ) ,
                            * ( * ( 2 . 5 , ^ ( 10 , -6 ) ) , 20_Ne )
              ) ) )
          )
        )
      )
    )

    ( Jupiter =
      (
        ( m = * ( * ( 8 . 72388 , ^ ( 10 , 34 ) ) , m_pl ) )
        ( T = * ( * ( 1 . 16461 , ^ ( 10 , -30 ) ) , T_pl ) )
        ( r <= * ( * ( 4 . 42332 , ^ ( 10 , 42 ) ) , l_pl ) )
        ( r >= * ( * ( 4 . 13636 , ^ ( 10 , 42 ) ) , l_pl ) )
        ( semimaj_a = * ( * ( 4 . 81699, ^ ( 10 , 46 ) ) , l_pl ) )
        ( ecc = 0 . 04839 )
        ( year = * ( * ( 6 . 94178 , ^ ( 10 , 51 ) ) , t_pl ) )
        ( day = * ( * ( 6 . 62742 , ^ ( 10 , 47 ) ) , t_pl ) )
        ( Orbits ( Sun ) )
        ( m = * ( m , + ( * ( 0 . 71 , H2 ) , * ( 0 . 24 , 4_He ) ,
                         0.05 ) ) )
      )
    )

    ( Saturn =
      (
        ( m = * ( * ( 2 . 61211 , ^ ( 10 , 34 ) ) , m_pl ) )
        ( T = * ( * ( 9 . 45803 , ^ ( 10 , -31 ) ) , T_pl ) )
        ( r <= * ( * ( 3 . 72887 , ^ ( 10 , 42 ) ) , l_pl ) )
        ( r >= * ( * ( 3 . 36358 , ^ ( 10 , 42 ) ) , l_pl ) )
        ( semimaj_a = * ( * ( 8 . 86897, ^ ( 10 , 46 ) ) , l_pl ) )
        ( ecc = 0 . 05415 )
        ( year = * ( * ( 1 . 73599 , ^ ( 10 , 52 ) ) , t_pl ) )
        ( day = * ( * ( 7 . 11554 , ^ ( 10 , 47 ) ) , t_pl ) )
        ( Orbits ( Sun ) )
```

```
      )
    )

    ( Uranus =
      (
        ( m = * ( * ( 3 . 99042 , ^ ( 10 , 33 ) ) , m_pl ) )
        ( T = * ( * ( 5 . 36426 , ^ ( 10 , -31 ) ) , T_pl ) )
        ( r <= * ( * ( 1 . 58137 , ^ ( 10 , 42 ) ) , l_pl ) )
        ( r >= * ( * ( 1 . 54512 , ^ ( 10 , 42 ) ) , l_pl ) )
        ( semimaj_a = * ( * ( 1 . 77985 , ^ ( 10 , 47 ) ) , l_pl ) )
        ( ecc = 0 . 04717 )
        ( year = * ( * ( 4 . 93586 , ^ ( 10 , 52 ) ) , t_pl ) )
        ( day = * ( * ( 1 . 1512 , ^ ( 10 , 48 ) ) , t_pl ) )
        ( Orbits ( Sun ) )
      )
    )

    ( Neptune =
      (
        ( m = * ( * ( 4 . 70677 , ^ ( 10 , 33 ) ) , m_pl ) )
        ( T = * ( * ( 5 . 08193 , ^ ( 10 , -31 ) ) , T_pl ) )
        ( r <= * ( * ( 1 . 53219 , ^ ( 10 , 42 ) ) , l_pl ) )
        ( r >= * ( * ( 1 . 50602 , ^ ( 10 , 42 ) ) , l_pl ) )
        ( semimaj_ax = * ( * ( 2 . 78635 , ^ ( 10 , 47 ) ) , l_pl ) )
        ( ecc = 0 . 00859 )
        ( year = * ( * ( 9 . 64605 , ^ ( 10 , 52 ) ) , t_pl ) )
        ( day = * ( * ( 1 . 07583 , ^ ( 10 , 48 ) ) , t_pl ) )
        ( Orbits ( Sun ) )
        ( m = * ( m , 1 ) )
      )
    )

    ( H2 =
      (
        ( * ( 2 , ( 1_H || 2_H ) ) )
      )
    )

    ( H2O =
      (
        ( + ( * ( 2 , ( 1_H || 2_H ) ) , ( 16_O || 17_O || 18_O ) ) )
      )
    )

    ( O2 =
      (
        ( * ( 2 , ( 16_O || 17_O || 18_O ) ) )
      )
    )

    ( O3 =
      (
        ( * ( 3 , ( 16_O || 17_O || 18_O ) ) )
      )
    )

    ( N2 =
      (
```

```
      ( * ( 2 , ( 14_N || 15_N ) ) )
    )
)

( NO =
    (
      ( + ( ( 14_N || 15_N ) , ( 16_O || 17_O || 18_O ) ) )
    )
)

( CO =
    (
      ( + ( ( 12_C || 13_C ) , ( 16_O || 17_O || 18_O ) ) )
    )
)

( CO2 =
    (
      ( + ( ( 12_C || 13_C ) , * ( 2, ( 16_O || 17_O || 18_O ) ) ) )
    )
)

( SO2 =
    (
      ( + ( 32_S , * ( 2, ( 16_O || 17_O || 18_O ) ) ) )
    )
)

( NH3 =
    (
      ( + ( ( 14_N || 15_N ) , * ( 3 , ( 1_H || 2_H ) ) ) )
    )
)

( CH4 =
    (
      ( + ( ( 12_C || 13_C ) , * ( 4 , ( 1_H || 2_H ) ) ) )
    )
)

( C2H6 =
    (
      ( + ( * ( 2 , ( 12_C || 13_C ) ) , * ( 6 , ( 1_H || 2_H ) ) ) )
    )
)

( NSH5 =
    (
      ( + ( ( 14_N || 15_N ) , 32_S , * ( 5 , ( 1_H || 2_H ) ) ) )
    )
)
```

# Supplementary Material 2: Key for Encoding the Test Message

| **Delimiters:** | | | | | |
|---|---|---|---|---|---|
| 00000000 | ( | 01111110 | ⇐ | **Numbers:** | |
| 00010000 | ) | 01222210 | \|\| | 10000001 | +1 |
| 00020000 | , | 01001000 | + | 10000003 | + |
| | | 01010000 | * | 12313101 | +1729 |
| **Operators:** | | 01100000 | - | 20000001 | -1 |
| 01000001 | = | 01000002 | ^ | 20000022 | -10 |
| 01000100 | ≠ | 01000200 | / | | |
| 01001001 | >= | 01020000 | . | | |
| 01001010 | <= | 01000030 | i | | |

**Nouns:**

| | | | | | |
|---|---|---|---|---|---|
| 30xxxxxx | variables | 31003111 | Life | 31130303 | Moon |
| 31000001 | e | 31003112 | Humanity | 31131000 | Mars |
| 31000002 | π | | | 31131100 | Jupiter |
| 31002000 | α | 31020000 | 1_H | 31131200 | Saturn |
| 31002001 | μ | 31020001 | 2_H | 31131300 | Uranus |
| 31002002 | αg | 31020002 | 3_He | 31132000 | Neptune |
| 31002003 | m_e | 31020003 | 4_He | | |
| 31002010 | m_p | 31020010 | 12_C | | |
| 31002011 | m_n | 31020011 | 13_C | | |
| 31002012 | m_pl | 31020012 | 14_N | | |
| 31002013 | h | 31020013 | 15_N | | |
| 31002020 | c | 31020020 | 16_O | | |
| 31002021 | G | 31020021 | 17_O | | |
| 31002022 | q | 31020022 | 18_O | | |
| 31002023 | h_b | 31020023 | 20_Ne | | |
| 31002030 | e_0 | 31020030 | 21_Ne | | |
| 31002031 | l_pl | 31020031 | 22_Ne | | |
| 31002032 | q_pl | 31020032 | 23_Na | | |
| 31002033 | e_pl | 31020033 | 24_Mg | | |
| 31002100 | T_pl | 31020100 | 28_Si | | |
| 31002101 | k | 31020101 | 32_S | | |
| 31002102 | P_pl | 31020102 | 40_Ar | | |
| 31002103 | t_pl | 31020103 | 56_Fe | | |
| 31003001 | proton | | | | |
| 31003002 | neutron | 31021001 | H2 | | |
| 31003003 | electron | 31021002 | H20 | | |
| 31003010 | m | 31021003 | O2 | | |
| 31003011 | charge | 31021010 | O3 | | |
| 31003012 | r | 31021011 | N2 | | |
| 31003013 | r_ground | 31021012 | NO | | |
| 31003020 | e_binding | 31021013 | CO | | |
| 31003021 | n_binding | 31021020 | CO2 | | |
| 31003022 | Lines | 31021021 | SO2 | | |
| 31003023 | T | 31021022 | NH3 | | |
| 31003030 | P | 31021023 | CH4 | | |
| 31003031 | age | 31021030 | C2H6 | | |
| 31003032 | Orbited_by | 31021031 | NSH5 | | |
| 31003033 | semimaj_a | | | | |
| 31003100 | ecc | 31130000 | Sun | | |
| 31003101 | year | 31130100 | Mercury | | |
| 31003102 | day | 31130200 | Venus | | |
| 31003103 | Orbits | 31130300 | Earth | | |
| 31003110 | atmosphere | | | | |

# Supplementary Material 3: Test Message, Encoded

```
00000000 10000000 01000001 10000000 00010000 00000000 10000001
01000001 10000001 00010000 00000000 10000002 01000001 10000002
00010000 00000000 10000003 01000001 10000003 00010000 00000000
10022133 01000001 10022133 00010000 00000000 10000000 01000001
20000000 00010000 00000000 10000001 01000100 20000001 00010000
00000000 10000001 01000100 10000000 00010000 00000000 10031212
01000001 10031212 00010000 00000000 10031212 01000100 10031213
00010000 00000000 10000002 01001001 10000001 00010000 00000000
10000001 01001001 10000001 00010000 00000000 10000101 01001001
10000001 00010000 00000000 10011111 01001001 10011111 00010000
00000000 10100220 01001001 10011111 00010000 00000000 10000001
01001010 10000001 00010000 00000000 10000000 01001010 10000001
00010000 00000000 20000101 01001010 10000001 00010000 00000000
20000313 01001010 10000000 00010000 00000000 10001001 01001010
10001001 00010000 00000000 20002013 01001010 10000000 00010000
00000000 01001000 00000000 10000000 00020000 10000000 00010000
01000001 10000000 00010000 00000000 01001000 00000000 10000000
00020000 10000001 00010000 01000001 10000001 00010000 00000000
01001000 00000000 10000000 00020000 10033123 00010000 01000001
10033123 00010000 00000000 01001000 00000000 10000001 00020000
10000001 00010000 01000001 10000002 00010000 00000000 01001000
00000000 10000312 00020000 10000001 00010000 01000001 10000313
00010000 00000000 01001000 00000000 10002133 00020000 10022023
00010000 01000001 10030222 00010000 00000000 01001000 00000000
10000002 00020000 10000002 00010000 01000001 10000010 00010000
00000000 01001000 00000000 10000002 00020000 20000002 00010000
01000001 10000000 00010000 00000000 01001000 00000000 01001000
00000000 10000011 00020000 10000221 00010000 00020000 10000012
00010000 01000001 10000310 00010000 00000000 01001000 00000000
01001000 00000000 10000221 00020000 10000012 00010000 00020000
10000011 00010000 01000001 10000310 00010000 00000000 01001000
00000000 10020303 00020000 10023211 00010000 01000001 01001000
00000000 10023211 00020000 10020303 00010000 00010000 00000000
00000000 30002002 01111110 10000001 00010000 00000000 30002002
01000001 10000001 00010000 00000000 01001000 00000000 30002002
00020000 10000001 00010000 01000001 10000002 00010000 00000000
30002002 01000100 10001221 00010000 00010000 00000000 00000000
30002002 01111110 10001221 00010000 00000000 30002002 01000100
10000001 00010000 00000000 01001000 00000000 30002002 00020000
10000001 00010000 01000001 10001222 00010000 00000000 30002002
01000001 10001221 00010000 00010000 00000000 00000000 30002002
01111110 10020301 00010000 00000000 30012002 01111110 30002002
00010000 00000000 30012002 01000001 10020301 00010000 00000000
01001000 00000000 30012002 00020000 20000023 00010000 01000001
10020212 00010000 00000000 30012002 01111110 01001000 00000000
30002002 00020000 10000302 00010000 00010000 00000000 30012002
```

```
01000001 10021203 00010000 00010000 00000000 01010000 00000000
10000000 00020000 10000000 00010000 01000001 10000000 00010000
00000000 01010000 00000000 10000001 00020000 10000000 00010000
01000001 10000000 00010000 00000000 01010000 00000000 10000001
00020000 10000002 00010000 01000001 10000002 00010000 00000000
01010000 00000000 10000002 00020000 10000002 00010000 01000001
10000010 00010000 00000000 01010000 00000000 10000003 00020000
10000010 00010000 01000001 10000030 00010000 00000000 01010000
00000000 30111111 00020000 10000000 00010000 01000001 10000000
00010000 00000000 01010000 00000000 30111111 00020000 10000002
00010000 01000001 01001000 00000000 30111111 00020000 30111111
00010000 00010000 00000000 01010000 00000000 30111111 00020000
10000002 00010000 01000001 01010000 00000000 10000002 00020000
30111111 00010000 00010000 00000000 01010000 00000000 30111111
00020000 30222222 00010000 01000001 01010000 00000000 30222222
00020000 30111111 00010000 00010000 00000000 01010000 00000000
01001000 00000000 30111111 00020000 30222222 00010000 00020000
30333333 00010000 01000001 01001000 00000000 01010000 00000000
30111111 00020000 30333333 00010000 00020000 01010000 00000000
30222222 00020000 30333333 00010000 00010000 00010000 00000000
01010000 00000000 01010000 00000000 30111111 00020000 30222222
00010000 00020000 30333333 00010000 01000001 01010000 00000000
01010000 00000000 30111111 00020000 30333333 00010000 00020000
30222222 00010000 00010000 00000000 10000002 01001001 10000001
00010000 00000000 10000001 01001001 10000001 00010000 00000000
10000101 01001001 10000001 00010000 00000000 10011111 01001001
10011111 00010000 00000000 10100220 01001001 10011111 00010000
00000000 10000001 01001010 10000001 00010000 00000000 10000000
01001010 10000001 00010000 00000000 20000101 01001010 10000001
00010000 00000000 20000313 01001010 10000000 00010000 00000000
10001001 01001010 10001001 00010000 00000000 20002013 01001010
10000000 00010000 00000000 01100000 00000000 30111111 00020000
30212121 00010000 01000001 01001000 00000000 30111111 00020000
01010000 00000000 20000001 00020000 30212121 00010000 00010000
00010000 00000000 01100000 00000000 10000010 00020000 20000011
00010000 01000001 10000021 00010000 00000000 01100000 00000000
10000032 00020000 10000021 00010000 01000001 10000011 00010000
00000000 01100000 00000000 01100000 00000000 30111111 00020000
30212121 00010000 00020000 30323232 00010000 01000001 01100000
00000000 30111111 00020000 01001000 00000000 30212121 00020000
30323232 00010000 00010000 00010000 00000000 01100000 00000000
20001003 00020000 10001013 00010000 01000001 20002022 00010000
00000000 01000200 00000000 10000010 00020000 10000010 00010000
01000001 10000001 00010000 00000000 01000200 00000000 10000010
00020000 10000002 00010000 01000001 10000002 00010000 00000000
01000200 00000000 01001000 00000000 30111111 00020000 30232323
00010000 00020000 30232323 00010000 01000001 30111111 00010000
00000000 01000200 00000000 01001000 00000000 30111111 00020000
```

```
30232323 00010000 00020000 30111111 00010000 01000001 30232323
00010000 00000000 01000200 00000000 01000200 00000000 30111111
00020000 30232323 00010000 00020000 30123123 00010000 01000001
01000200 00000000 30111111 00020000 01001000 00000000 30232323
00020000 30123123 00010000 00010000 00010000 00000000 01000200
00000000 01000200 00000000 30111111 00020000 30232323 00010000
00020000 01000200 00000000 30123123 00020000 30321321 00010000
00010000 01000001 01000200 00000000 01001000 00000000 30111111
00020000 30321321 00010000 00020000 01001000 00000000 30232323
00020000 30123123 00010000 00010000 00010000 00000000 01000200
00000000 10000333 00020000 10000021 00010000 01000001 10000013
00010000 00000000 01000200 00000000 10001331 00020000 10000121
00010000 01000001 10000011 00010000 00000000 01000200 00000000
10000003 00020000 10000002 00010000 01000001 10000001 01020000
12000000 00010000 00000000 01000200 00000000 10000001 00020000
10000003 00010000 01000001 10000000 01020000 11111111 00010000
00000000 01000200 00000000 10000033 00020000 10000010 00010000
01000001 10000003 01020000 13000000 00010000 00000000 01000200
00000000 10000020 00020000 10000013 00010000 01000001 10000001
01020000 10210210 00010000 00000000 01010000 00000000 10000001
01020000 11111111 00020000 10000201 00010000 01000001 10000230
00010000 00000000 01010000 00000000 10000210 00020000 10000012
01020000 10222223 00010000 01000001 10003132 00010000 00000000
01001000 00000000 10000002 01020000 11111111 00020000 10000011
01020000 12222222 00010000 01000001 10000120 00010000 00000000
01001000 00000000 10000013 01020000 10312313 00020000 10000023
01020000 11021021 00010000 01000001 10000102 01020000 12000000
00010000 00000000 01000002 00000000 10000001 00020000 30333333
00010000 01000001 10000001 00010000 00000000 01000002 00000000
10000010 00020000 10000000 01020000 12000000 00010000 01000001
10000002 00010000 00000000 01000002 00000000 10000021 00020000
10000000 01020000 12000000 00010000 01000001 10000003 00010000
00000000 01000002 00000000 01010000 00000000 30333333 00020000
30333333 00010000 00020000 10000000 01020000 12000000 00010000
01000001 30333333 00010000 00000000 01000002 00000000 30333333
00020000 10000002 00010000 01000001 01010000 00000000 30333333
00020000 30333333 00010000 00010000 00000000 01000002 00000000
30333333 00020000 10000003 00010000 01000001 01010000 00000000
01010000 00000000 30333333 00020000 30333333 00010000 00020000
30333333 00010000 00010000 00000000 01000002 00000000 01000002
00000000 30333333 00020000 30232323 00010000 00020000 30222111
00010000 01000001 01000002 00000000 30333333 00020000 01010000
00000000 30232323 00020000 30222111 00010000 00010000 00010000
00000000 01000002 00000000 10000003 00020000 10000010 00010000
01000001 10001101 00010000 00000000 01000002 00000000 10000010
00020000 10000001 01020000 12000000 00010000 01000001 10000020
00010000 00000000 01000002 00000000 10000002 00020000 10000000
01020000 12000000 00010000 01000001 10000001 01020000 11222002
```

```
00010000 00000000 01000002 00000000 10000011 00020000 10000000
01020000 12000000 00010000 01000001 10000002 01020000 10330123
00010000 00000000 01000002 00000000 20000001 00020000 10000000
01020000 12000000 00010000 01000001 01000030 00000000 10000001
00010000 00010000 00000000 01000002 00000000 20000010 00020000
10000000 01020000 12000000 00010000 01000001 01000030 00000000
10000002 00010000 00010000 00000000 01010000 00000000 01000030
00000000 10000001 00010000 00020000 01000030 00000000 10000001
00010000 00010000 01000001 20000001 00010000 00000000 01010000
00000000 01001000 00000000 30111222 00020000 01000030 00000000
30222000 00010000 00010000 00020000 01001000 00000000 30222111
00020000 01000030 00000000 30333000 00010000 00010000 00010000
01000001 01001000 00000000 01010000 00000000 30111222 00020000
30222111 00010000 01100000 01010000 00000000 30222000 00020000
30333000 00010000 00020000 01000030 00000000 01001000 00000000
01010000 00000000 30222000 00020000 30222111 00010000 00020000
01010000 00000000 30111222 00020000 30333000 00010000 00010000
00010000 00010000 00010000 00000000 00000000 01000002 00000000
30231231 00020000 10000002 00010000 01111110 10000001 00010000
00000000 30231231 01000001 10000001 00010000 01222210 00000000
30231231 01000001 20000001 00010000 00010000 00000000 00000000
01010000 00000000 01001000 00000000 30231231 00020000 10000001
00010000 00020000 01001000 00000000 30231231 00020000 20000001
00010000 00010000 01111110 10000000 00010000 00000000 30231231
01000001 10000001 00010000 01222210 00000000 30231231 01000001
20000001 00010000 00010000 00000000 00000000 30222222 01111110
10000000 00010000 00000000 30333333 01000001 01000200 00000000
01001000 00000000 01010000 00000000 20000001 00020000 30232323
00010000 00020000 01000002 00000000 01100000 00000000 01010000
00000000 30232323 00020000 30232323 00010000 00020000 01010000
00000000 10000010 00020000 01010000 00000000 30121212 00020000
30303030 00010000 00010000 00020000 10000000 01020000 12000000
00010000 00010000 00010000 00020000 01010000 00000000 10000002
00020000 30121212 00010000 00010000 00010000 01222210 00000000
30333333 01000001 01000200 00000000 01100000 00000000 01010000
00000000 20000001 00020000 30232323 00010000 00020000 01000002
00000000 01100000 00000000 01010000 00000000 30232323 00020000
30232323 00010000 00020000 01010000 00000000 10000010 00020000
01010000 00000000 30121212 00020000 30303030 00010000 00010000
00020000 10000000 01020000 12000000 00010000 00010000 00010000
00020000 01010000 00000000 10000002 00020000 30121212 00010000
00010000 00010000 00000000 30222222 01000001 01001000 00000000
01010000 00000000 30121212 00020000 01010000 00000000 30333333
00020000 30333333 00010000 00010000 00020000 01001000 00000000
01010000 00000000 30232323 00020000 30333333 00010000 00020000
30303030 00010000 00010000 00010000 00010000 00000000 01000002
00000000 30101010 00020000 01010000 00000000 20000001 00020000
30313131 00010000 00010000 01000001 01000200 00000000 10000001
```

```
00020000 01000002 00000000 30101010 00020000 30313131 00010000
00010000 00010000 00000000 01000002 00000000 01000030 00000000
10000001 00010000 00020000 10000000 01020000 12000000 00010000
01000001 01001000 00000000 01000002 00000000 10000002 00020000
20000000 01020000 12000000 00010000 00020000 01000030 00000000
01000002 00000000 10000002 00020000 20000000 01020000 12000000
00010000 00010000 00010000 00010000 00000000 01000002 00000000
31000001 00020000 01000030 00000000 31000002 00010000 00010000
01000001 20000001 00010000 00000000 31000001 01000001 10000002
01020000 12313320 00010000 00000000 31000002 01000001 10000003
01020000 10210033 00010000 00000000 01000002 00000000 31000001
00020000 01000030 00000000 01000200 00000000 31000002 00020000
10000002 00010000 00010000 00010000 01000001 01000030 00000000
20000001 00010000 00010000 00000000 01000002 00000000 31000001
00020000 10000002 00010000 01000001 10000013 01020000 11203212
00010000 00000000 01000002 00000000 31000002 00020000 31000001
00010000 01000001 10000112 01020000 11311202 00010000 00000000
10000000 01000001 10000000 00010000 00000000 10000001 01000001
10000001 00010000 00000000 10000002 01000001 10000002 00010000
00000000 10000003 01000001 10000003 00010000 00000000 10022133
01000001 10022133 00010000 00000000 10000000 01000001 20000000
00010000 00000000 10000001 01000100 20000001 00010000 00000000
10000001 01000100 10000000 00010000 00000000 10031212 01000001
10031212 00010000 00000000 10031212 01000100 10031213 00010000
00000000 01010000 00000000 10000000 00020000 10000000 00010000
01000001 10000000 00010000 00000000 01010000 00000000 10000001
00020000 10000000 00010000 01000001 10000000 00010000 00000000
01010000 00000000 10000001 00020000 10000002 00010000 01000001
10000002 00010000 00000000 01010000 00000000 10000002 00020000
10000002 00010000 01000001 10000010 00010000 00000000 01010000
00000000 10000003 00020000 10000010 00010000 01000001 10000030
00010000 00000000 01010000 00000000 30111111 00020000 10000000
00010000 01000001 10000000 00010000 00000000 01010000 00000000
30111111 00020000 10000002 00010000 01000001 01001000 00000000
30111111 00020000 30111111 00010000 00010000 00000000 01010000
00000000 30111111 00020000 10000002 00010000 01000001 01010000
00000000 10000002 00020000 30111111 00010000 00010000 00000000
01010000 00000000 30111111 00020000 30222222 00010000 01000001
01010000 00000000 30222222 00020000 30111111 00010000 00010000
00000000 01010000 00000000 01001000 00000000 30111111 00020000
30222222 00010000 00020000 30333333 00010000 01000001 01001000
00000000 01010000 00000000 30111111 00020000 30333333 00010000
00020000 01010000 00000000 30222222 00020000 30333333 00010000
00010000 00010000 00000000 01010000 00000000 01010000 00000000
30111111 00020000 30222222 00010000 00020000 30333333 00010000
01000001 01010000 00000000 01010000 00000000 30111111 00020000
30333333 00010000 00020000 30222222 00010000 00010000 00000000
10000000 01000001 10000000 00010000 00000000 10000001 01000001
```

```
10000001 00010000 00000000 10000002 01000001 10000002 00010000
00000000 10000003 01000001 10000003 00010000 00000000 10022133
01000001 10022133 00010000 00000000 10000000 01000001 20000000
00010000 00000000 10000001 01000100 20000001 00010000 00000000
10000001 01000100 10000000 00010000 00000000 10031212 01000001
10031212 00010000 00000000 10031212 01000100 10031213 00010000
00000000 01001000 00000000 10000000 00020000 10000000 00010000
01000001 10000000 00010000 00000000 01001000 00000000 10000000
00020000 10000001 00010000 01000001 10000001 00010000 00000000
01001000 00000000 10000000 00020000 10033123 00010000 01000001
10033123 00010000 00000000 01001000 00000000 10000001 00020000
10000001 00010000 01000001 10000002 00010000 00000000 01001000
00000000 10000312 00020000 10000001 00010000 01000001 10000313
00010000 00000000 01001000 00000000 10002133 00020000 10022023
00010000 01000001 10030222 00010000 00000000 01001000 00000000
10000002 00020000 10000002 00010000 01000001 10000010 00010000
00000000 01001000 00000000 10000002 00020000 20000002 00010000
01000001 10000000 00010000 00000000 01001000 00000000 01001000
00000000 10000011 00020000 10000221 00010000 00020000 10000012
00010000 01000001 10000310 00010000 00000000 01001000 00000000
01001000 00000000 10000221 00020000 10000012 00010000 00020000
10000011 00010000 01000001 10000310 00010000 00000000 01001000
00000000 10020303 00020000 10023211 00010000 01000001 01001000
00000000 10023211 00020000 10020303 00010000 00010000 00000000
01000200 00000000 10000010 00020000 10000010 00010000 01000001
10000001 00010000 00000000 01000200 00000000 10000010 00020000
10000002 00010000 01000001 10000002 00010000 00000000 01000200
00000000 01001000 00000000 30111111 00020000 30232323 00010000
00020000 30232323 00010000 01000001 30111111 00010000 00000000
01000200 00000000 01001000 00000000 30111111 00020000 30232323
00010000 00020000 30111111 00010000 01000001 30232323 00010000
00000000 01000200 00000000 01000200 00000000 30111111 00020000
30232323 00010000 00020000 30123123 00010000 01000001 01000200
00000000 30111111 00020000 01001000 00000000 30232323 00020000
30123123 00010000 00010000 00010000 00000000 01000200 00000000
01000200 00000000 30111111 00020000 30232323 00010000 00020000
01000200 00000000 30123123 00020000 30321321 00010000 00010000
01000001 01000200 00000000 01001000 00000000 30111111 00020000
30321321 00010000 00020000 01001000 00000000 30232323 00020000
30123123 00010000 00010000 00010000 00000000 01000200 00000000
10000333 00020000 10000021 00010000 01000001 10000013 00010000
00000000 01000200 00000000 10001331 00020000 10000121 00010000
01000001 10000011 00010000 00000000 01000200 00000000 10000003
00020000 10000002 00010000 01000001 10000001 01020000 12000000
00010000 00000000 01000200 00000000 10000001 00020000 10000003
00010000 01000001 10000000 01020000 11111111 00010000 00000000
01000200 00000000 10000033 00020000 10000010 00010000 01000001
10000003 01020000 13000000 00010000 00000000 01000200 00000000
```

```
10000020 00020000 10000013 00010000 01000001 10000001 01020000
10210210 00010000 00000000 01010000 00000000 10000001 01020000
11111111 00020000 10000201 00010000 01000001 10000230 00010000
00000000 01010000 00000000 10000210 00020000 10000012 01020000
10222223 00010000 01000001 10003132 00010000 00000000 01001000
00000000 10000002 01020000 11111111 00020000 10000011 01020000
12222222 00010000 01000001 10000120 00010000 00000000 01001000
00000000 10000013 01020000 10312313 00020000 10000023 01020000
11021021 00010000 01000001 10000102 01020000 12000000 00010000
00000000 00000000 30002002 01111110 10000001 00010000 00000000
30002002 01000001 10000001 00010000 00000000 01001000 00000000
30002002 00020000 10000001 00010000 01000001 10000002 00010000
00000000 30002002 01000100 10001221 00010000 00010000 00000000
00000000 30002002 01111110 10001221 00010000 00000000 30002002
01000100 10000001 00010000 00000000 01001000 00000000 30002002
00020000 10000001 00010000 01000001 10001222 00010000 00000000
30002002 01000001 10001221 00010000 00010000 00000000 00000000
30002002 01111110 10020301 00010000 00000000 30012002 01111110
30002002 00010000 00000000 30012002 01000001 10020301 00010000
00000000 01001000 00000000 30012002 00020000 20000023 00010000
01000001 10020212 00010000 00000000 30012002 01111110 01001000
00000000 30002002 00020000 10000302 00010000 00010000 00000000
30012002 01000001 10021203 00010000 00010000 00000000 01010000
00000000 10000001 01020000 11111111 00020000 10000201 00010000
01000001 10000230 00010000 00000000 01010000 00000000 10000210
00020000 10000012 01020000 10222223 00010000 01000001 10003132
00010000 00000000 01001000 00000000 10000002 01020000 11111111
00020000 10000011 01020000 12222222 00010000 01000001 10000120
00010000 00000000 01001000 00000000 10000013 01020000 10312313
00020000 10000023 01020000 11021021 00010000 01000001 10000102
01020000 12000000 00010000 00000000 01000002 00000000 10000001
00020000 30333333 00010000 01000001 10000001 00010000 00000000
01000002 00000000 10000010 00020000 10000000 01020000 12000000
00010000 01000001 10000002 00010000 00000000 01000002 00000000
10000021 00020000 10000000 01020000 12000000 00010000 01000001
10000003 00010000 00000000 01000002 00000000 01010000 00000000
30333333 00020000 30333333 00010000 00020000 10000000 01020000
12000000 00010000 01000001 30333333 00010000 00000000 01000002
00000000 30333333 00020000 10000002 00010000 01000001 01010000
00000000 30333333 00020000 30333333 00010000 00010000 00000000
01000002 00000000 30333333 00020000 10000003 00010000 01000001
01010000 00000000 01010000 00000000 30333333 00020000 30333333
00010000 00020000 30333333 00010000 00010000 00000000 01000002
00000000 01000002 00000000 30333333 00020000 30232323 00010000
00020000 30222111 00010000 01000001 01000002 00000000 30333333
00020000 01010000 00000000 30232323 00020000 30222111 00010000
00010000 00010000 00000000 01000002 00000000 10000003 00020000
10000010 00010000 01000001 10001101 00010000 00000000 01000002
```

```
00000000 10000010 00020000 10000001 01020000 12000000 00010000
01000001 10000020 00010000 00000000 01000002 00000000 10000002
00020000 10000000 01020000 12000000 00010000 01000001 10000001
01020000 11222002 00010000 00000000 01000002 00000000 10000011
00020000 10000000 01020000 12000000 00010000 01000001 10000002
01020000 10330123 00010000 00000000 01000002 00000000 20000001
00020000 10000000 01020000 12000000 00010000 01000001 01000030
00000000 10000001 00010000 00010000 00000000 01000002 00000000
20000010 00020000 10000000 01020000 12000000 00010000 01000001
01000030 00000000 10000002 00010000 00010000 00000000 01010000
00000000 01000030 00000000 10000001 00010000 00020000 01000030
00000000 10000001 00010000 00010000 01000001 20000001 00010000
00000000 01010000 00000000 01001000 00000000 30111222 00020000
01000030 00000000 30222000 00010000 00010000 00020000 01001000
00000000 30222111 00020000 01000030 00000000 30333000 00010000
00010000 00010000 01000001 01001000 00000000 01010000 00000000
30111222 00020000 30222111 00010000 01100000 01010000 00000000
30222000 00020000 30333000 00010000 00020000 01000030 00000000
01001000 00000000 01010000 00000000 30222000 00020000 30222111
00010000 00020000 01010000 00000000 30111222 00020000 30333000
00010000 00010000 00010000 00010000 00010000 00000000 00000000
01000002 00000000 30231231 00020000 10000002 00010000 01111110
10000001 00010000 00000000 30231231 01000001 10000001 00010000
01222210 00000000 30231231 01000001 20000001 00010000 00010000
00000000 00000000 01010000 00000000 01001000 00000000 30231231
00020000 10000001 00010000 00020000 01001000 00000000 30231231
00020000 20000001 00010000 00010000 01111110 10000000 00010000
00000000 30231231 01000001 10000001 00010000 01222210 00000000
30231231 01000001 20000001 00010000 00010000 00000000 00000000
30222222 01111110 10000000 00010000 00000000 30333333 01000001
01000200 00000000 01001000 00000000 01010000 00000000 20000001
00020000 30232323 00010000 00020000 01000002 00000000 01100000
00000000 01010000 00000000 30232323 00020000 30232323 00010000
00020000 01010000 00000000 10000010 00020000 01010000 00000000
30121212 00020000 30303030 00010000 00010000 00020000 10000000
01020000 12000000 00010000 00010000 00010000 00020000 01010000
00000000 10000002 00020000 30121212 00010000 00010000 00010000
01222210 00000000 30333333 01000001 01000200 00000000 01100000
00000000 01010000 00000000 20000001 00020000 30232323 00010000
00020000 01000002 00000000 01100000 00000000 01010000 00000000
30232323 00020000 30232323 00010000 00020000 01010000 00000000
10000010 00020000 01010000 00000000 30121212 00020000 30303030
00010000 00010000 00020000 10000000 01020000 12000000 00010000
00010000 00010000 00020000 01010000 00000000 10000002 00020000
30121212 00010000 00010000 00010000 00000000 30222222 01000001
01001000 00000000 01010000 00000000 30121212 00020000 01010000
00000000 30333333 00020000 30333333 00010000 00010000 00020000
01001000 00000000 01010000 00000000 30232323 00020000 30333333
```

```
00010000  00020000  30303030  00010000  00010000  00010000  00010000
00000000  10000000  01000001  10000000  00010000  00000000  10000001
01000001  10000001  00010000  00000000  10000002  01000001  10000002
00010000  00000000  10000003  01000001  10000003  00010000  00000000
10022133  01000001  10022133  00010000  00000000  10000000  01000001
20000000  00010000  00000000  10000001  01000100  20000001  00010000
00000000  10000001  01000100  10000000  00010000  00000000  10031212
01000001  10031212  00010000  00000000  10031212  01000100  10031213
00010000  00000000  01100000  00000000  30111111  00020000  30212121
00010000  01000001  01001000  00000000  30111111  00020000  01010000
00000000  20000001  00020000  30212121  00010000  00010000  00010000
00000000  01100000  00000000  10000010  00020000  20000011  00010000
01000001  10000021  00010000  00000000  01100000  00000000  10000032
00020000  10000021  00010000  01000001  10000011  00010000  00000000
01100000  00000000  01100000  00000000  30111111  00020000  30212121
00010000  00020000  30323232  00010000  01000001  01100000  00000000
30111111  00020000  01001000  00000000  30212121  00020000  30323232
00010000  00010000  00010000  00000000  01100000  00000000  20001003
00020000  10001013  00010000  01000001  20002022  00010000  00000000
01000002  00000000  30101010  00020000  01010000  00000000  20000001
00020000  30313131  00010000  00010000  01000001  01000200  00000000
10000001  00020000  01000002  00000000  30101010  00020000  30313131
00010000  00010000  00010000  00000000  01000002  00000000  01000030
00000000  10000001  00010000  00020000  10000000  01020000  12000000
00010000  01000001  01001000  00000000  01000002  00000000  10000002
00020000  20000000  01020000  12000000  00010000  00020000  01000030
00000000  01000002  00000000  10000002  00020000  20000000  01020000
12000000  00010000  00010000  00010000  00010000  00000000  01000002
00000000  31000001  00020000  01000030  00000000  31000002  00010000
00010000  01000001  20000001  00010000  00000000  31000001  01000001
10000002  01020000  12313320  00010000  00000000  31000002  01000001
10000003  01020000  10210033  00010000  00000000  01000002  00000000
31000001  00020000  01000030  00000000  01000200  00000000  31000002
00020000  10000002  00010000  00010000  00010000  01000001  01000030
00000000  20000001  00010000  00010000  00000000  01000002  00000000
31000001  00020000  10000002  00010000  01000001  10000013  01020000
11203212  00010000  00000000  01000002  00000000  31000002  00020000
31000001  00010000  01000001  10000112  01020000  11311202  00010000
00000000  31002000  01000001  01010000  00000000  10000013  01020000
11030013  00020000  01000002  00000000  10000022  00020000  20000003
00010000  00010000  00010000  00000000  31002000  01000001  01000200
00000000  10000001  00020000  10002021  01020000  10021100  00010000
00010000  00000000  31002001  01000001  10130101  01020000  10213011
00010000  00000000  31002002  01000001  01010000  00000000  10000001
01020000  13000131  00020000  01000002  00000000  10000022  00020000
20000231  00010000  00010000  00010000  00000000  31002003  01000001
01010000  00000000  01000002  00000000  31002002  00020000  10000000
01020000  12000000  00010000  00020000  31002012  00010000  00010000
```

```
00000000 31002010 01000001 01010000 00000000 31002001 00020000
31002003 00010000 00010000 00000000 31002010 01000001 01010000
00000000 10130101 01020000 10213011 00020000 31002003 00010000
00010000 00000000 31002011 01000001 01010000 00000000 10130233
01020000 12233001 00020000 31002003 00010000 00010000 00000000
31002011 01000001 01010000 00000000 10000001 01020000 10000113
00020000 31002010 00010000 00010000 00000000 31002010 01000001
01010000 00000000 01010000 00000000 10000013 01020000 22233133
00020000 01000002 00000000 10000022 00020000 20000110 00010000
00010000 00020000 31002012 00010000 00010000 00000000 31002012
01000001 01000002 00000000 01010000 00000000 31002013 00020000
01000200 00000000 31002020 00020000 31002021 00010000 00010000
00020000 10000000 01020000 12000000 00010000 00010000 00000000
31002000 01000001 01000200 00000000 01010000 00000000 31002022
00020000 31002022 00010000 00020000 01010000 00000000 31002023
00020000 01010000 00000000 31002020 00020000 01010000 00000000
10000010 00020000 01010000 00000000 31000002 00020000 31002030
00010000 00010000 00010000 00010000 00010000 00010000 00000000
31002031 01000001 01000002 00000000 01000200 00000000 01010000
00000000 31002023 00020000 31002021 00010000 00020000 01000002
00000000 31002020 00020000 10000003 00010000 00010000 00020000
10000000 01020000 12000000 00010000 00010000 00000000 31002031
01000001 01000200 00000000 01010000 00000000 10000002 00020000
01010000 00000000 31002021 00020000 31002012 00010000 00010000
00020000 01000002 00000000 31002020 00020000 10000002 00010000
00010000 00010000 00000000 31002032 01000001 01010000 00000000
31002012 00020000 01010000 00000000 10000002 00020000 01010000
00000000 31000002 00020000 01000002 00000000 01010000 00000000
31002021 00020000 31002030 00010000 00020000 10000000 01020000
10000011 00010000 00010000 00010000 00010000 00010000 00000000
31002032 01000001 01010000 00000000 10000023 01020000 12310303
00020000 31002022 00010000 00010000 00000000 31002033 01000001
01000002 00000000 01010000 00000000 31002013 00020000 01000200
00000000 01000002 00000000 31002020 00020000 10000011 00010000
00020000 31002021 00010000 00010000 00020000 10000000 01020000
12000000 00010000 00010000 00000000 31002033 01000001 01010000
00000000 31002012 00020000 01010000 00000000 31002020 00020000
31002020 00010000 00010000 00010000 00000000 31002100 01000001
01000200 00000000 31002033 00020000 31002101 00010000 00010000
00000000 31002102 01000001 01000200 00000000 01000002 00000000
31002020 00020000 10000011 00010000 00020000 31002021 00010000
00010000 00000000 31002103 01000001 01000200 00000000 31002031
00020000 31002020 00010000 00010000 00000000 31130000 01000001
00000000 00000000 31003010 01000001 01010000 00000000 01010000
00000000 10000021 01020000 10203130 00020000 01000002 00000000
10000022 00020000 20000211 00010000 00010000 00020000 31002012
00010000 00010000 00000000 31003023 01000001 01010000 00000000
01010000 00000000 10000003 01020000 13233322 00020000 01000002
```

```
00000000 10000022 00020000 20000131 00010000 00010000 00020000
31002100 00010000 00010000 00000000 31003030 01000001 01010000
00000000 01010000 00000000 10000001 01020000 10033113 00020000
01000002 00000000 10000022 00020000 20000122 00010000 00010000
00020000 31002102 00010000 00010000 00000000 31003012 01000001
01010000 00000000 01010000 00000000 10000010 01020000 11031213
00020000 01000002 00000000 10000022 00020000 10000223 00010000
00010000 00020000 31002031 00010000 00010000 00000000 31003031
01000001 01010000 00000000 01010000 00000000 10000002 01020000
12230303 00020000 01000002 00000000 10000022 00020000 10000330
00010000 00010000 00020000 31002103 00010000 00010000 00000000
31003032 00000000 31130100 00020000 31130200 00020000 31130300
00020000 31131000 00020000 31131100 00020000 31131200 00020000
31131300 00020000 31132000 00010000 00010000 00000000 31003010
01000001 01010000 00000000 31003010 00020000 01001000 00000000
01010000 00000000 10000000 01020000 12330002 00020000 31020000
00010000 00020000 01010000 00000000 10000000 01020000 10333213
00020000 31020003 00010000 00020000 01010000 00000000 10000000
01020000 10001332 00020000 31020020 00010000 00020000 01010000
00000000 10000000 01020000 10000300 00020000 31020010 00010000
00020000 01010000 00000000 10000000 01020000 10000121 00020000
31020103 00010000 00020000 01010000 00000000 10000000 01020000
10000102 00020000 31020101 00010000 00020000 01010000 00000000
10000000 01020000 10000033 00020000 31020012 00010000 00020000
01010000 00000000 10000000 01020000 10000023 00020000 31020100
00010000 00020000 01010000 00000000 10000000 01020000 10000002
00020000 31020001 00010000 00020000 01010000 00000000 10000000
01020000 10000103 00020000 31020023 00010000 00020000 01010000
00000000 01010000 00000000 10000003 01020000 11002203 00020000
01000002 00000000 10000022 00020000 20000005 00010000 00010000
00020000 31020011 00010000 00020000 01010000 00000000 01010000
00000000 10000001 01020000 12001323 00020000 01000002 00000000
10000022 00020000 20000005 00010000 00010000 00020000 31020022
00010000 00020000 01010000 00000000 01010000 00000000 10000002
01020000 13130022 00020000 01000002 00000000 10000022 00020000
20000006 00010000 00010000 00020000 31020021 00010000 00020000
01010000 00000000 01010000 00000000 10000003 01020000 11030303
00020000 01000002 00000000 10000022 00020000 20000006 00010000
00010000 00020000 31020013 00010000 00020000 01010000 00000000
01010000 00000000 10000001 01020000 10303031 00020000 01000002
00000000 10000022 00020000 20000010 00010000 00010000 00020000
31020030 00010000 00020000 01010000 00000000 01010000 00000000
10000003 01020000 12000000 00020000 01000002 00000000 10000022
00020000 20000006 00010000 00010000 00020000 31020031 00010000
00020000 10000000 01020000 10000031 00010000 00010000 00010000
00010000 00010000 00000000 01001000 00000000 30111111 00020000
30222222 00020000 30333333 00010000 01000001 01001000 00000000
01001000 00000000 30111111 00020000 30222222 00010000 00020000
```

```
30333333 00010000 00010000 00000000 01001000 00000000 30111111
00020000 30222222 00020000 30333333 00020000 30000000 00010000
01000001 01001000 00000000 01001000 00000000 30111111 00020000
30222222 00010000 00020000 01001000 00000000 30333333 00020000
30000000 00010000 00010000 00010000 00000000 31003001 01000001
00000000 00000000 31003010 01000001 31002010 00010000 00000000
31003011 01000001 01001000 00000000 20000001 00020000 31002022
00010000 00010000 00000000 31003012 01000001 01001000 00000000
01001000 00000000 10000011 01000030 11221330 00020000 01000002
00000000 10000022 00020000 10000103 00010000 00010000 00020000
31002031 00010000 00010000 00010000 00010000 00000000 31003003
01000001 00000000 00000000 31003010 01000001 31002003 00010000
00000000 31003011 01000001 31002022 00010000 00000000 31003012
01000001 10000000 00010000 00010000 00010000 00000000 31003002
01000001 00000000 00000000 31003010 01000001 31002011 00010000
00000000 31003011 01000001 10000000 00010000 00010000 00010000
00000000 31020000 01000001 00000000 00000000 01001000 00000000
31003001 00020000 31003003 00010000 00010000 00000000 31003010
01000001 01010000 00000000 01010000 00000000 10000013 01020000
12232132 00020000 01000002 00000000 10000022 00020000 20000110
00010000 00010000 00020000 31002012 00010000 00010000 00000000
31003013 01000001 01010000 00000000 01010000 00000000 10000003
01020000 11012100 00020000 01000002 00000000 10000022 00020000
10000120 00010000 00010000 00020000 31002031 00010000 00010000
00000000 31003020 01000001 01010000 00000000 01010000 00000000
10000001 01020000 10131022 00020000 01000002 00000000 10000022
00020000 20000123 00010000 00010000 00020000 31002033 00010000
00010000 00000000 31003021 01000001 10000000 00010000 00000000
31003022 01000001 00000000 01010000 00000000 01010000 00000000
10000001 01020000 11032103 00020000 01000002 00000000 10000022
00020000 10000202 00010000 00010000 00020000 31002031 00010000
00020000 01010000 00000000 01010000 00000000 10000010 01020000
10033210 00020000 01000002 00000000 10000022 00020000 10000130
00010000 00010000 00020000 31002031 00010000 00020000 01010000
00000000 01010000 00000000 10000003 01020000 10001330 00020000
01000002 00000000 10000022 00020000 10000130 00010000 00010000
00020000 31002031 00010000 00020000 01010000 00000000 01010000
00000000 10000002 01020000 12233211 00020000 01000002 00000000
10000022 00020000 10000130 00010000 00010000 00020000 31002031
00010000 00020000 01010000 00000000 01010000 00000000 10000002
01020000 12021232 00020000 01000002 00000000 10000022 00020000
10000130 00010000 00010000 00020000 31002031 00010000 00010000
00010000 00010000 00010000 00000000 31020001 01000001 00000000
00000000 01001000 00000000 31003001 00020000 31003002 00020000
31003003 00010000 00010000 00000000 31003010 01000001 01010000
00000000 01010000 00000000 10000001 01020000 12020321 00010000
00020000 01000002 00000000 10000022 00020000 20000103 00010000
00010000 00020000 31002012 00010000 00000000 31003021 01000001
```

```
01010000 00000000 01010000 00000000 10000001 01020000 13102000
00010000 00020000 01000002 00000000 10000022 00020000 20000112
00010000 00010000 00020000 31002033 00010000 00010000 00010000
00000000 31020002 01000001 00000000 00000000 01001000 00000000
01010000 00000000 10000002 00020000 31003001 00010000 00020000
31003002 00020000 01010000 00000000 10000002 00020000 31003003
00010000 00010000 00010000 00000000 31003010 01000001 01010000
00000000 01010000 00000000 10000002 01020000 11031011 00010000
00020000 01000002 00000000 10000022 00020000 20000103 00010000
00010000 00020000 31002012 00010000 00010000 00010000 00000000
31020003 01000001 00000000 00000000 01001000 00000000 01010000
00000000 10000002 00020000 31003001 00010000 00020000 01010000
00000000 10000002 00020000 31003002 00010000 00020000 01010000
00000000 10000002 00020000 31003003 00010000 00010000 00010000
00000000 31003010 01000001 01010000 00000000 01010000 00000000
10000003 01020000 10031301 00010000 00020000 01000002 00000000
10000022 00020000 20000103 00010000 00010000 00020000 31002012
00010000 00010000 00010000 00000000 31020010 01000001 00000000
00000000 01001000 00000000 01010000 00000000 10000012 00020000
31003001 00010000 00020000 01010000 00000000 10000012 00020000
31003002 00010000 00020000 01010000 00000000 10000012 00020000
31003003 00010000 00010000 00010000 00000000 31003010 01000001
01010000 00000000 01010000 00000000 10000021 01020000 10213310
00020000 01000002 00000000 10000022 00020000 20000103 00010000
00010000 00020000 31002012 00010000 00010000 00000000 31003021
01000001 01010000 00000000 01010000 00000000 10000013 01020000
12030132 00020000 01000002 00000000 10000022 00020000 20000111
00010000 00010000 00020000 31002033 00010000 00010000 00010000
00010000 00000000 31020011 01000001 00000000 00000000 01001000
00000000 01010000 00000000 10000012 00020000 31003001 00010000
00020000 01010000 00000000 10000013 00020000 31003002 00010000
00020000 01010000 00000000 10000012 00020000 31003003 00010000
00010000 00010000 00000000 31003010 01000001 01010000 00000000
01010000 00000000 10000021 01020000 13223303 00020000 01000002
00000000 10000022 00020000 20000103 00010000 00010000 00020000
31002012 00010000 00010000 00010000 00010000 00000000 31020012
01000001 00000000 00000000 01001000 00000000 01010000 00000000
10000013 00020000 31003001 00010000 00020000 01010000 00000000
10000013 00020000 31003002 00010000 00020000 01010000 00000000
10000013 00020000 31003003 00010000 00010000 00010000 00000000
31003010 01000001 01010000 00000000 01010000 00000000 10000001
01020000 10101200 00020000 01000002 00000000 10000022 00020000
20000102 00010000 00010000 00020000 31002012 00010000 00010000
00010000 00010000 00000000 31020013 01000001 00000000 00000000
01001000 00000000 01010000 00000000 10000013 00020000 31003001
00010000 00020000 01010000 00000000 10000020 00020000 31003002
00010000 00020000 01010000 00000000 10000013 00020000 31003003
00010000 00010000 00010000 00000000 31003010 01000001 01010000
```

```
00000000 01010000 00000000 10000001 01020000 10210333 00020000
01000002 00000000 10000022 00020000 20000102 00010000 00010000
00020000 31002012 00010000 00010000 00010000 00010000 00000000
31020020 01000001 00000000 00000000 01001000 00000000 01010000
00000000 10000020 00020000 31003001 00010000 00020000 01010000
00000000 10000020 00020000 31003002 00010000 00020000 01010000
00000000 10000020 00020000 31003003 00010000 00010000 00010000
00000000 31003010 01000001 01010000 00000000 01010000 00000000
10000001 01020000 10320122 00020000 01000002 00000000 10000022
00020000 20000102 00010000 00010000 00020000 31002012 00010000
00010000 00010000 00010000 00000000 31020021 01000001 00000000
00000000 01001000 00000000 01010000 00000000 10000020 00020000
31003001 00010000 00020000 01010000 00000000 10000021 00020000
31003002 00010000 00020000 01010000 00000000 10000020 00020000
31003003 00010000 00010000 00010000 00000000 31003010 01000001
01010000 00000000 01010000 00000000 10000001 01020000 11320003
00020000 01000002 00000000 10000022 00020000 20000102 00010000
00010000 00020000 31002012 00010000 00010000 00010000 00010000
00000000 31020022 01000001 00000000 00000000 01001000 00000000
01010000 00000000 10000020 00020000 31003001 00010000 00020000
01010000 00000000 10000022 00020000 31003002 00010000 00020000
01010000 00000000 10000020 00020000 31003003 00010000 00010000
00010000 00000000 31003010 01000001 01010000 00000000 01010000
00000000 10000001 01020000 11133210 00020000 01000002 00000000
10000022 00020000 20000102 00010000 00010000 00020000 31002012
00010000 00010000 00010000 00010000 00000000 31020023 01000001
00000000 00000000 01001000 00000000 01010000 00000000 10000022
00020000 31003001 00010000 00020000 01010000 00000000 10000022
00020000 31003002 00010000 00020000 01010000 00000000 10000022
00020000 31003003 00010000 00010000 00010000 00000000 31003010
01000001 01010000 00000000 01010000 00000000 10000001 01020000
12012200 00020000 01000002 00000000 10000022 00020000 20000102
00010000 00010000 00020000 31002012 00010000 00010000 00010000
00010000 00000000 31020030 01000001 00000000 00000000 01001000
00000000 01010000 00000000 10000022 00020000 31003001 00010000
00020000 01010000 00000000 10000023 00020000 31003002 00010000
00020000 01010000 00000000 10000022 00020000 31003003 00010000
00010000 00010000 00000000 31003010 01000001 01010000 00000000
01010000 00000000 10000001 01020000 12122003 00020000 01000002
00000000 10000022 00020000 20000102 00010000 00010000 00020000
31002012 00010000 00010000 00010000 00010000 00000000 31020031
01000001 00000000 00000000 01001000 00000000 01010000 00000000
10000022 00020000 31003001 00010000 00020000 01010000 00000000
10000030 00020000 31003002 00010000 00020000 01010000 00000000
10000022 00020000 31003003 00010000 00010000 00010000 00000000
31003010 01000001 01010000 00000000 01010000 00000000 10000001
01020000 12231202 00020000 01000002 00000000 10000022 00020000
20000102 00010000 00010000 00020000 31002012 00010000 00010000
```

```
00010000 00010000 00000000 31020032 01000001 00000000 00000000
01001000 00000000 01010000 00000000 10000022 00020000 31003001
00010000 00020000 01010000 00000000 10000031 00020000 31003002
00010000 00020000 01010000 00000000 10000022 00020000 31003003
00010000 00010000 00010000 00000000 31003010 01000001 01010000
00000000 01010000 00000000 10000001 01020000 13001003 00020000
01000002 00000000 10000022 00020000 20000102 00010000 00010000
00020000 31002012 00010000 00010000 00010000 00010000 00000000
31020033 01000001 00000000 00000000 01001000 00000000 01010000
00000000 10000030 00020000 31003001 00010000 00020000 01010000
00000000 10000030 00020000 31003002 00010000 00020000 01010000
00000000 10000030 00020000 31003003 00010000 00010000 00010000
00000000 31003010 01000001 01010000 00000000 01010000 00000000
10000001 01020000 13110132 00020000 01000002 00000000 10000022
00020000 20000102 00010000 00010000 00020000 31002012 00010000
00010000 00010000 00010000 00000000 31020100 01000001 00000000
00000000 01001000 00000000 01010000 00000000 10000032 00020000
31003001 00010000 00020000 01010000 00000000 10000032 00020000
31003002 00010000 00020000 01010000 00000000 10000032 00020000
31003003 00010000 00010000 00010000 00000000 31003010 01000001
01010000 00000000 01010000 00000000 10000002 01020000 10202130
00020000 01000002 00000000 10000022 00020000 20000102 00010000
00010000 00020000 31002012 00010000 00010000 00010000 00010000
00000000 31020101 01000001 00000000 00000000 01001000 00000000
01010000 00000000 10000100 00020000 31003001 00010000 00020000
01010000 00000000 10000100 00020000 31003002 00010000 00020000
01010000 00000000 10000100 00020000 31003003 00010000 00010000
00010000 00000000 31003010 01000001 01010000 00000000 01010000
00000000 10000002 01020000 11300132 00020000 01000002 00000000
10000022 00020000 20000102 00010000 00010000 00020000 31002012
00010000 00010000 00010000 00010000 00000000 31020102 01000001
00000000 00000000 01001000 00000000 01010000 00000000 10000102
00020000 31003001 00010000 00020000 01010000 00000000 10000112
00020000 31003002 00010000 00020000 01010000 00000000 10000102
00020000 31003003 00010000 00010000 00010000 00000000 31003010
01000001 01010000 00000000 01010000 00000000 10000003 01020000
10030202 00020000 01000002 00000000 10000022 00020000 20000102
00010000 00010000 00020000 31002012 00010000 00010000 00010000
00010000 00000000 31020103 01000001 00000000 00000000 01001000
00000000 01010000 00000000 10000122 00020000 31003001 00010000
00020000 01010000 00000000 10000132 00020000 31003002 00010000
00020000 01010000 00000000 10000122 00020000 31003003 00010000
00010000 00010000 00000000 31003010 01000001 01010000 00000000
01010000 00000000 10000010 01020000 11010201 00020000 01000002
00000000 10000022 00020000 20000102 00010000 00010000 00020000
31002012 00000000 31021001 01000001 00000000 00000000 01010000
00000000 10000002 00020000 00000000 31020000 01222210 31020001
00010000 00010000 00010000 00010000 00010000 00000000 31021002
```

```
01000001 00000000 00000000 01001000 00000000 01010000 00000000
10000002 00020000 00000000 31020000 01222210 31020001 00010000
00010000 00020000 00000000 31020020 01222210 31020021 01222210
31020022 00010000 00010000 00010000 00010000 00010000 00000000
31021003 01000001 00000000 00000000 01010000 00000000 10000002
00020000 00000000 31020020 01222210 31020021 01222210 31020022
00010000 00010000 00010000 00010000 00010000 00000000 31021010
01000001 00000000 00000000 01010000 00000000 10000003 00020000
00000000 31020020 01222210 31020021 01222210 31020022 00010000
00010000 00010000 00010000 00010000 00000000 31021011 01000001
00000000 00000000 01010000 00000000 10000002 00020000 00000000
31020012 01222210 31020013 00010000 00010000 00010000 00010000
00010000 00000000 31021012 01000001 00000000 00000000 01001000
00000000 00000000 31020012 01222210 31020013 00010000 00020000
00000000 31020020 01222210 31020021 01222210 31020022 00010000
00010000 00010000 00010000 00010000 00000000 31021013 01000001
00000000 00000000 01001000 00000000 00000000 31020010 01222210
31020011 00010000 00020000 00000000 31020020 01222210 31020021
01222210 31020022 00010000 00010000 00010000 00010000 00010000
00000000 31021020 01000001 00000000 00000000 01001000 00000000
00000000 31020010 01222210 31020011 00010000 00020000 01010000
00000000 10000002 00020000 00000000 31020020 01222210 31020021
01222210 31020022 00010000 00010000 00010000 00010000 00010000
00010000 00000000 31021021 01000001 00000000 00000000 01001000
00000000 31020101 00020000 01010000 00000000 10000002 00020000
00000000 31020020 01222210 31020021 01222210 31020022 00010000
00010000 00010000 00010000 00010000 00010000 00000000 31021022
01000001 00000000 00000000 01001000 00000000 00000000 31020012
01222210 31020013 00010000 00020000 01010000 00000000 10000003
00020000 00000000 31020000 01222210 31020001 00010000 00010000
00010000 00010000 00010000 00010000 00000000 31021023 01000001
00000000 00000000 01001000 00000000 00000000 31020010 01222210
31020011 00010000 00020000 01010000 00000000 10000010 00020000
00000000 31020000 01222210 31020001 00010000 00010000 00010000
00010000 00010000 00010000 00000000 31021030 01000001 00000000
00000000 01001000 00000000 01010000 00000000 10000002 00020000
00000000 31020010 01222210 31020011 00010000 00010000 00020000
01010000 00000000 10000012 00020000 00000000 31020000 01222210
31020001 00010000 00010000 00010000 00010000 00010000 00010000
00000000 31021031 01000001 00000000 00000000 01001000 00000000
00000000 31020012 01222210 31020013 00010000 00020000 31020101
00020000 01010000 00000000 10000011 00020000 00000000 31020000
01222210 31020001 00010000 00010000 00010000 00010000 00010000
00010000 00000000 10000000 01000001 10000000 00010000 00000000
10000001 01000001 10000001 00010000 00000000 10000002 01000001
10000002 00010000 00000000 10000003 01000001 10000003 00010000
00000000 10022133 01000001 10022133 00010000 00000000 10000000
01000001 20000000 00010000 00000000 10000001 01000100 20000001
```

```
00010000 00000000 10000001 01000100 10000000 00010000 00000000
10031212 01000001 10031212 00010000 00000000 10031212 01000100
10031213 00010000 00000000 10000002 01001001 10000001 00010000
00000000 10000001 01001001 10000001 00010000 00000000 10000101
01001001 10000001 00010000 00000000 10011111 01001001 10011111
00010000 00000000 10100220 01001001 10011111 00010000 00000000
10000001 01001010 10000001 00010000 00000000 10000000 01001010
10000001 00010000 00000000 20000101 01001010 10000001 00010000
00000000 20000313 01001010 10000000 00010000 00000000 10001001
01001010 10001001 00010000 00000000 20002013 01001010 10000000
00010000 00000000 01001000 00000000 10000000 00020000 10000000
00010000 01000001 10000000 00010000 00000000 01001000 00000000
10000000 00020000 10000001 00010000 01000001 10000001 00010000
00000000 01001000 00000000 10000000 00020000 10033123 00010000
01000001 10033123 00010000 00000000 01001000 00000000 10000001
00020000 10000001 00010000 01000001 10000002 00010000 00000000
01001000 00000000 10000312 00020000 10000001 00010000 01000001
10000313 00010000 00000000 01001000 00000000 10002133 00020000
10022023 00010000 01000001 10030222 00010000 00000000 01001000
00000000 10000002 00020000 10000002 00010000 01000001 10000010
00010000 00000000 01001000 00000000 10000002 00020000 20000002
00010000 01000001 10000000 00010000 00000000 01001000 00000000
01001000 00000000 10000011 00020000 10000221 00010000 00020000
10000012 00010000 01000001 10000310 00010000 00000000 01001000
00000000 01001000 00000000 10000221 00020000 10000012 00010000
00020000 10000011 00010000 01000001 10000310 00010000 00000000
01001000 00000000 10020303 00020000 10023211 00010000 01000001
01001000 00000000 10023211 00020000 10020303 00010000 00010000
00000000 00000000 30002002 01111110 10000001 00010000 00000000
30002002 01000001 10000001 00010000 00000000 01001000 00000000
30002002 00020000 10000001 00010000 01000001 10000002 00010000
00000000 30002002 01000100 10001221 00010000 00010000 00000000
00000000 30002002 01111110 10001221 00010000 00000000 30002002
01000100 10000001 00010000 00000000 01001000 00000000 30002002
00020000 10000001 00010000 01000001 10001222 00010000 00000000
30002002 01000001 10001221 00010000 00010000 00000000 00000000
30002002 01111110 10020301 00010000 00000000 30012002 01111110
30002002 00010000 00000000 30012002 01000001 10020301 00010000
00000000 01001000 00000000 30012002 00020000 20000023 00010000
01000001 10020212 00010000 00000000 30012002 01111110 01001000
00000000 30002002 00020000 10000302 00010000 00010000 00000000
30012002 01000001 10021203 00010000 00010000 00000000 01010000
00000000 10000000 00020000 10000000 00010000 01000001 10000000
00010000 00000000 01010000 00000000 10000001 00020000 10000000
00010000 01000001 10000000 00010000 00000000 01010000 00000000
10000001 00020000 10000002 00010000 01000001 10000002 00010000
00000000 01010000 00000000 10000002 00020000 10000002 00010000
01000001 10000010 00010000 00000000 01010000 00000000 10000003
```

```
00020000 10000010 00010000 01000001 10000030 00010000 00000000
01010000 00000000 30111111 00020000 10000000 00010000 01000001
10000000 00010000 00000000 01010000 00000000 30111111 00020000
10000002 00010000 01000001 01001000 00000000 30111111 00020000
30111111 00010000 00010000 00000000 01010000 00000000 30111111
00020000 10000002 00010000 01000001 01010000 00000000 10000002
00020000 30111111 00010000 00010000 00000000 01010000 00000000
30111111 00020000 30222222 00010000 01000001 01010000 00000000
30222222 00020000 30111111 00010000 00010000 00000000 01010000
00000000 01001000 00000000 30111111 00020000 30222222 00010000
00020000 30333333 00010000 01000001 01001000 00000000 01010000
00000000 30111111 00020000 30333333 00010000 00020000 01010000
00000000 30222222 00020000 30333333 00010000 00010000 00010000
00000000 01010000 00000000 01010000 00000000 30111111 00020000
30222222 00010000 00020000 30333333 00010000 01000001 01010000
00000000 01010000 00000000 30111111 00020000 30333333 00010000
00020000 30222222 00010000 00010000 00000000 31002000 01000001
01010000 00000000 10000013 01020000 11030013 00020000 01000002
00000000 10000022 00020000 20000003 00010000 00010000 00010000
00000000 31002000 01000001 01000200 00000000 10000001 00020000
10002021 01020000 10021100 00010000 00010000 00000000 31002001
01000001 10130101 01020000 10213011 00010000 00000000 31002002
01000001 01010000 00000000 10000001 01020000 13000131 00020000
01000002 00000000 10000022 00020000 20000231 00010000 00010000
00010000 00000000 31002003 01000001 01010000 00000000 01000002
00000000 31002002 00020000 10000000 01020000 12000000 00010000
00020000 31002012 00010000 00010000 00000000 31002010 01000001
01010000 00000000 31002001 00020000 31002003 00010000 00010000
00000000 31002010 01000001 01010000 00000000 10130101 01020000
10213011 00020000 31002003 00010000 00010000 00000000 31002011
01000001 01010000 00000000 10130233 01020000 12233001 00020000
31002003 00010000 00010000 00000000 31002011 01000001 01010000
00000000 10000001 01020000 10000113 00020000 31002010 00010000
00010000 00000000 31002010 01000001 01010000 00000000 01010000
00000000 10000013 01020000 22233133 00020000 01000002 00000000
10000022 00020000 20000110 00010000 00010000 00020000 31002012
00010000 00010000 00000000 31002012 01000001 01000002 00000000
01010000 00000000 31002013 00020000 01000200 00000000 31002020
00020000 31002021 00010000 00010000 00020000 10000000 01020000
12000000 00010000 00010000 00000000 31002000 01000001 01000200
00000000 01010000 00000000 31002022 00020000 31002022 00010000
00020000 01010000 00000000 31002023 00020000 01010000 00000000
31002020 00020000 01010000 00000000 10000010 00020000 01010000
00000000 31000002 00020000 31002030 00010000 00010000 00010000
00010000 00010000 00010000 00000000 31002031 01000001 01000002
00000000 01000200 00000000 01010000 00000000 31002023 00020000
31002021 00010000 00020000 01000002 00000000 31002020 00020000
10000003 00010000 00010000 00020000 10000000 01020000 12000000
```

```
00010000 00010000 00000000 31002031 01000001 01000200 00000000
01010000 00000000 10000002 00020000 01010000 00000000 31002021
00020000 31002012 00010000 00010000 00020000 01000002 00000000
31002020 00020000 10000002 00010000 00010000 00010000 00000000
31002032 01000001 01010000 00000000 31002012 00020000 01010000
00000000 10000002 00020000 01010000 00000000 31000002 00020000
01000002 00000000 01010000 00000000 31002021 00020000 31002030
00010000 00020000 10000000 01020000 10000011 00010000 00010000
00010000 00010000 00010000 00000000 31002032 01000001 01010000
00000000 10000023 01020000 12310303 00020000 31002022 00010000
00010000 00000000 31002033 01000001 01000002 00000000 01010000
00000000 31002013 00020000 01000200 00000000 01000002 00000000
31002020 00020000 10000011 00010000 00020000 31002021 00010000
00010000 00020000 10000000 01020000 12000000 00010000 00010000
00000000 31002033 01000001 01010000 00000000 31002012 00020000
01010000 00000000 31002020 00020000 31002020 00010000 00010000
00010000 00000000 31002100 01000001 01000200 00000000 31002033
00020000 31002101 00010000 00010000 00000000 31002102 01000001
01000200 00000000 01000002 00000000 31002020 00020000 10000011
00010000 00020000 31002021 00010000 00010000 00000000 31002103
01000001 01000200 00000000 31002031 00020000 31002020 00010000
00010000 00000000 01100000 00000000 30111111 00020000 30212121
00010000 01000001 01001000 00000000 30111111 00020000 01010000
00000000 20000001 00020000 30212121 00010000 00010000 00010000
00000000 01100000 00000000 10000010 00020000 20000011 00010000
01000001 10000021 00010000 00000000 01100000 00000000 10000032
00020000 10000021 00010000 01000001 10000011 00010000 00000000
01100000 00000000 01100000 00000000 30111111 00020000 30212121
00010000 00020000 30323232 00010000 01000001 01100000 00000000
30111111 00020000 01001000 00000000 30212121 00020000 30323232
00010000 00010000 00010000 00000000 01100000 00000000 20001003
00020000 10001013 00010000 01000001 20002022 00010000 00000000
01000200 00000000 10000010 00020000 10000010 00010000 01000001
10000001 00010000 00000000 01000200 00000000 10000010 00020000
10000002 00010000 01000001 10000002 00010000 00000000 01000200
00000000 01001000 00000000 30111111 00020000 30232323 00010000
00020000 30232323 00010000 01000001 30111111 00010000 00000000
01000200 00000000 01001000 00000000 30111111 00020000 30232323
00010000 00020000 30111111 00010000 01000001 30232323 00010000
00000000 01000200 00000000 01000200 00000000 30111111 00020000
30232323 00010000 00020000 30123123 00010000 01000001 01000200
00000000 30111111 00020000 01001000 00000000 30232323 00020000
30123123 00010000 00010000 00010000 00000000 01000200 00000000
01000200 00000000 30111111 00020000 30232323 00010000 00020000
01000200 00000000 30123123 00020000 30321321 00010000 00010000
01000001 01000200 00000000 01001000 00000000 30111111 00020000
30321321 00010000 00020000 01001000 00000000 30232323 00020000
30123123 00010000 00010000 00010000 00000000 01000200 00000000
```

```
10000333 00020000 10000021 00010000 01000001 10000013 00010000
00000000 01000200 00000000 10001331 00020000 10000121 00010000
01000001 10000011 00010000 00000000 01000200 00000000 10000003
00020000 10000002 00010000 01000001 10000001 01020000 12000000
00010000 00000000 01000200 00000000 10000001 00020000 10000003
00010000 01000001 10000000 01020000 11111111 00010000 00000000
01000200 00000000 10000033 00020000 10000010 00010000 01000001
10000003 01020000 13000000 00010000 00000000 01000200 00000000
10000020 00020000 10000013 00010000 01000001 10000001 01020000
10210210 00010000 00000000 01001000 00000000 30111111 00020000
30222222 00020000 30333333 00010000 01000001 01001000 00000000
01001000 00000000 30111111 00020000 30222222 00010000 00020000
30333333 00010000 00010000 00000000 01001000 00000000 30111111
00020000 30222222 00020000 30333333 00020000 30000000 00010000
01000001 01001000 00000000 01001000 00000000 30111111 00020000
30222222 00010000 00020000 01001000 00000000 30333333 00020000
30000000 00010000 00010000 00010000 00000000 01010000 00000000
10000001 01020000 11111111 00020000 10000201 00010000 01000001
10000230 00010000 00000000 01010000 00000000 10000210 00020000
10000012 01020000 10222223 00010000 01000001 10003132 00010000
00000000 01001000 00000000 10000002 01020000 11111111 00020000
10000011 01020000 12222222 00010000 01000001 10000120 00010000
00000000 01001000 00000000 10000013 01020000 10312313 00020000
10000023 01020000 11021021 00010000 01000001 10000102 01020000
12000000 00010000 00000000 10000000 01000001 10000000 00010000
00000000 10000001 01000001 10000001 00010000 00000000 10000002
01000001 10000002 00010000 00000000 10000003 01000001 10000003
00010000 00000000 10022133 01000001 10022133 00010000 00000000
10000000 01000001 20000000 00010000 00000000 10000001 01000100
20000001 00010000 00000000 10000001 01000100 10000000 00010000
00000000 10031212 01000001 10031212 00010000 00000000 10031212
01000100 10031213 00010000 00000000 01000002 00000000 10000001
00020000 30333333 00010000 01000001 10000001 00010000 00000000
01000002 00000000 10000010 00020000 10000000 01020000 12000000
00010000 01000001 10000002 00010000 00000000 01000002 00000000
10000021 00020000 10000000 01020000 12000000 00010000 01000001
10000003 00010000 00000000 01000002 00000000 01010000 00000000
30333333 00020000 30333333 00010000 00020000 10000000 01020000
12000000 00010000 01000001 30333333 00010000 00000000 01000002
00000000 30333333 00020000 10000002 00010000 01000001 01010000
00000000 30333333 00020000 30333333 00010000 00010000 00000000
01000002 00000000 30333333 00020000 10000003 00010000 01000001
01010000 00000000 01010000 00000000 30333333 00020000 30333333
00010000 00020000 30333333 00010000 00010000 00000000 01000002
00000000 01000002 00000000 30333333 00020000 30232323 00010000
00020000 30222111 00010000 01000001 01000002 00000000 30333333
00020000 01010000 00000000 30232323 00020000 30222111 00010000
00010000 00010000 00000000 01000002 00000000 10000003 00020000
```

```
10000010 00010000 01000001 10001101 00010000 00000000 01000002
00000000 10000010 00020000 10000001 01020000 12000000 00010000
01000001 10000020 00010000 00000000 01000002 00000000 10000002
00020000 10000000 01020000 12000000 00010000 01000001 10000001
01020000 11222002 00010000 00000000 01000002 00000000 10000011
00020000 10000000 01020000 12000000 00010000 01000001 10000002
01020000 10330123 00010000 00000000 01000002 00000000 20000001
00020000 10000000 01020000 12000000 00010000 01000001 01000030
00000000 10000001 00010000 00010000 00000000 01000002 00000000
20000010 00020000 10000000 01020000 12000000 00010000 01000001
01000030 00000000 10000002 00010000 00010000 00000000 01010000
00000000 01000030 00000000 10000001 00010000 00020000 01000030
00000000 10000001 00010000 00010000 01000001 20000001 00010000
00000000 01010000 00000000 01001000 00000000 30111222 00020000
01000030 00000000 30222000 00010000 00010000 00020000 01001000
00000000 30222111 00020000 01000030 00000000 30333000 00010000
00010000 00010000 01000001 01001000 00000000 01010000 00000000
30111222 00020000 30222111 00010000 01100000 01010000 00000000
30222000 00020000 30333000 00010000 00020000 01000030 00000000
01001000 00000000 01010000 00000000 30222000 00020000 30222111
00010000 00020000 01010000 00000000 30111222 00020000 30333000
00010000 00010000 00010000 00010000 00010000 00000000 31130000
01000001 00000000 00000000 31003010 01000001 01010000 00000000
01010000 00000000 10000021 01020000 10203130 00020000 01000002
00000000 10000022 00020000 20000211 00010000 00010000 00020000
31002012 00010000 00010000 00000000 31003023 01000001 01010000
00000000 01010000 00000000 10000003 01020000 13233322 00020000
01000002 00000000 10000022 00020000 20000131 00010000 00010000
00020000 31002100 00010000 00010000 00000000 31003030 01000001
01010000 00000000 01010000 00000000 10000001 01020000 10033113
00020000 01000002 00000000 10000022 00020000 20000122 00010000
00010000 00020000 31002102 00010000 00010000 00000000 31003012
01000001 01010000 00000000 01010000 00000000 10000010 01020000
11031213 00020000 01000002 00000000 10000022 00020000 10000223
00010000 00010000 00020000 31002031 00010000 00010000 00000000
31003031 01000001 01010000 00000000 01010000 00000000 10000002
01020000 12230303 00020000 01000002 00000000 10000022 00020000
10000330 00010000 00010000 00020000 31002103 00010000 00010000
00000000 31003032 00000000 31130100 00020000 31130200 00020000
31130300 00020000 31131000 00020000 31131100 00020000 31131200
00020000 31131300 00020000 31132000 00010000 00010000 00000000
31003010 01000001 01010000 00000000 31003010 00020000 01001000
00000000 01010000 00000000 10000000 01020000 12330002 00020000
31020000 00010000 00020000 01010000 00000000 10000000 01020000
10333213 00020000 31020003 00010000 00020000 01010000 00000000
10000000 01020000 10001332 00020000 31020020 00010000 00020000
01010000 00000000 10000000 01020000 10000300 00020000 31020010
00010000 00020000 01010000 00000000 10000000 01020000 10000121
```

```
00020000 31020103 00010000 00020000 01010000 00000000 10000000
01020000 10000102 00020000 31020101 00010000 00020000 01010000
00000000 10000000 01020000 10000033 00020000 31020012 00010000
00020000 01010000 00000000 10000000 01020000 10000023 00020000
31020100 00010000 00020000 01010000 00000000 10000000 01020000
10000002 00020000 31020001 00010000 00020000 01010000 00000000
10000000 01020000 10000103 00020000 31020023 00010000 00020000
01010000 00000000 01010000 00000000 10000003 01020000 11002203
00020000 01000002 00000000 10000022 00020000 20000005 00010000
00010000 00020000 31020011 00010000 00020000 01010000 00000000
01010000 00000000 10000001 01020000 12001323 00020000 01000002
00000000 10000022 00020000 20000005 00010000 00010000 00020000
31020022 00010000 00020000 01010000 00000000 01010000 00000000
10000002 01020000 13130022 00020000 01000002 00000000 10000022
00020000 20000006 00010000 00010000 00020000 31020021 00010000
00020000 01010000 00000000 01010000 00000000 10000003 01020000
11030303 00020000 01000002 00000000 10000022 00020000 20000006
00010000 00010000 00020000 31020013 00010000 00020000 01010000
00000000 01010000 00000000 10000001 01020000 10303031 00020000
01000002 00000000 10000022 00020000 20000010 00010000 00010000
00020000 31020030 00010000 00020000 01010000 00000000 01010000
00000000 10000003 01020000 12000000 00020000 01000002 00000000
10000022 00020000 20000006 00010000 00010000 00020000 31020031
00010000 00020000 10000000 01020000 10000031 00010000 00010000
00010000 00010000 00010000 00000000 31330100 01000001 00000000
00000000 31003010 01000001 01001000 00000000 01001000 00000000
10000001 01020000 12010123 00020000 01000002 00000000 10000022
00020000 10000133 00010000 00010000 00020000 31002012 00010000
00010000 00000000 31003023 01001010 01001000 00000000 01001000
00000000 10000010 01020000 13300311 00020000 01000002 00000000
10000022 00020000 20000132 00010000 00010000 00020000 31002100
00010000 00010000 00000000 31003023 01001001 01001000 00000000
01001000 00000000 10000011 01020000 12211202 00020000 01000002
00000000 10000022 00020000 20000133 00010000 00010000 00020000
31002100 00010000 00010000 00000000 31003012 01000001 01001000
00000000 01001000 00000000 10000001 01020000 12002123 00020000
01000002 00000000 10000022 00020000 10000221 00010000 00010000
00020000 31002031 00010000 00010000 00000000 31003033 01000001
01001000 00000000 01001000 00000000 10000003 01020000 12111033
00020000 01000002 00000000 10000022 00020000 20000203 00010000
00010000 00020000 31002031 00010000 00010000 00000000 31003100
01000001 10000000 01020000 10310221 00010000 00000000 31003101
01000001 01001000 00000000 01001000 00000000 10000001 01020000
11220322 00020000 01000002 00000000 10000022 00020000 20000302
00010000 00010000 00020000 31002100 00010000 00010000 00000000
31003102 01000001 01001000 00000000 01001000 00000000 10000021
01020000 11212003 00020000 01000002 00000000 10000022 00020000
20000301 00010000 00010000 00020000 31002100 00010000 00010000
```

```
00000000 31003103 00000000 31130000 00010000 00010000 00000000
31003110 01000001 00000000 00000000 31003010 01000001 01001000
00000000 01001000 00000000 10000021 01020000 10311301 00020000
01000002 00000000 10000022 00020000 10000022 00010000 00010000
00020000 31002012 00010000 00010000 00000000 31003010 01000001
01001000 00000000 31003010 00020000 01001000 00000000 01001000
00000000 10000000 01020000 11223201 00020000 31021003 00010000
00020000 01001000 00000000 10000000 01020000 11022033 00020000
31020032 00010000 00020000 01001000 00000000 10000000 01020000
10320110 00020000 31021001 00010000 00020000 01001000 00000000
10000000 01020000 10033113 00020000 31020003 00010000 00020000
10000000 01020000 10002210 00010000 00010000 00010000 00010000
00010000 00010000 00010000 00000000 31130200 01000001 00000000
00000000 31003010 01000001 01001000 00000000 01001000 00000000
10000002 01020000 10330231 00020000 01000002 00000000 10000022
00020000 10000200 00010000 00010000 00020000 31002012 00010000
00010000 00000000 31003023 01000001 01001000 00000000 01001000
00000000 10000011 01020000 10300011 00020000 01000002 00000000
10000022 00020000 20000132 00010000 00010000 00020000 31002100
00010000 00010000 00000000 31003012 01000001 01001000 00000000
01001000 00000000 10000003 01020000 12332203 00020000 01000002
00000000 10000022 00020000 20000221 00010000 00010000 00020000
31002031 00010000 00010000 00000000 31003033 01000001 01001000
00000000 01001000 00000000 10000012 01020000 12301330 00020000
01000002 00000000 10000022 00020000 20000231 00010000 00010000
00020000 31002031 00010000 00010000 00000000 31003100 01000001
10000000 01020000 10001233 00010000 00000000 31003101 01000001
01001000 00000000 01001000 00000000 10000003 01020000 10121320
00020000 01000002 00000000 10000022 00020000 10000302 00010000
00010000 00020000 31002100 00010000 00010000 00000000 31003102
01000001 01001000 00000000 01001000 00000000 10000003 01020000
13211001 00020000 01000002 00000000 10000022 00020000 10000302
00010000 00010000 00020000 31002100 00010000 00010000 00000000
31003103 00000000 31130000 00010000 00010000 00000000 31003110
01000001 00000000 00000000 31003010 01000001 01001000 00000000
01001000 00000000 10000002 01020000 10310211 00020000 01000002
00000000 10000022 00020000 10000130 00010000 00010000 00020000
31002012 00010000 00010000 00000000 31003010 01000001 01001000
00000000 31003010 00020000 01001000 00000000 01001000 00000000
10000000 01020000 13313003 00020000 31021020 00010000 00020000
01001000 00000000 10000000 01020000 10020331 00020000 31021011
00010000 00020000 01001000 00000000 01001000 00000000 10000001
01020000 12000000 00020000 01000002 00000000 10000022 00020000
20000010 00010000 00010000 00020000 31021021 00010000 00020000
01001000 00000000 01001000 00000000 10000013 00020000 01000002
00000000 10000022 00020000 20000011 00010000 00010000 00020000
31020102 00010000 00020000 01001000 00000000 01001000 00000000
10000001 01020000 12303031 00020000 01000002 00000000 10000022
```

```
00020000 20000011 00010000 00010000 00020000 31021013 00010000
00020000 01001000 00000000 01001000 00000000 10000001 01020000
10303031 00020000 01000002 00000000 10000022 00020000 20000011
00010000 00010000 00020000 31020003 00010000 00020000 01001000
00000000 01001000 00000000 10000013 00020000 01000002 00000000
10000022 00020000 20000012 00010000 00010000 00020000 31020023
00010000 00010000 00010000 00010000 00010000 00010000 00010000
00010000 00000000 31130300 01000001 00000000 00000000 31003010
01000001 01001000 00000000 01001000 00000000 10000002 01020000
12332231 00020000 01000002 00000000 10000022 00020000 10000200
00010000 00010000 00020000 31002012 00010000 00010000 00000000
31003023 01001010 01001000 00000000 01001000 00000000 10000002
01020000 11112012 00020000 01000002 00000000 10000022 00020000
20000132 00010000 00010000 00020000 31002100 00010000 00010000
00000000 31003023 01001001 01001000 00000000 01001000 00000000
10000001 01020000 11030132 00020000 01000002 00000000 10000022
00020000 20000132 00010000 00010000 00020000 31002100 00010000
00010000 00000000 31003012 01001010 01001000 00000000 01001000
00000000 10000003 01020000 13302033 00020000 01000002 00000000
10000022 00020000 10000230 00010000 00010000 00020000 31002031
00010000 00010000 00000000 31003012 01001001 01001000 00000000
01001000 00000000 10000003 01020000 13232313 00020000 01000002
00000000 10000022 00020000 10000230 00010000 00010000 00020000
31002031 00010000 00010000 00000000 31003033 01000001 01001000
00000000 01001000 00000000 10000021 01020000 11001200 00020000
01000002 00000000 10000022 00020000 10000231 00010000 00010000
00020000 31002031 00010000 00010000 00000000 31003100 01000001
10000000 01020000 10010102 00010000 00000000 31003101 01000001
01001000 00000000 01001000 00000000 10000011 01020000 13122201
00020000 01000002 00000000 10000022 00020000 10000302 00010000
00010000 00020000 31002100 00010000 00010000 00000000 31003102
01000001 01001000 00000000 01001000 00000000 10000001 01020000
12121021 00020000 01000002 00000000 10000022 00020000 10000300
00010000 00010000 00020000 31002100 00010000 00010000 00000000
31003103 00000000 31130000 00010000 00010000 00000000 31003032
00000000 31130303 00010000 00010000 00000000 31003010 01000001
01001000 00000000 31003010 00020000 01001000 00000000 01001000
00000000 10000000 01020000 11023122 00020000 31020103 00010000
00020000 01001000 00000000 10000000 01020000 11030303 00020000
31020020 00010000 00020000 01001000 00000000 10000000 01020000
10203311 00020000 31020100 00010000 00020000 01001000 00000000
10000000 01020000 10130022 00020000 31020033 00010000 00020000
01001000 00000000 10000000 01020000 10013023 00020000 31020101
00010000 00020000 10000000 01020000 10211220 00010000 00010000
00010000 00000000 31003110 01000001 00000000 00000000 31003010
01001010 01001000 00000000 01001000 00000000 10000011 01020000
10211231 00020000 01000002 00000000 10000022 00020000 10000102
00010000 00010000 00020000 31002012 00010000 00010000 00000000
```

```
31003010 01001001 01001000 00000000 01001000 00000000 10000011
01020000 10212011 00020000 01000002 00000000 10000022 00020000
10000102 00010000 00010000 00020000 31002012 00010000 00010000
00000000 31003010 01000001 01001000 00000000 31003010 00020000
01001000 00000000 01001000 00000000 10000000 01020000 12013012
00020000 31021011 00010000 00020000 01001000 00000000 10000000
01020000 10311120 00020000 31021003 00010000 00020000 01001000
00000000 10000000 01020000 10002103 00020000 31020102 00010000
00020000 01001000 00000000 10000000 01020000 10001001 00020000
31021002 00010000 00020000 01001000 00000000 01001000 00000000
01001001 10000003 01020000 13110133 00020000 01000002 00000000
10000022 00020000 20000010 00010000 00010000 00020000 31021020
00010000 00020000 01001000 00000000 01001000 00000000 10000001
01020000 13033213 00020000 01000002 00000000 10000022 00020000
20000011 00010000 00010000 00020000 31020023 00010000 00020000
01001000 00000000 01001000 00000000 10000011 01020000 10320110
00020000 01000002 00000000 10000022 00020000 20000012 00010000
00010000 00020000 31020003 00010000 00020000 01001000 00000000
01001000 00000000 01001001 10000001 01020000 12332232 00020000
01000002 00000000 10000022 00020000 20000012 00010000 00010000
00020000 31021023 00010000 00020000 01001000 00000000 01001000
00000000 10000003 01020000 12000000 00020000 01000002 00000000
10000022 00020000 20000020 00010000 00010000 00020000 31021010
00010000 00010000 00010000 00010000 00010000 00010000 00000000
31003111 00010000 00000000 31003112 00010000 00010000 00010000
00000000 31130303 01000001 00000000 00000000 31003010 01000001
01001000 00000000 01001000 00000000 10000003 01020000 11200101
00020000 01000002 00000000 10000022 00020000 10000132 00010000
00010000 00020000 31002012 00010000 00010000 00000000 31003023
01001010 01001000 00000000 01001000 00000000 10000002 01020000
13000231 00020000 01000002 00000000 10000022 00020000 10000132
00010000 00010000 00020000 31002100 00010000 00010000 00000000
31003023 01001001 01001000 00000000 01001000 00000000 10000010
01020000 13300311 00020000 01000002 00000000 10000022 00020000
20000133 00010000 00010000 00020000 31002100 00010000 00010000
00000000 31003012 01001001 01001000 00000000 01001000 00000000
10000001 01020000 10102332 00020000 01000002 00000000 10000022
00020000 10000221 00010000 00010000 00020000 31002031 00010000
00010000 00000000 31003012 01001010 01001000 00000000 01001000
00000000 10000001 01020000 10103110 00020000 01000002 00000000
10000022 00020000 10000221 00010000 00010000 00020000 31002031
00010000 00010000 00000000 31003033 01000001 01001000 00000000
01001000 00000000 10000002 01020000 11200312 00020000 01000002
00000000 10000022 00020000 10000223 00010000 00010000 00020000
31002031 00010000 00010000 00000000 31003100 01000001 10000000
01020000 10032003 00010000 00000000 31003101 01000001 01001000
00000000 01001000 00000000 10000010 01020000 11200323 00020000
01000002 00000000 10000022 00020000 10000301 00010000 00010000
```

```
00020000 31002100 00010000 00010000 00000000 31003102 01000001
01001000 00000000 01001000 00000000 10000010 01020000 11200323
00020000 01000002 00000000 10000022 00020000 10000301 00010000
00010000 00020000 31002100 00010000 00010000 00000000 31003103
00000000 31130300 00010000 00010000 00000000 31003110 01000001
00000000 00000000 31003010 01000001 01001000 00000000 01001000
00000000 10000003 01020000 11010012 00020000 01000002 00000000
10000022 00020000 10000023 00010000 00010000 00020000 31002012
00010000 00010000 00010000 00010000 00010000 00010000 00000000
31131000 01000001 00000000 00000000 31003010 01000001 01001000
00000000 01001000 00000000 10000002 01020000 13303002 00020000
01000002 00000000 10000022 00020000 10000133 00010000 00010000
00020000 31002012 00010000 00010000 00000000 31003023 01001010
01001000 00000000 01001000 00000000 10000001 01020000 13210100
00020000 01000002 00000000 10000022 00020000 20000132 00010000
00010000 00020000 31002100 00010000 00010000 00000000 31003023
01001001 01001000 00000000 01001000 00000000 10000001 01020000
11100011 00020000 01000002 00000000 10000022 00020000 20000132
00010000 00010000 00020000 31002100 00010000 00010000 00000000
31003012 01001010 01001000 00000000 01001000 00000000 10000002
01020000 10121323 00020000 01000002 00000000 10000022 00020000
10000221 00010000 00010000 00020000 31002031 00010000 00010000
00000000 31003012 01001001 01001000 00000000 01001000 00000000
10000002 01020000 10112301 00020000 01000002 00000000 10000022
00020000 10000221 00010000 00010000 00020000 31002031 00010000
00010000 00000000 31003033 01000001 01001000 00000000 01001000
00000000 10000001 01020000 11221002 00020000 01000002 00000000
10000022 00020000 10000232 00010000 00010000 00020000 31002031
00010000 00010000 00000000 31003100 01000001 10000000 01020000
10113322 00010000 00000000 31003101 01000001 01001000 00000000
01001000 00000000 10000001 01020000 10121312 00020000 01000002
00000000 10000022 00020000 10000303 00010000 00010000 00020000
31002100 00010000 00010000 00000000 31003102 01000001 01001000
00000000 01001000 00000000 10000001 01020000 12210323 00020000
01000002 00000000 10000022 00020000 10000300 00010000 00010000
00020000 31002100 00010000 00010000 00000000 31003103 00000000
31130000 00010000 00010000 00000000 31003110 01000001 00000000
00000000 31003010 01000001 01001000 00000000 01001000 00000000
10000001 01020000 10212010 00020000 01000002 00000000 10000022
00020000 10000120 00010000 00010000 00020000 31002012 00010000
00010000 00000000 31003010 01000001 01001000 00000000 31003010
00020000 01001000 00000000 01001000 00000000 10000000 01020000
13310001 00020000 31021020 00010000 00020000 01001000 00000000
10000000 01020000 10010013 00020000 31021011 00010000 00020000
01001000 00000000 10000000 01020000 10010012 00020000 31020102
00010000 00020000 01001000 00000000 10000000 01020000 10000111
00020000 31021003 00010000 00020000 01001000 00000000 10000000
01020000 10000023 00020000 31021013 00010000 00020000 01001000
```

```
00000000 10000000 01020000 10000011 00020000 31021002 00010000
00020000 01001000 00000000 01001000 00000000 10000001 01020000
11030303 00020000 01000002 00000000 10000022 00020000 20000010
00010000 00010000 00020000 31021012 00010000 00020000 01001000
00000000 01001000 00000000 10000002 01020000 12000000 00020000
01000002 00000000 10000022 00020000 20000012 00010000 00010000
00020000 31020023 00010000 00010000 00010000 00010000 00010000
00010000 00010000 00010000 00000000 31131100 01000001 00000000
00000000 31003010 01000001 01001000 00000000 01001000 00000000
10000020 01020000 12321110 00020000 01000002 00000000 10000022
00020000 10000202 00010000 00010000 00020000 31002012 00010000
00010000 00000000 31003023 01000001 01001000 00000000 01001000
00000000 10000001 01020000 10222021 00020000 01000002 00000000
10000022 00020000 20000132 00010000 00010000 00020000 31002100
00010000 00010000 00000000 31003012 01001010 01001000 00000000
01001000 00000000 10000010 01020000 11230120 00020000 01000002
00000000 10000022 00020000 10000222 00010000 00010000 00020000
31002031 00010000 00010000 00000000 31003012 01001001 01001000
00000000 01001000 00000000 10000010 01020000 10202322 00020000
01000002 00000000 10000022 00020000 10000222 00010000 00010000
00020000 31002031 00010000 00010000 00000000 31003033 01000001
01001000 00000000 01001000 00000000 10000010 01020000 13101022
00020000 01000002 00000000 10000022 00020000 10000232 00010000
00010000 00020000 31002031 00010000 00010000 00000000 31003100
01000001 10000000 01020000 10030121 00010000 00000000 31003101
01000001 01001000 00000000 01001000 00000000 10000012 01020000
13301012 00020000 01000002 00000000 10000022 00020000 10000303
00010000 00010000 00020000 31002100 00010000 00010000 00000000
31003102 01000001 01001000 00000000 01001000 00000000 10000012
01020000 12200220 00020000 01000002 00000000 10000022 00020000
10000233 00010000 00010000 00020000 31002100 00010000 00010000
00000000 31003103 00000000 31130000 00010000 00010000 00000000
31003010 01000001 01001000 00000000 31003010 00020000 01001000
00000000 01001000 00000000 10000000 01020000 12311311 00020000
31021001 00010000 00020000 01001000 00000000 10000000 01020000
10331130 00020000 31020003 00010000 00020000 10000000 01020000
10030303 00010000 00010000 00010000 00010000 00010000 00000000
31131200 01000001 00000000 00000000 31003010 01000001 01001000
00000000 01001000 00000000 10000002 01020000 12130231 00020000
01000002 00000000 10000022 00020000 10000202 00010000 00010000
00020000 31002012 00010000 00010000 00000000 31003023 01000001
01001000 00000000 01001000 00000000 10000021 01020000 11311100
00020000 01000002 00000000 10000022 00020000 20000133 00010000
00010000 00020000 31002100 00010000 00010000 00000000 31003012
01001010 01001000 00000000 01001000 00000000 10000003 01020000
12322212 00020000 01000002 00000000 10000022 00020000 10000222
00010000 00010000 00020000 31002031 00010000 00010000 00000000
31003012 01001001 01001000 00000000 01001000 00000000 10000003
```

```
01020000 11131011 00020000 01000002 00000000 10000022 00020000
10000222 00010000 00010000 00020000 31002031 00010000 00010000
00000000 31003033 01000001 01001000 00000000 01001000 00000000
10000020 01020000 13132131 00020000 01000002 00000000 10000022
00020000 10000232 00010000 00010000 00020000 31002031 00010000
00010000 00000000 31003100 01000001 10000000 01020000 31131200
00010000 00000000 31003101 01000001 01001000 00000000 01001000
00000000 10000001 01020000 12330122 00020000 01000002 00000000
10000022 00020000 10000310 00010000 00010000 00020000 31002100
00010000 00010000 00000000 31003102 01000001 01001000 00000000
01001000 00000000 10000013 01020000 10131211 00020000 01000002
00000000 10000022 00020000 10000233 00010000 00010000 00020000
31002100 00010000 00010000 00000000 31003103 00000000 31130000
00010000 00010000 00010000 00010000 00000000 31131300 01000001
00000000 00000000 31003010 01000001 01001000 00000000 01001000
00000000 10000003 01020000 13331203 00020000 01000002 00000000
10000002 00020000 10000201 00010000 00010000 00020000 31002012
00010000 00010000 00000000 31003023 01000001 01001000 00000000
01001000 00000000 10000011 01020000 11131100 00020000 01000002
00000000 10000022 00020000 20000133 00010000 00010000 00020000
31002100 00010000 00010000 00000000 31003012 01001010 01001000
00000000 01001000 00000000 10000001 01020000 12110311 00020000
01000002 00000000 10000022 00020000 10000222 00010000 00010000
00020000 31002031 00010000 00010000 00000000 31003012 01001001
01001000 00000000 01001000 00000000 10000001 01020000 12023203
00020000 01000002 00000000 10000022 00020000 10000222 00010000
00010000 00020000 31002031 00010000 00010000 00000000 31003033
01000001 01001000 00000000 01001000 00000000 10000001 01020000
13013221 00020000 01000002 00000000 10000022 00020000 10000233
00010000 00010000 00020000 31002031 00010000 00010000 00000000
31003100 01000001 10000000 01020000 10030011 00010000 00000000
31003101 01000001 01001000 00000000 01001000 00000000 10000010
01020000 13233211 00020000 01000002 00000000 10000022 00020000
10000310 00010000 00010000 00020000 31002100 00010000 00010000
00000000 31003102 01000001 01001000 00000000 01001000 00000000
10000001 01020000 10212231 00020000 01000002 00000000 10000022
00020000 10000300 00010000 00010000 00020000 31002100 00010000
00010000 00000000 31003103 00000000 31130000 00010000 00010000
00010000 00010000 00000000 31132000 01000001 00000000 00000000
31003010 01000001 01001000 00000000 01001000 00000000 10000010
01020000 12310330 00020000 01000002 00000000 10000022 00020000
10000201 00010000 00010000 00020000 31002012 00010000 00010000
00000000 31003023 01000001 01001000 00000000 01001000 00000000
10000011 01020000 10110332 00020000 01000002 00000000 10000022
00020000 20000133 00010000 00010000 00020000 31002100 00010000
00010000 00000000 31003012 01001010 01001000 00000000 01001000
00000000 10000001 01020000 12020033 00020000 01000002 00000000
10000022 00020000 10000222 00010000 00010000 00020000 31002031
```

```
00010000 00010000 00000000 31003012 01001001 01001000 00000000
01001000 00000000 10000001 01020000 12001203 00020000 01000002
00000000 10000022 00020000 10000222 00010000 00010000 00020000
31002031 00010000 00010000 00000000 31003033 01000001 01001000
00000000 01001000 00000000 10000002 01020000 13021110 00020000
01000002 00000000 10000022 00020000 10000233 00010000 00010000
00020000 31002031 00010000 00010000 00000000 31003100 01000001
10000000 01020000 10002031 00010000 00000000 31003101 01000001
01001000 00000000 01001000 00000000 10000021 01020000 12211121
00020000 01000002 00000000 10000022 00020000 10000310 00010000
00010000 00020000 31002100 00010000 00010000 00000000 31003102
01000001 01001000 00000000 01001000 00000000 10000001 01020000
10103122 00020000 01000002 00000000 10000022 00020000 10000300
00010000 00010000 00020000 31002100 00010000 00010000 00000000
31003103 00000000 31130000 00010000 00010000 00010000 00010000
00000000 00000000 30002002 01111110 10000001 00010000 00000000
30002002 01000001 10000001 00010000 00000000 01001000 00000000
30002002 00020000 10000001 00010000 01000001 10000002 00010000
00000000 30002002 01000100 10001221 00010000 00010000 00000000
00000000 30002002 01111110 10001221 00010000 00000000 30002002
01000100 10000001 00010000 00000000 01001000 00000000 30002002
00020000 10000001 00010000 01000001 10001222 00010000 00000000
30002002 01000001 10001221 00010000 00010000 00000000 00000000
30002002 01111110 10020301 00010000 00000000 30012002 01111110
30002002 00010000 00000000 30012002 01000001 10020301 00010000
00000000 01001000 00000000 30012002 00020000 20000023 00010000
01000001 10020212 00010000 00000000 30012002 01111110 01001000
00000000 30002002 00020000 10000302 00010000 00010000 00000000
30012002 01000001 10021203 00010000 00010000 00000000 00000000
30222222 01111110 10000000 00010000 00000000 30333333 01000001
01000200 00000000 01001000 00000000 01010000 00000000 20000001
00020000 30232323 00010000 00020000 01000002 00000000 01100000
00000000 01010000 00000000 30232323 00020000 30232323 00010000
00020000 01010000 00000000 10000010 00020000 01010000 00000000
30121212 00020000 30303030 00010000 00010000 00020000 10000000
01020000 12000000 00010000 00010000 00010000 00020000 01010000
00000000 10000002 00020000 30121212 00010000 00010000 00010000
01222210 00000000 30333333 01000001 01000200 00000000 01100000
00000000 01010000 00000000 20000001 00020000 30232323 00010000
00020000 01000002 00000000 01100000 00000000 01010000 00000000
30232323 00020000 30232323 00010000 00020000 01010000 00000000
10000010 00020000 01010000 00000000 30121212 00020000 30303030
00010000 00010000 00020000 10000000 01020000 12000000 00010000
00010000 00010000 00020000 01010000 00000000 10000002 00020000
30121212 00010000 00010000 00010000 00000000 30222222 01000001
01001000 00000000 01010000 00000000 30121212 00020000 01010000
00000000 30333333 00020000 30333333 00010000 00010000 00020000
01001000 00000000 01010000 00000000 30232323 00020000 30333333
```

```
00010000  00020000  30303030  00010000  00010000  00010000  00010000
00000000  01000002  00000000  30101010  00020000  01010000  00000000
20000001  00020000  30313131  00010000  00010000  01000001  01000200
00000000  10000001  00020000  01000002  00000000  30101010  00020000
30313131  00010000  00010000  00010000  00000000  01000002  00000000
01000030  00000000  10000001  00010000  00020000  10000000  01020000
12000000  00010000  01000001  01001000  00000000  01000002  00000000
10000002  00020000  20000000  01020000  12000000  00010000  00020000
01000030  00000000  01000002  00000000  10000002  00020000  20000000
01020000  12000000  00010000  00010000  00010000  00010000  00000000
01000002  00000000  31000001  00020000  01000030  00000000  31000002
00010000  00010000  01000001  20000001  00010000  00000000  31000001
01000001  10000002  01020000  12313320  00010000  00000000  31000002
01000001  10000003  01020000  10210033  00010000  00000000  01000002
00000000  31000001  00020000  01000030  00000000  01000200  00000000
31000002  00020000  10000002  00010000  00010000  00010000  01000001
01000030  00000000  20000001  00010000  00010000  00000000  01000002
00000000  31000001  00020000  10000002  00010000  01000001  10000013
01020000  11203212  00010000  00000000  01000002  00000000  31000002
00020000  31000001  00010000  01000001  10000112  01020000  11311202
00010000  00000000  31002000  01000001  01010000  00000000  10000013
01020000  11030013  00020000  01000002  00000000  10000022  00020000
20000003  00010000  00010000  00010000  00000000  31002000  01000001
01000200  00000000  10000001  00020000  10002021  01020000  10021100
00010000  00010000  00000000  31002001  01000001  10130101  01020000
10213011  00010000  00000000  31002002  01000001  01010000  00000000
10000001  01020000  13000131  00020000  01000002  00000000  10000022
00020000  20000231  00010000  00010000  00010000  00000000  31002003
01000001  01010000  00000000  01000002  00000000  31002002  00020000
10000000  01020000  12000000  00010000  00020000  31002012  00010000
00010000  00000000  31002010  01000001  01010000  00000000  31002001
00020000  31002003  00010000  00010000  00000000  31002010  01000001
01010000  00000000  10130101  01020000  10213011  00020000  31002003
00010000  00010000  00000000  31002011  01000001  01010000  00000000
10130233  01020000  12233001  00020000  31002003  00010000  00010000
00000000  31002011  01000001  01010000  00000000  10000001  01020000
10000113  00020000  31002010  00010000  00010000  00000000  31002010
01000001  01010000  00000000  01010000  00000000  10000013  01020000
22233133  00020000  01000002  00000000  10000022  00020000  20000110
00010000  00010000  00020000  31002012  00010000  00010000  00000000
31002012  01000001  01000002  00000000  01010000  00000000  31002013
00020000  01000200  00000000  31002020  00020000  31002021  00010000
00010000  00020000  10000000  01020000  12000000  00010000  00010000
00000000  31002000  01000001  01000200  00000000  01010000  00000000
31002022  00020000  31002022  00010000  00020000  01010000  00000000
31002023  00020000  01010000  00000000  31002020  00020000  01010000
00000000  10000010  00020000  01010000  00000000  31000002  00020000
31002030  00010000  00010000  00010000  00010000  00010000  00010000
```

```
00000000 31002031 01000001 01000002 00000000 01000200 00000000
01010000 00000000 31002023 00020000 31002021 00010000 00020000
01000002 00000000 31002020 00020000 10000003 00010000 00010000
00020000 10000000 01020000 12000000 00010000 00010000 00000000
31002031 01000001 01000200 00000000 01010000 00000000 10000002
00020000 01010000 00000000 31002021 00020000 31002012 00010000
00010000 00020000 01000002 00000000 31002020 00020000 10000002
00010000 00010000 00010000 00000000 31002032 01000001 01010000
00000000 31002012 00020000 01010000 00000000 10000002 00020000
01010000 00000000 31000002 00020000 01000002 00000000 01010000
00000000 31002021 00020000 31002030 00010000 00020000 10000000
01020000 10000011 00010000 00010000 00010000 00010000 00010000
00000000 31002032 01000001 01010000 00000000 10000023 01020000
12310303 00020000 31002022 00010000 00010000 00000000 31002033
01000001 01000002 00000000 01010000 00000000 31002013 00020000
01000200 00000000 01000002 00000000 31002020 00020000 10000011
00010000 00020000 31002021 00010000 00010000 00020000 10000000
01020000 12000000 00010000 00010000 00000000 31002033 01000001
01010000 00000000 31002012 00020000 01010000 00000000 31002020
00020000 31002020 00010000 00010000 00010000 00000000 31002100
01000001 01000200 00000000 31002033 00020000 31002101 00010000
00010000 00000000 31002102 01000001 01000200 00000000 01000002
00000000 31002020 00020000 10000011 00010000 00020000 31002021
00010000 00010000 00000000 31002103 01000001 01000200 00000000
31002031 00020000 31002020 00010000 00010000 00000000 31003001
01000001 00000000 00000000 31003010 01000001 31002010 00010000
00000000 31003011 01000001 01001000 00000000 20000001 00020000
31002022 00010000 00010000 00000000 31003012 01000001 01001000
00000000 01001000 00000000 10000011 01000030 11221330 00020000
01000002 00000000 10000022 00020000 10000103 00010000 00010000
00020000 31002031 00010000 00010000 00010000 00010000 00000000
31003003 01000001 00000000 00000000 31003010 01000001 31002003
00010000 00000000 31003011 01000001 31002022 00010000 00000000
31003012 01000001 10000000 00010000 00010000 00010000 00000000
31003002 01000001 00000000 00000000 31003010 01000001 31002011
00010000 00000000 31003011 01000001 10000000 00010000 00010000
00010000 00000000 01001000 00000000 30111111 00020000 30222222
00020000 30333333 00010000 01000001 01001000 00000000 01001000
00000000 30111111 00020000 30222222 00010000 00020000 30333333
00010000 00010000 00000000 01001000 00000000 30111111 00020000
30222222 00020000 30333333 00020000 30000000 00010000 01000001
01001000 00000000 01001000 00000000 30111111 00020000 30222222
00010000 00020000 01001000 00000000 30333333 00020000 30000000
00010000 00010000 00010000 00000000 31020000 01000001 00000000
00000000 01001000 00000000 31003001 00020000 31003003 00010000
00010000 00000000 31003010 01000001 01010000 00000000 01010000
00000000 10000013 01020000 12232132 00020000 01000002 00000000
10000022 00020000 20000110 00010000 00010000 00020000 31002012
```

```
00010000 00010000 00000000 31003013 01000001 01010000 00000000
01010000 00000000 10000003 01020000 11012100 00020000 01000002
00000000 10000022 00020000 10000120 00010000 00010000 00020000
31002031 00010000 00010000 00000000 31003020 01000001 01010000
00000000 01010000 00000000 10000001 01020000 10131022 00020000
01000002 00000000 10000022 00020000 20000123 00010000 00010000
00020000 31002033 00010000 00010000 00000000 31003021 01000001
10000000 00010000 00000000 31003022 01000001 00000000 01010000
00000000 01010000 00000000 10000001 01020000 11032103 00020000
01000002 00000000 10000022 00020000 10000202 00010000 00010000
00020000 31002031 00010000 00020000 01010000 00000000 01010000
00000000 10000001 01020000 10033210 00020000 01000002 00000000
10000022 00020000 10000130 00010000 00010000 00020000 31002031
00010000 00020000 01010000 00000000 01010000 00000000 10000003
01020000 10001330 00020000 01000002 00000000 10000022 00020000
10000130 00010000 00010000 00020000 31002031 00010000 00020000
01010000 00000000 01010000 00000000 10000002 01020000 12233211
00020000 01000002 00000000 10000022 00020000 10000130 00010000
00010000 00020000 31002031 00010000 00020000 01010000 00000000
01010000 00000000 10000002 01020000 12021232 00020000 01000002
00000000 10000022 00020000 10000130 00010000 00010000 00020000
31002031 00010000 00010000 00010000 00010000 00010000 00000000
31020001 01000001 00000000 00000000 01001000 00000000 31003001
00020000 31003002 00020000 31003003 00010000 00010000 00000000
31003010 01000001 01010000 00000000 01010000 00000000 10000001
01020000 12020321 00010000 00020000 01000002 00000000 10000022
00020000 20000103 00010000 00010000 00020000 31002012 00010000
00000000 31003021 01000001 01010000 00000000 01010000 00000000
10000001 01020000 13102000 00010000 00020000 01000002 00000000
10000022 00020000 20000112 00010000 00010000 00020000 31002033
00010000 00010000 00010000 00000000 31020002 01000001 00000000
00000000 01001000 00000000 01010000 00000000 10000002 00020000
31003001 00010000 00020000 31003002 00020000 01010000 00000000
10000002 00020000 31003003 00010000 00010000 00010000 00000000
31003010 01000001 01010000 00000000 01010000 00000000 10000002
01020000 11031011 00010000 00020000 01000002 00000000 10000022
00020000 20000103 00010000 00010000 00020000 31002012 00010000
00010000 00010000 00000000 31020003 01000001 00000000 00000000
01001000 00000000 01010000 00000000 10000002 00020000 31003001
00010000 00020000 01010000 00000000 10000002 00020000 31003002
00010000 00020000 01010000 00000000 10000002 00020000 31003003
00010000 00010000 00010000 00000000 31003010 01000001 01010000
00000000 01010000 00000000 10000003 01020000 10031301 00010000
00020000 01000002 00000000 10000022 00020000 20000103 00010000
00010000 00020000 31002012 00010000 00010000 00010000 00000000
31020010 01000001 00000000 00000000 01001000 00000000 01010000
00000000 10000012 00020000 31003001 00010000 00020000 01010000
00000000 10000012 00020000 31003002 00010000 00020000 01010000
```

```
00000000 10000012 00020000 31003003 00010000 00010000 00010000
00000000 31003010 01000001 01010000 00000000 01010000 00000000
10000021 01020000 10213310 00020000 01000002 00000000 10000022
00020000 20000103 00010000 00010000 00020000 31002012 00010000
00010000 00000000 31003021 01000001 01010000 00000000 01010000
00000000 10000013 01020000 12030132 00020000 01000002 00000000
10000022 00020000 20000111 00010000 00010000 00020000 31002033
00010000 00010000 00010000 00010000 00000000 31020011 01000001
00000000 00000000 01001000 00000000 01010000 00000000 10000012
00020000 31003001 00010000 00020000 01010000 00000000 10000013
00020000 31003002 00010000 00020000 01010000 00000000 10000012
00020000 31003003 00010000 00010000 00010000 00000000 31003010
01000001 01010000 00000000 01010000 00000000 10000021 01020000
13223303 00020000 01000002 00000000 10000022 00020000 20000103
00010000 00010000 00020000 31002012 00010000 00010000 00010000
00010000 00000000 31020012 01000001 00000000 00000000 01001000
00000000 01010000 00000000 10000013 00020000 31003001 00010000
00020000 01010000 00000000 10000013 00020000 31003002 00010000
00020000 01010000 00000000 10000013 00020000 31003003 00010000
00010000 00010000 00000000 31003010 01000001 01010000 00000000
01010000 00000000 10000001 01020000 10101200 00020000 01000002
00000000 10000022 00020000 20000102 00010000 00010000 00020000
31002012 00010000 00010000 00010000 00010000 00000000 31020013
01000001 00000000 00000000 01001000 00000000 01010000 00000000
10000013 00020000 31003001 00010000 00020000 01010000 00000000
10000020 00020000 31003002 00010000 00020000 01010000 00000000
10000013 00020000 31003003 00010000 00010000 00010000 00000000
31003010 01000001 01010000 00000000 01010000 00000000 10000001
01020000 10210333 00020000 01000002 00000000 10000022 00020000
20000102 00010000 00010000 00020000 31002012 00010000 00010000
00010000 00010000 00000000 31020020 01000001 00000000 00000000
01001000 00000000 01010000 00000000 10000020 00020000 31003001
00010000 00020000 01010000 00000000 10000020 00020000 31003002
00010000 00020000 01010000 00000000 10000020 00020000 31003003
00010000 00010000 00010000 00000000 31003010 01000001 01010000
00000000 01010000 00000000 10000001 01020000 10320122 00020000
01000002 00000000 10000022 00020000 20000102 00010000 00010000
00020000 31002012 00010000 00010000 00010000 00010000 00000000
31020021 01000001 00000000 00000000 01001000 00000000 01010000
00000000 10000020 00020000 31003001 00010000 00020000 01010000
00000000 10000021 00020000 31003002 00010000 00020000 01010000
00000000 10000020 00020000 31003003 00010000 00010000 00010000
00000000 31003010 01000001 01010000 00000000 01010000 00000000
10000001 01020000 11320003 00020000 01000002 00000000 10000022
00020000 20000102 00010000 00010000 00020000 31002012 00010000
00010000 00010000 00010000 00000000 31020022 01000001 00000000
00000000 01001000 00000000 01010000 00000000 10000020 00020000
31003001 00010000 00020000 01010000 00000000 10000022 00020000
```

```
31003002 00010000 00020000 01010000 00000000 10000020 00020000
31003003 00010000 00010000 00010000 00000000 31003010 01000001
01010000 00000000 01010000 00000000 10000001 01020000 11133210
00020000 01000002 00000000 10000022 00020000 20000102 00010000
00010000 00020000 31002012 00010000 00010000 00010000 00010000
00000000 31020023 01000001 00000000 00000000 01001000 00000000
01010000 00000000 10000022 00020000 31003001 00010000 00020000
01010000 00000000 10000022 00020000 31003002 00010000 00020000
01010000 00000000 10000022 00020000 31003003 00010000 00010000
00010000 00000000 31003010 01000001 01010000 00000000 01010000
00000000 10000001 01020000 12012200 00020000 01000002 00000000
10000022 00020000 20000102 00010000 00010000 00020000 31002012
00010000 00010000 00010000 00010000 00000000 31020030 01000001
00000000 00000000 01001000 00000000 01010000 00000000 10000022
00020000 31003001 00010000 00020000 01010000 00000000 10000023
00020000 31003002 00010000 00020000 01010000 00000000 10000022
00020000 31003003 00010000 00010000 00010000 00000000 31003010
01000001 01010000 00000000 01010000 00000000 10000001 01020000
12122003 00020000 01000002 00000000 10000022 00020000 20000102
00010000 00010000 00020000 31002012 00010000 00010000 00010000
00010000 00000000 31020031 01000001 00000000 00000000 01001000
00000000 01010000 00000000 10000022 00020000 31003001 00010000
00020000 01010000 00000000 10000030 00020000 31003002 00010000
00020000 01010000 00000000 10000022 00020000 31003003 00010000
00010000 00010000 00000000 31003010 01000001 01010000 00000000
01010000 00000000 10000001 01020000 12231202 00020000 01000002
00000000 10000022 00020000 20000102 00010000 00010000 00020000
31002012 00010000 00010000 00010000 00010000 00000000 31020032
01000001 00000000 00000000 01001000 00000000 01010000 00000000
10000022 00020000 31003001 00010000 00020000 01010000 00000000
10000031 00020000 31003002 00010000 00020000 01010000 00000000
10000022 00020000 31003003 00010000 00010000 00010000 00000000
31003010 01000001 01010000 00000000 01010000 00000000 10000001
01020000 13001003 00020000 01000002 00000000 10000022 00020000
20000102 00010000 00010000 00020000 31002012 00010000 00010000
00010000 00010000 00000000 31020033 01000001 00000000 00000000
01001000 00000000 01010000 00000000 10000030 00020000 31003001
00010000 00020000 01010000 00000000 10000030 00020000 31003002
00010000 00020000 01010000 00000000 10000030 00020000 31003003
00010000 00010000 00010000 00000000 31003010 01000001 01010000
00000000 01010000 00000000 10000001 01020000 13110132 00020000
01000002 00000000 10000022 00020000 20000102 00010000 00010000
00020000 31002012 00010000 00010000 00010000 00010000 00000000
31020100 01000001 00000000 00000000 01001000 00000000 01010000
00000000 10000032 00020000 31003001 00010000 00020000 01010000
00000000 10000032 00020000 31003002 00010000 00020000 01010000
00000000 10000032 00020000 31003003 00010000 00010000 00010000
00000000 31003010 01000001 01010000 00000000 01010000 00000000
```

```
10000002 01020000 10202130 00020000 01000002 00000000 10000022
00020000 20000102 00010000 00010000 00020000 31002012 00010000
00010000 00010000 00010000 00000000 31020101 01000001 00000000
00000000 01001000 00000000 01010000 00000000 10000100 00020000
31003001 00010000 00020000 01010000 00000000 10000100 00020000
31003002 00010000 00020000 01010000 00000000 10000100 00020000
31003003 00010000 00010000 00010000 00000000 31003010 01000001
01010000 00000000 01010000 00000000 10000002 01020000 11300132
00020000 01000002 00000000 10000022 00020000 20000102 00010000
00010000 00020000 31002012 00010000 00010000 00010000 00010000
00000000 31020102 01000001 00000000 00000000 01001000 00000000
01010000 00000000 10000102 00020000 31003001 00010000 00020000
01010000 00000000 10000112 00020000 31003002 00010000 00020000
01010000 00000000 10000102 00020000 31003003 00010000 00010000
00010000 00000000 31003010 01000001 01010000 00000000 01010000
00000000 10000003 01020000 10030202 00020000 01000002 00000000
10000022 00020000 20000102 00010000 00010000 00020000 31002012
00010000 00010000 00010000 00010000 00000000 31020103 01000001
00000000 00000000 01001000 00000000 01010000 00000000 10000122
00020000 31003001 00010000 00020000 01010000 00000000 10000132
00020000 31003002 00010000 00020000 01010000 00000000 10000122
00020000 31003003 00010000 00010000 00010000 00000000 31003010
01000001 01010000 00000000 01010000 00000000 10000010 01020000
11010201 00020000 01000002 00000000 10000022 00020000 20000102
00010000 00010000 00020000 31002012 00000000 10000002 01001001
10000001 00010000 00000000 10000001 01001001 10000001 00010000
00000000 10000101 01001001 10000001 00010000 00000000 10011111
01001001 10011111 00010000 00000000 10100220 01001001 10011111
00010000 00000000 10000001 01001010 10000001 00010000 00000000
10000000 01001010 10000001 00010000 00000000 20000101 01001010
10000001 00010000 00000000 20000313 01001010 10000000 00010000
00000000 10001001 01001010 10001001 00010000 00000000 20002013
01001010 10000000 00010000 00000000 31330100 01000001 00000000
00000000 31003010 01000001 01001000 00000000 01001000 00000000
10000001 01020000 12010123 00020000 01000002 00000000 10000022
00020000 10000133 00010000 00010000 00020000 31002012 00010000
00010000 00000000 31003023 01001010 01001000 00000000 01001000
00000000 10000010 01020000 13300311 00020000 01000002 00000000
10000022 00020000 20000132 00010000 00010000 00020000 31002100
00010000 00010000 00000000 31003023 01001001 01001000 00000000
01001000 00000000 10000011 01020000 12211202 00020000 01000002
00000000 10000022 00020000 20000133 00010000 00010000 00020000
31002100 00010000 00010000 00000000 31003012 01000001 01001000
00000000 01001000 00000000 10000001 01020000 12002123 00020000
01000002 00000000 10000022 00020000 10000221 00010000 00010000
00020000 31002031 00010000 00010000 00000000 31003033 01000001
01001000 00000000 01001000 00000000 10000003 01020000 12111033
00020000 01000002 00000000 10000022 00020000 20000203 00010000
```

```
00010000 00020000 31002031 00010000 00010000 00000000 31003100
01000001 10000000 01020000 10310221 00010000 00000000 31003101
01000001 01001000 00000000 01001000 00000000 10000001 01020000
11220322 00020000 01000002 00000000 10000022 00020000 20000302
00010000 00010000 00020000 31002100 00010000 00010000 00000000
31003102 01000001 01001000 00000000 01001000 00000000 10000021
01020000 11212003 00020000 01000002 00000000 10000022 00020000
20000301 00010000 00010000 00020000 31002100 00010000 00010000
00000000 31003103 00000000 31130000 00010000 00010000 00000000
31003110 01000001 00000000 00000000 31003010 01000001 01001000
00000000 01001000 00000000 10000021 01020000 10311301 00020000
01000002 00000000 10000022 00020000 10000022 00010000 00010000
00020000 31002012 00010000 00010000 00000000 31003010 01000001
01001000 00000000 31003010 00020000 01001000 00000000 01001000
00000000 10000000 01020000 11223201 00020000 31021003 00010000
00020000 01001000 00000000 10000000 01020000 11022033 00020000
31020032 00010000 00020000 01001000 00000000 10000000 01020000
10320110 00020000 31021001 00010000 00020000 01001000 00000000
10000000 01020000 10033113 00020000 31020003 00010000 00020000
10000000 01020000 10002210 00010000 00010000 00010000 00010000
00010000 00010000 00010000 00000000 31130200 01000001 00000000
00000000 31003010 01000001 01001000 00000000 01001000 00000000
10000002 01020000 10330231 00020000 01000002 00000000 10000022
00020000 10000200 00010000 00010000 00020000 31002012 00010000
00010000 00000000 31003023 01000001 01001000 00000000 01001000
00000000 10000011 01020000 10300011 00020000 01000002 00000000
10000022 00020000 20000132 00010000 00010000 00020000 31002100
00010000 00010000 00000000 31003012 01000001 01001000 00000000
01001000 00000000 10000003 01020000 12332203 00020000 01000002
00000000 10000022 00020000 20000221 00010000 00010000 00020000
31002031 00010000 00010000 00000000 31003033 01000001 01001000
00000000 01001000 00000000 10000012 01020000 12301330 00020000
01000002 00000000 10000022 00020000 20000231 00010000 00010000
00020000 31002031 00010000 00010000 00000000 31003100 01000001
10000000 01020000 10001233 00010000 00000000 31003101 01000001
01001000 00000000 01001000 00000000 10000003 01020000 10121320
00020000 01000002 00000000 10000022 00020000 10000302 00010000
00010000 00020000 31002100 00010000 00010000 00000000 31003102
01000001 01001000 00000000 01001000 00000000 10000003 01020000
13211001 00020000 01000002 00000000 10000022 00020000 10000302
00010000 00010000 00020000 31002100 00010000 00010000 00000000
31003103 00000000 31130000 00010000 00010000 00000000 31003110
01000001 00000000 00000000 31003010 01000001 01001000 00000000
01001000 00000000 10000002 01020000 10310211 00020000 01000002
00000000 10000022 00020000 10000130 00010000 00010000 00020000
31002012 00010000 00010000 00000000 31003010 01000001 01001000
00000000 31003010 00020000 01001000 00000000 01001000 00000000
10000000 01020000 13313003 00020000 31021020 00010000 00020000
```

```
01001000 00000000 10000000 01020000 10020331 00020000 31021011
00010000 00020000 01001000 00000000 01001000 00000000 10000001
01020000 12000000 00020000 01000002 00000000 10000022 00020000
20000010 00010000 00010000 00020000 31021021 00010000 00020000
01001000 00000000 01001000 00000000 10000013 00020000 01000002
00000000 10000022 00020000 20000011 00010000 00010000 00020000
31020102 00010000 00020000 01001000 00000000 01001000 00000000
10000001 01020000 12303031 00020000 01000002 00000000 10000022
00020000 20000011 00010000 00010000 00020000 31021013 00010000
00020000 01001000 00000000 01001000 00000000 10000001 01020000
10303031 00020000 01000002 00000000 10000022 00020000 20000011
00010000 00010000 00020000 31020003 00010000 00020000 01001000
00000000 01001000 00000000 10000013 00020000 01000002 00000000
10000022 00020000 20000012 00010000 00010000 00020000 31020023
00010000 00010000 00010000 00010000 00010000 00010000 00010000
00010000 00000000 31130300 01000001 00000000 00000000 31003010
01000001 01001000 00000000 01001000 00000000 10000002 01020000
12332231 00020000 01000002 00000000 10000022 00020000 10000200
00010000 00010000 00020000 31002012 00010000 00010000 00000000
31003023 01001010 01001000 00000000 01001000 00000000 10000002
01020000 11112012 00020000 01000002 00000000 10000022 00020000
20000132 00010000 00010000 00020000 31002100 00010000 00010000
00000000 31003023 01001001 01001000 00000000 01001000 00000000
10000001 01020000 11030132 00020000 01000002 00000000 10000022
00020000 20000132 00010000 00010000 00020000 31002100 00010000
00010000 00000000 31003012 01001010 01001000 00000000 01001000
00000000 10000003 01020000 13302033 00020000 01000002 00000000
10000022 00020000 10000230 00010000 00010000 00020000 31002031
00010000 00010000 00000000 31003012 01001001 01001000 00000000
01001000 00000000 10000003 01020000 13232313 00020000 01000002
00000000 10000022 00020000 10000230 00010000 00010000 00020000
31002031 00010000 00010000 00000000 31003033 01000001 01001000
00000000 01001000 00000000 10000021 01020000 11001200 00020000
01000002 00000000 10000022 00020000 10000231 00010000 00010000
00020000 31002031 00010000 00010000 00000000 31003100 01000001
10000000 01020000 10010102 00010000 00000000 31003101 01000001
01001000 00000000 01001000 00000000 10000011 01020000 13122201
00020000 01000002 00000000 10000022 00020000 10000302 00010000
00010000 00020000 31002100 00010000 00010000 00000000 31003102
01000001 01001000 00000000 01001000 00000000 10000001 01020000
12121021 00020000 01000002 00000000 10000022 00020000 10000300
00010000 00010000 00020000 31002100 00010000 00010000 00000000
31003103 00000000 31130000 00010000 00010000 00000000 31003032
00000000 31130303 00010000 00010000 00000000 31003010 01000001
01001000 00000000 31003010 00020000 01001000 00000000 01001000
00000000 10000000 01020000 11023122 00020000 31020103 00010000
00020000 01001000 00000000 10000000 01020000 11030303 00020000
31020020 00010000 00020000 01001000 00000000 10000000 01020000
```

```
10203311 00020000 31020100 00010000 00020000 01001000 00000000
10000000 01020000 10130022 00020000 31020033 00010000 00020000
01001000 00000000 10000000 01020000 10013023 00020000 31020101
00010000 00020000 10000000 01020000 10211220 00010000 00010000
00010000 00000000 31003110 01000001 00000000 00000000 31003010
01001010 01001000 00000000 01001000 00000000 10000011 01020000
10211231 00020000 01000002 00000000 10000022 00020000 10000102
00010000 00010000 00020000 31002012 00010000 00010000 00000000
31003010 01001001 01001000 00000000 01001000 00000000 10000011
01020000 10212011 00020000 01000002 00000000 10000022 00020000
10000102 00010000 00010000 00020000 31002012 00010000 00010000
00000000 31003010 01000001 01001000 00000000 31003010 00020000
01001000 00000000 01001000 00000000 10000000 01020000 12013012
00020000 31021011 00010000 00020000 01001000 00000000 10000000
01020000 10311120 00020000 31021003 00010000 00020000 01001000
00000000 10000000 01020000 10002103 00020000 31020102 00010000
00020000 01001000 00000000 10000000 01020000 10001001 00020000
31021002 00010000 00020000 01001000 00000000 01001000 00000000
01001001 10000003 01020000 13110133 00020000 01000002 00000000
10000022 00020000 20000010 00010000 00010000 00020000 31021020
00010000 00020000 01001000 00000000 01001000 00000000 10000001
01020000 13033213 00020000 01000002 00000000 10000022 00020000
20000011 00010000 00010000 00020000 31020023 00010000 00020000
01001000 00000000 01001000 00000000 10000011 01020000 10320110
00020000 01000002 00000000 10000022 00020000 20000012 00010000
00010000 00020000 31020003 00010000 00020000 01001000 00000000
01001000 00000000 01001001 10000001 01020000 12332232 00020000
01000002 00000000 10000022 00020000 20000012 00010000 00010000
00020000 31021023 00010000 00020000 01001000 00000000 01001000
00000000 10000003 01020000 12000000 00020000 01000002 00000000
10000022 00020000 20000020 00010000 00010000 00020000 31021010
00010000 00010000 00010000 00010000 00010000 00010000 00000000
31003111 00010000 00000000 31003112 00010000 00010000 00010000
00000000 31130303 01000001 00000000 00000000 31003010 01000001
01001000 00000000 01001000 00000000 10000003 01020000 11200101
00020000 01000002 00000000 10000022 00020000 10000132 00010000
00010000 00020000 31002012 00010000 00010000 00000000 31003023
01001010 01001000 00000000 01001000 00000000 10000002 01020000
13000231 00020000 01000002 00000000 10000022 00020000 10000132
00010000 00010000 00020000 31002100 00010000 00010000 00000000
31003023 01001001 01001000 00000000 01001000 00000000 10000010
01020000 13300311 00020000 01000002 00000000 10000022 00020000
20000133 00010000 00010000 00020000 31002100 00010000 00010000
00000000 31003012 01001001 01001000 00000000 01001000 00000000
10000001 01020000 10102332 00020000 01000002 00000000 10000022
00020000 10000221 00010000 00010000 00020000 31002031 00010000
00010000 00000000 31003012 01001010 01001000 00000000 01001000
00000000 10000001 01020000 10103110 00020000 01000002 00000000
```

```
10000022 00020000 10000221 00010000 00010000 00020000 31002031
00010000 00010000 00000000 31003033 01000001 01001000 00000000
01001000 00000000 10000002 01020000 11200312 00020000 01000002
00000000 10000022 00020000 10000223 00010000 00010000 00020000
31002031 00010000 00010000 00000000 31003100 01000001 10000000
01020000 10032003 00010000 00000000 31003101 01000001 01001000
00000000 01001000 00000000 10000010 01020000 11200323 00020000
01000002 00000000 10000022 00020000 10000301 00010000 00010000
00020000 31002100 00010000 00010000 00000000 31003102 01000001
01001000 00000000 01001000 00000000 10000010 01020000 11200323
00020000 01000002 00000000 10000022 00020000 10000301 00010000
00010000 00020000 31002100 00010000 00010000 00000000 31003103
00000000 31130300 00010000 00010000 00000000 31003110 01000001
00000000 00000000 31003010 01000001 01001000 00000000 01001000
00000000 10000003 01020000 11010012 00020000 01000002 00000000
10000022 00020000 10000023 00010000 00010000 00020000 31002012
00010000 00010000 00010000 00010000 00010000 00010000 00000000
31131000 01000001 00000000 00000000 31003010 01000001 01001000
00000000 01001000 00000000 10000002 01020000 13303002 00020000
01000002 00000000 10000022 00020000 10000133 00010000 00010000
00020000 31002012 00010000 00010000 00000000 31003023 01001010
01001000 00000000 01001000 00000000 10000001 01020000 13210100
00020000 01000002 00000000 10000022 00020000 20000132 00010000
00010000 00020000 31002100 00010000 00010000 00000000 31003023
01001001 01001000 00000000 01001000 00000000 10000001 01020000
11100011 00020000 01000002 00000000 10000022 00020000 20000132
00010000 00010000 00020000 31002100 00010000 00010000 00000000
31003012 01001010 01001000 00000000 01001000 00000000 10000002
01020000 10121323 00020000 01000002 00000000 10000022 00020000
10000221 00010000 00010000 00020000 31002031 00010000 00010000
00000000 31003012 01001001 01001000 00000000 01001000 00000000
10000002 01020000 10112301 00020000 01000002 00000000 10000022
00020000 10000221 00010000 00010000 00020000 31002031 00010000
00010000 00000000 31003033 01000001 01001000 00000000 01001000
00000000 10000001 01020000 11221002 00020000 01000002 00000000
10000022 00020000 10000232 00010000 00010000 00020000 31002031
00010000 00010000 00000000 31003100 01000001 10000000 01020000
10113322 00010000 00000000 31003101 01000001 01001000 00000000
01001000 00000000 10000001 01020000 10121312 00020000 01000002
00000000 10000022 00020000 10000303 00010000 00010000 00020000
31002100 00010000 00010000 00000000 31003102 01000001 01001000
00000000 01001000 00000000 10000001 01020000 12210323 00020000
01000002 00000000 10000022 00020000 10000300 00010000 00010000
00020000 31002100 00010000 00010000 00000000 31003103 00000000
31130000 00010000 00010000 00000000 31003110 01000001 00000000
00000000 31003010 01000001 01001000 00000000 01001000 00000000
10000001 01020000 10212010 00020000 01000002 00000000 10000022
00020000 10000120 00010000 00010000 00020000 31002012 00010000
```

```
00010000 00000000 31003010 01000001 01001000 00000000 31003010
00020000 01001000 00000000 01001000 00000000 10000000 01020000
13310001 00020000 31021020 00010000 00020000 01001000 00000000
10000000 01020000 10010013 00020000 31021011 00010000 00020000
01001000 00000000 10000000 01020000 10010012 00020000 31020102
00010000 00020000 01001000 00000000 10000000 01020000 10000111
00020000 31021003 00010000 00020000 01001000 00000000 10000000
01020000 10000023 00020000 31021013 00010000 00020000 01001000
00000000 10000000 01020000 10000011 00020000 31021002 00010000
00020000 01001000 00000000 01001000 00000000 10000001 01020000
11030303 00020000 01000002 00000000 10000022 00020000 20000010
00010000 00010000 00020000 31021012 00010000 00020000 01001000
00000000 01001000 00000000 10000002 01020000 12000000 00020000
01000002 00000000 10000022 00020000 20000012 00010000 00010000
00020000 31020023 00010000 00010000 00010000 00010000 00010000
00010000 00010000 00010000 00000000 31131100 01000001 00000000
00000000 31003010 01000001 01001000 00000000 01001000 00000000
10000020 01020000 12321110 00020000 01000002 00000000 10000022
00020000 10000202 00010000 00010000 00020000 31002012 00010000
00010000 00000000 31003023 01000001 01001000 00000000 01001000
00000000 10000001 01020000 10222021 00020000 01000002 00000000
10000022 00020000 20000132 00010000 00010000 00020000 31002100
00010000 00010000 00000000 31003012 01001010 01001000 00000000
01001000 00000000 10000010 01020000 11230120 00020000 01000002
00000000 10000022 00020000 10000222 00010000 00010000 00020000
31002031 00010000 00010000 00000000 31003012 01001001 01001000
00000000 01001000 00000000 10000010 01020000 10202322 00020000
01000002 00000000 10000022 00020000 10000222 00010000 00010000
00020000 31002031 00010000 00010000 00000000 31003033 01000001
01001000 00000000 01001000 00000000 10000010 01020000 13101022
00020000 01000002 00000000 10000022 00020000 10000232 00010000
00010000 00020000 31002031 00010000 00010000 00000000 31003100
01000001 10000000 01020000 10030121 00010000 00000000 31003101
01000001 01001000 00000000 01001000 00000000 10000012 01020000
13301012 00020000 01000002 00000000 10000022 00020000 10000303
00010000 00010000 00020000 31002100 00010000 00010000 00000000
31003102 01000001 01001000 00000000 01001000 00000000 10000012
01020000 12200220 00020000 01000002 00000000 10000022 00020000
10000233 00010000 00010000 00020000 31002100 00010000 00010000
00000000 31003103 00000000 31130000 00010000 00010000 00000000
31003010 01000001 01001000 00000000 31003010 00020000 01001000
00000000 01001000 00000000 10000000 01020000 12311311 00020000
31021001 00010000 00020000 01001000 00000000 10000000 01020000
10331130 00020000 31020003 00010000 00020000 10000000 01020000
10030303 00010000 00010000 00010000 00010000 00010000 00000000
31131200 01000001 00000000 00000000 31003010 01000001 01001000
00000000 01001000 00000000 10000002 01020000 12130231 00020000
01000002 00000000 10000022 00020000 10000202 00010000 00010000
```

```
00020000  31002012  00010000  00010000  00000000  31003023  01000001
01001000  00000000  01001000  00000000  10000021  01020000  11311100
00020000  01000002  00000000  10000022  00020000  20000133  00010000
00010000  00020000  31002100  00010000  00010000  00000000  31003012
01001010  01001000  00000000  01001000  00000000  10000003  01020000
12322212  00020000  01000002  00000000  10000022  00020000  10000222
00010000  00010000  00020000  31002031  00010000  00010000  00000000
31003012  01001001  01001000  00000000  01001000  00000000  10000003
01020000  11131011  00020000  01000002  00000000  10000022  00020000
10000222  00010000  00010000  00020000  31002031  00010000  00010000
00000000  31003033  01000001  01001000  00000000  01001000  00000000
10000020  01020000  13132131  00020000  01000002  00000000  10000022
00020000  10000232  00010000  00010000  00020000  31002031  00010000
00010000  00000000  31003100  01000001  10000000  01020000  31131200
00010000  00000000  31003101  01000001  01001000  00000000  01001000
00000000  10000001  01020000  12330122  00020000  01000002  00000000
10000022  00020000  10000310  00010000  00010000  00020000  31002100
00010000  00010000  00000000  31003102  01000001  01001000  00000000
01001000  00000000  10000013  01020000  10131211  00020000  01000002
00000000  10000022  00020000  10000233  00010000  00010000  00020000
31002100  00010000  00010000  00000000  31003103  00000000  31130000
00010000  00010000  00010000  00010000  00000000  31131300  01000001
00000000  00000000  31003010  01000001  01001000  00000000  01001000
00000000  10000003  01020000  13331203  00020000  01000002  00000000
10000002  00020000  10000201  00010000  00010000  00020000  31002012
00010000  00010000  00000000  31003023  01000001  01001000  00000000
01001000  00000000  10000011  01020000  11131100  00020000  01000002
00000000  10000022  00020000  20000133  00010000  00010000  00020000
31002100  00010000  00010000  00000000  31003012  01001010  01001000
00000000  01001000  00000000  10000001  01020000  12110311  00020000
01000002  00000000  10000022  00020000  10000222  00010000  00010000
00020000  31002031  00010000  00010000  00000000  31003012  01001001
01001000  00000000  01001000  00000000  10000001  01020000  12023203
00020000  01000002  00000000  10000022  00020000  10000222  00010000
00010000  00020000  31002031  00010000  00010000  00000000  31003033
01000001  01001000  00000000  01001000  00000000  10000001  01020000
13013221  00020000  01000002  00000000  10000022  00020000  10000233
00010000  00010000  00020000  31002031  00010000  00010000  00000000
31003100  01000001  10000000  01020000  10030011  00010000  00000000
31003101  01000001  01001000  00000000  01001000  00000000  10000010
01020000  13233211  00020000  01000002  00000000  10000022  00020000
10000310  00010000  00010000  00020000  31002100  00010000  00010000
00000000  31003102  01000001  01001000  00000000  01001000  00000000
10000001  01020000  10212231  00020000  01000002  00000000  10000022
00020000  10000300  00010000  00010000  00020000  31002100  00010000
00010000  00000000  31003103  00000000  31130000  00010000  00010000
00010000  00010000  00000000  31132000  01000001  00000000  00000000
31003010  01000001  01001000  00000000  01001000  00000000  10000010
```

```
01020000 12310330 00020000 01000002 00000000 10000022 00020000
10000201 00010000 00010000 00020000 31002012 00010000 00010000
00000000 31003023 01000001 01001000 00000000 01001000 00000000
10000011 01020000 10110332 00020000 01000002 00000000 10000022
00020000 20000133 00010000 00010000 00020000 31002100 00010000
00010000 00000000 31003012 01001010 01001000 00000000 01001000
00000000 10000001 01020000 12020033 00020000 01000002 00000000
10000022 00020000 10000222 00010000 00010000 00020000 31002031
00010000 00010000 00000000 31003012 01001001 01001000 00000000
01001000 00000000 10000001 01020000 12001203 00020000 01000002
00000000 10000022 00020000 10000222 00010000 00010000 00020000
31002031 00010000 00010000 00000000 31003033 01000001 01001000
00000000 01001000 00000000 10000002 01020000 13021110 00020000
01000002 00000000 10000022 00020000 10000233 00010000 00010000
00020000 31002031 00010000 00010000 00000000 31003100 01000001
10000000 01020000 10002031 00010000 00000000 31003101 01000001
01001000 00000000 01001000 00000000 10000021 01020000 12211121
00020000 01000002 00000000 10000022 00020000 10000310 00010000
00010000 00020000 31002100 00010000 00010000 00000000 31003102
01000001 01001000 00000000 01001000 00000000 10000001 01020000
10103122 00020000 01000002 00000000 10000022 00020000 10000300
00010000 00010000 00020000 31002100 00010000 00010000 00000000
31003103 00000000 31130000 00010000 00010000 00010000 00010000
00000000 31021001 01000001 00000000 00000000 01010000 00000000
10000002 00020000 00000000 31020000 01222210 31020001 00010000
00010000 00010000 00010000 00010000 00000000 31021002 01000001
00000000 00000000 01001000 00000000 01010000 00000000 10000002
00020000 00000000 31020000 01222210 31020001 00010000 00010000
00020000 00000000 31020020 01222210 31020021 01222210 31020022
00010000 00010000 00010000 00010000 00010000 00000000 31021003
01000001 00000000 00000000 01010000 00000000 10000002 00020000
00000000 31020020 01222210 31020021 01222210 31020022 00010000
00010000 00010000 00010000 00010000 00000000 31021010 01000001
00000000 00000000 01010000 00000000 10000003 00020000 00000000
31020020 01222210 31020021 01222210 31020022 00010000 00010000
00010000 00010000 00010000 00000000 31021011 01000001 00000000
00000000 01010000 00000000 10000002 00020000 00000000 31020012
01222210 31020013 00010000 00010000 00010000 00010000 00010000
00000000 31021012 01000001 00000000 00000000 01001000 00000000
00000000 31020012 01222210 31020013 00010000 00020000 00000000
31020020 01222210 31020021 01222210 31020022 00010000 00010000
00010000 00010000 00010000 00000000 31021013 01000001 00000000
00000000 01001000 00000000 00000000 31020010 01222210 31020011
00010000 00020000 00000000 31020020 01222210 31020021 01222210
31020022 00010000 00010000 00010000 00010000 00010000 00000000
31021020 01000001 00000000 00000000 01001000 00000000 00000000
31020010 01222210 31020011 00010000 00020000 01010000 00000000
10000002 00020000 00000000 31020020 01222210 31020021 01222210
```

```
31020022 00010000 00010000 00010000 00010000 00010000 00010000
00000000 31021021 01000001 00000000 00000000 01001000 00000000
31020101 00020000 01010000 00000000 10000002 00020000 00000000
31020020 01222210 31020021 01222210 31020022 00010000 00010000
00010000 00010000 00010000 00010000 00000000 31021022 01000001
00000000 00000000 01001000 00000000 00000000 31020012 01222210
31020013 00010000 00020000 01010000 00000000 10000003 00020000
00000000 31020000 01222210 31020001 00010000 00010000 00010000
00010000 00010000 00010000 00000000 31021023 01000001 00000000
00000000 01001000 00000000 00000000 31020010 01222210 31020011
00010000 00020000 01010000 00000000 10000010 00020000 00000000
31020000 01222210 31020001 00010000 00010000 00010000 00010000
00010000 00010000 00000000 31021030 01000001 00000000 00000000
01001000 00000000 01010000 00000000 10000002 00020000 00000000
31020010 01222210 31020011 00010000 00010000 00020000 01010000
00000000 10000012 00020000 00000000 31020000 01222210 31020001
00010000 00010000 00010000 00010000 00010000 00010000 00000000
31021031 01000001 00000000 00000000 01001000 00000000 00000000
31020012 01222210 31020013 00010000 00020000 31020101 00020000
01010000 00000000 10000011 00020000 00000000 31020000 01222210
31020001 00010000 00010000 00010000 00010000 00010000 00010000
```